\documentclass[a4paper,12pt]{article}
\usepackage[a4paper, left=3cm, right=1cm, top=2.5cm, bottom=2.5cm]{geometry}
\usepackage[utf8]{inputenc}
\usepackage[T1]{fontenc}
\usepackage{lipsum} % pour générer du texte factice
\usepackage{subcaption}
\usepackage{amsmath, amsfonts, amsthm, amssymb}
\usepackage{soul}
\usepackage{dsfont}
\usepackage{graphicx}
\usepackage{tikz}
\usepackage{xcolor}
\usepackage{hyperref}
\usepackage{float}
\usepackage{pdflscape} 
\usepackage[toc,page]{appendix}
 
\usepackage{dsfont}
\usepackage{rotating}
\usepackage[authoryear]{natbib}
\newtheorem{defi}{Definition}[section]

\newtheorem{prop}[defi]{Proposition}

\newtheorem{rk}[defi]{Remark}

\newcommand{\Exp}{\mathbb{E}}
\newcommand{\Var}{\mathrm{Var}}
\renewcommand{\Pr}{\mathbb{P}}
\newcommand{\R}{\mathbb{R}}
\newcommand{\N}{\mathbb{N}}

\newcommand{\ind}[1]{\mathds{1}_{\{#1\}}}

\newcommand{\BBB}[1]{{\color{gray}#1}}% Retirer du texte
\title{A tree-based Polynomial Chaos expansion for surrogate modeling and sensitivity analysis of complex numerical models} 

\usepackage{authblk}

\author[1,2]{Faten Ben Said}
\author[1,4]{Aurélien Alfonsi}
\author[5]{Anne Dutfoy}
\author[2,3]{Cédric Goeury}
\author[2,3]{Magali Jodeau}
\author[1]{Julien Reygner}
\author[2]{Fabrice Zaoui}

\affil[1]{CERMICS, ENPC, Institut Polytechnique de Paris, Marne-la-Vallée, France}

\affil[2]{\'{E}lectricité de France (EDF), Research and Development Division, National Laboratory for Hydraulics and Environment (LNHE), Chatou, France %6 Quai Watier, 78400 Chatou, France.
}
\affil[3]{Laboratoire d’Hydraulique Saint-Venant (LHSV), ENPC, Institut Polytechnique de Paris, EDF R\&D, Chatou, France %6 Quai Watier, Chatou, 78400, France
}
\affil[4]{MathRisk, INRIA, Paris, France}
\affil[5]{\'{E}lectricité de France (EDF), Research and Development Division, PErformance et prévention des Risques Industriels du parC par la simuLation et les EtudeS (PERICLES), Paris-Saclay, France.}
\date{}

\begin{document}

\maketitle
\abstract{
This paper introduces Tree-based Polynomial Chaos Expansion (Tree-PCE), a novel surrogate modeling technique designed to efficiently approximate complex numerical models exhibiting nonlinearities and discontinuities. Tree-PCE combines the expressive power of Polynomial Chaos Expansion (PCE) with an adaptive partitioning strategy inspired by regression trees. By recursively dividing the input space into hyperrectangular subdomains and fitting localized PCEs, Tree-PCE constructs a piecewise polynomial surrogate that improves both accuracy and computational efficiency. The method is particularly well-suited for global sensitivity analysis, enabling direct computation of Sobol' indices from local expansion coefficients and introducing a new class of sensitivity indices derived from the tree structure itself. Numerical experiments on synthetic and real-world models, including a 2D morphodynamic case, demonstrate that Tree-PCE offers a favorable balance between accuracy and complexity, especially in the presence of discontinuities. While its performance depends on the compromise between the number of subdomains and the degree of local polynomials, this trade-off can be explored using automated hyperparameter optimization frameworks. This opens promising perspectives for systematically identifying optimal configurations and enhancing the robustness of surrogate modeling in complex systems. 
}
\section{Introduction}

Many physical systems in environmental and engineering sciences exhibit abrupt or discontinuous behavior due to threshold effects, phase transitions, or nonlinear dynamics. Such systems, common in fields like fluid dynamics and fracture mechanics for instance, can respond unpredictably to small input variations, making them particularly difficult to model and analyze. These challenges are especially pronounced when attempting to quantify uncertainty, as traditional uncertainty quantification methods often rely on assumptions of smoothness and regularity in model responses.

Modeling hydro-morphodynamic processes is no exception to these challenges. The uncertainties, associated with morphodynamic models, arise from the complexity of sediment transport dynamics, limited field measurements, data inaccuracies, and difficulties in parameterization \citep{Oliveira_2021}. Addressing them is crucial for improving model reliability and supporting robust decision-making in environmental and engineering contexts.

Among the various uncertainty quantification techniques, Polynomial Chaos Expansion (PCE) has emerged as a powerful tool for propagating uncertainties and performing global sensitivity analysis. Its ability to compute Sobol' indices directly from expansion coefficients makes it particularly attractive for assessing the influence of uncertain parameters \citep{Sudret2008, Crestaux2009}. In parallel, several efficient sampling-based strategies have been developed to improve Sobol’ indices computation, such as the adaptive use of replicated Latin Hypercube Designs \citep{Damblin2021} and modularized first-order estimators \citep{Li2016}. When a reliable surrogate can be constructed, PCE remains among the most efficient options for estimating Sobol’ indices.  Despite its strengths, the practical application of classical PCE methods can be limited by computational challenges, particularly in high-dimensional and non-smooth problems. To address this, the development of adaptive-sparse PCE has significantly extended the method’s applicability to complex, real-world problems by improving computational efficiency \citep{Baltman2010}. Yet, applying PCE to sediment transport models remains challenging due to the presence of nonlinearities and discontinuities, which violate the smoothness assumptions underlying classical PCE. Capturing such behavior often requires high-order expansions, increasing computational cost and the risk of overfitting, while abrupt transitions can degrade the accuracy of global polynomial approximations.

Several PCE extensions of varying complexity have been proposed in the literature to address the challenges of PCE-based metamodeling in the presence of nonlinearities. These methods can be broadly classified into two principal categories: Representation-Enhanced PCE Methods and Domain-Decomposed PCE Methods.

Representation-Enhanced PCE Methods aim to improve the expressiveness or adaptability of the PCE representation. \citet{LeMaitre2001} proposed a wavelet-based PCE using the Haar basis to better capture sharp or discontinuous features in model responses. While the method demonstrated robustness to parameter shocks, its convergence was slower and highly dependent on the alignment between discontinuities and the Haar basis intervals. \citet{PoetteLucor2012} introduced an iterative polynomial method where successive PCEs are performed between the metamodel from the previous iteration and the output variable, starting from a classical input-output PCE, and repeated until convergence. \citet{Feinberg_2018} tackled discontinuous variables by transforming them into smoother ones, but this often led to dependent parameters, adding complexity.

Domain-Decomposed PCE Methods address sharp variations by partitioning the input space into subdomains where local PCEs are applied. Machine learning-based approaches, such as the Mixture-of-Experts model as proposed by \citet{Elgarnoussi2020} and SVM-based partitioning by \citet{Meng2021}, identify smooth regions for localized surrogate modeling. Multi-Element PCE methods \citep{Wan2005,Foo2008,Dreau2023} divide the space into hyperrectangles or discontinuity-aligned subdomains using edge detection techniques. 

While these methods effectively capture local behavior and improve accuracy, they often sacrifice the direct computation of sensitivity measures, limiting their use in global sensitivity analysis. Recent efforts aim to reconcile the need for discontinuity handling with the preservation of sensitivity analysis capabilities. In this spirit, several advanced domain-decomposed approaches have emerged. \citet{Vauchel2022} introduced a non-intrusive multi-element PCE method that uses derivative-based agglomerative clustering to partition the parameter space and manage irregular or discontinuous outputs. Although it enables sensitivity analysis via Sobol' indices, it relies on canonical basis transformations, resulting in a complex analytical framework. Similarly, the Arbitrary Multi-Resolution Polynomial Chaos (aMR-PC) method by \citet{Kroker2023} employs hierarchical dyadic decomposition to enhance approximation accuracy and derive analytical Sobol' indices. However, its hierarchical structure leads to exponential growth in expansion terms, increasing computational cost. In the same vein, the Stochastic Spectral Embedding (SSE) method \citep{Marelli2021} recursively partitions the input space and construct local polynomial approximations, offering a balanced trade-off between accuracy and sensitivity analysis capabilities. However, its fixed binary splitting based on equal probability mass can hinder convergence near discontinuities and lead to excessive partitioning, increasing computational cost and complexity.

Despite these recent advances, significant challenges persist in balancing computational efficiency, model accuracy, and sensitivity analysis capabilities particularly when dealing with strong nonlinearities and irregular model behaviors. To address these issues, we introduce the Tree-based Polynomial Chaos Expansion (Tree-PCE), a novel metamodeling approach designed to efficiently and accurately approximate complex computational models while preserving robust global sensitivity analysis. Tree-PCE builds upon the expressive framework of Polynomial Chaos Expansion (PCE) and incorporates an adaptive partitioning strategy inspired by regression trees. The method adaptively partitions the input space into hyperrectangular subdomains, each reflecting distinct model behavior. At each step, it selects the split that most reduces global residual error, forming a binary tree of localized PCEs. This targeted refinement minimizes unnecessary complexity. The final surrogate model, composed of these local expansions, offers a piecewise polynomial approximation that improves both accuracy and computational efficiency. Tree-PCE  supports efficient global sensitivity analysis in two ways. First, it enables direct computation of Sobol' indices using analytical formulas derived from its orthogonal local expansions. Second, its hierarchical structure naturally provides sensitivity metrics that quantify each input variable’s contribution to output variability, allowing for effective input ranking and screening without additional computational cost.

The paper is organized as follows. Section \ref{sec2} recalls the fundamental principles of the standard Polynomial Chaos Expansion (PCE). In Section \ref{sec3}, we introduce the Tree-PCE approach and describe the algorithm used to detect irregularities and decompose the input domain. Section \ref{sec4} presents numerical examples and illustrations that highlight the behavior of the proposed method. Section \ref{sec5} presents formulas for computing Sobol' indices from the Tree-PCE surrogate and introduces a hierarchical sensitivity framework for efficient input ranking and screening. Finally, Section \ref{sec6} applies the proposed method to a hydro-morphodynamic case study, comparing the performance of standard PCE and Tree-PCE.

\section{Standard Polynomial Chaos Expansion (PCE)}\label{sec2}
\subsection{Model and Orthonormal Expansion}

Let \( \mathbf{X} = (X_1, X_2, \dots, X_d) \in \mathbb{R}^d \) denote the input vector, consisting of \( d \) random variables. Each component $X_i$ takes its values in some (possibly unbounded) interval $\mathcal{D}_{X_i}$, and we denote by
\[
    \mathcal{D}_\mathbf{X} = \prod_{i=1}^d \mathcal{D}_{X_i} \subset \mathbb{R}^d
\]
the (possibly unbounded) $d$-dimensional rectangle in which $\textbf{X}$ takes its values. The model function \( G \) maps the input \( \mathbf{X} \) to an output \( Y = G(\mathbf{X}) \), where
\[
G : \mathbf{x} \in \mathcal{D}_{\mathbf{X}} \longmapsto y = G(\mathbf{x}) \in \mathbb{R}.
\]
The input parameters are modeled as a random vector with a prescribed probability distribution $P_\mathbf{X}$, which is assumed to admit a joint probability density function (PDF) \( f_\mathbf{X} \) on $\mathcal{D}_\mathbf{X}$. Consequently, the quantity of interest, \( Y = G(\mathbf{X}) \), is a random variable obtained by propagating the uncertainty in \( \mathbf{X} \) through \( G \). Assuming that $\mathbb{E}[Y^2] < \infty$,
the Polynomial Chaos Expansion (PCE) approach based on the seminal work of~\citep{wiener1938homogeneous} and developed by~\cite{XiKa} in the context of uncertainty analysis relies on the expansion of the stochastic response of the model \(Y\) on a basis of functions
$(\Psi_{\boldsymbol{\alpha}})_{\boldsymbol{\alpha} \in \N^d}$ which is assumed to be orthonormal in $L^2(\R^d,P_\mathbf{X})$, namely:
\begin{equation}\label{eq:decompY}
    Y = G(\mathbf{X}) = \sum_{\boldsymbol{\alpha} \in \N^d} y_{\boldsymbol{\alpha}} \Psi_{\boldsymbol{\alpha}}(\mathbf{X}),
\end{equation}
where, for each $\boldsymbol{\alpha} \in \N^d$, $\Psi_{\boldsymbol{\alpha}}$ is a multivariate function parametrized by $\boldsymbol{\alpha}$ and $(y_{\boldsymbol{\alpha}})_{\boldsymbol{\alpha} \in \N^d}$ are the coefficients of $G$ in this basis, defined by $y_{\boldsymbol{\alpha}} = \mathbb{E}[Y \Psi_{\boldsymbol{\alpha}}(\mathbf{X})]$.

\subsection{Construction of the Multivariate Basis}
\label{ss:construction-multivariate-basis}

Let us first assume that the input components $X_i$ are independent, so that the joint PDF of $\mathbf{X}$ factorizes as:
\[
f_{\mathbf{X}}(\mathbf{x}) = \prod_{i=1}^d f_{X_i}(x_i), \qquad \mathbf{x} = (x_1, \ldots, x_d) \in \mathcal{D}_\mathbf{X},
\]
where \( f_{X_i} \) is the PDF of the marginal distribution $P_{X_i}$ of \( X_i \) over \(\mathcal{D}_{X_i}\). This factorization allows the construction of the multivariate orthonormal basis as a tensor product of univariate polynomials:
\[
\Psi_{\boldsymbol{\alpha}}(\mathbf{x}) = \prod_{i=1}^d \phi_{\alpha_i}^{(i)}(x_i), \qquad \boldsymbol{\alpha} = (\alpha_1, \ldots, \alpha_d) \in \N^d,
\]
where, for each $i \in \{1, \ldots, d\}$, $(\phi^{(i)}_\alpha)_{\alpha \in \N}$ is a basis of univariate polynomials which is orthonormal in $L^2(\R,P_{X_i})$, and such that $\phi^{(i)}_\alpha$ has degree $\alpha$. For most standard distributions, associated polynomial families are available in the literature \citep{xiu2002wiener}, such as Legendre polynomials for uniform distribution.

For dependent input components, the multivariate orthonormal basis can take the form \citep[Lemma 1]{Soize2004} and \citep[Eq.~(4.20)] {Sudret2007}:
\[
\Psi_{\boldsymbol{\alpha}}(\mathbf{x}) = \left(\prod_{i=1}^d \phi_{\alpha_i}^{(i)}(x_i)\right) \sqrt{\frac{f_{X_1}(x_1) \cdots f_{X_d}(x_d)}{f_{\mathbf{X}}(\mathbf{x})}} 
= \frac{\prod_{i=1}^d \phi_{\alpha_i}^{(i)}(x_i)}{\sqrt{c_\mathbf{X}(F_{X_1}(x_1), \dots, F_{X_d}(x_d))}},
\]
where, for each $i \in \{1, \ldots, d\}$, $(\phi^{(i)}_\alpha)_{\alpha \in \N}$ remains a basis of univariate polynomials which is orthonormal in $L^2(\R,P_{X_i})$, and  \( c_\mathbf{X} \) is the copula density defined as:
\[
c_\mathbf{X}(u_1, \dots, u_d) = \frac{\partial^d C_\mathbf{X}(u_1, \dots, u_d)}{\partial u_1 \cdots \partial u_d},
\]
with \( C_\mathbf{X} : [0,1]^d \rightarrow [0,1] \) being the copula function that captures the dependence structure between the input variables, namely the multivariate cumulative distribution function of the random vector $(F_{X_1}(X_1), \ldots, F_{X_d}(X_d))$. 
However, due to the copula correction, the function \( \Psi_{\boldsymbol{\alpha}} \) is no longer a polynomial in \( \mathbf{x} \). Alternatively, in order to build
the multivariate polynomial basis as the tensorized product of univariate polynomial basis, we can
map the dependent random vector X to a new random vector $\mathbf{Z} = T( \mathbf{X} )$ with independent components using the \citet{Nataf1962} or generalized Nataf \citep{DutfoyLebrun2009a}, \citep{DutfoyLebrun2009b}
or \citet{Rosenblatt1952} transformation. Then, we can build PCE in the transformed \( \mathbf{Z} \)-space using
this tensorized multivariate polynomials basis and compose the meta model with the isoprobabilistic
transformation to get the meta model in the original space.
  
\subsection{Truncation Scheme and Coefficients Computation}  
The expansion~\eqref{eq:decompY} contains infinitely many terms and is not computationally tractable. Therefore, the cornerstone of PCE is the approximation of $Y$ by a truncation of this series, retaining only finitely many terms:
\[
    Y \simeq \sum_{\boldsymbol{\alpha} \in \mathcal{A}} y_{\boldsymbol{\alpha}} \Psi_{\boldsymbol{\alpha}}(\mathbf{X}),
\]
where $\mathcal{A}$ is a finite subset of $\N^d$ called the \emph{truncation scheme}. To describe how to select this truncation scheme, let us first define, for any $\boldsymbol{\alpha} = (\alpha_1, \ldots, \alpha_d) \in \N^d$,
\[
|\boldsymbol{\alpha}| = \sum_{i=1}^d \alpha_i.
\]
The standard truncation scheme, known as the \textit{linear enumeration strategy}, retains all polynomials whose total degree \(|\boldsymbol{\alpha}|\) does not exceed a prescribed value \( p \). This leads to the set
\[
\mathcal{A}^{d,p} = \{\boldsymbol{\alpha} \in \mathbb{N}^d \mid |\boldsymbol{\alpha}| \leq p\},
\]
with cardinality
\[
\text{card}(\mathcal{A}^{d,p}) = \binom{d+p}{p} = \frac{(d+p)!}{d! \, p!}.
\]
Alternative chaos basis enumeration strategies exist, such as \emph{hyperbolic}, \emph{anisotropic}, \emph{infinity-norm}, and \emph{low-rank index sets}, which offer more flexible selection criteria depending on the problem at hand \citep{blatman2009}. Once the truncation scheme $\mathcal{A}$ is chosen, one needs to compute the associated coefficients $y_{\boldsymbol{\alpha}}$, $\boldsymbol{\alpha} \in \mathcal{A}$. This is done thanks to a dataset of input-output samples $\{(\mathbf{x}^{(k)}, y^{(k)})\}_{1 \leq k \leq N}$, and several methods may be employed, either \textit{intrusive} \citep{Ghanem1991} or \textit{non-intrusive}: in particular, direct Monte Carlo evaluation (or by quadrature method in low-dimensional cases) of each coefficient $\mathbb{E}[Y \Psi_{\boldsymbol{\alpha}}(\mathbf{X})]$, or joint estimation of the coefficients by least-square regression~\citep{blatman2009}. 
However, this method becomes computationally impractical in high-dimensional spaces due to the rapid growth in the number of coefficients. When the number of unknowns exceeds the available model evaluations, the problem becomes underdetermined. Sparse PCE techniques, such as LASSO, Forward Stagewise Regression, and Least Angle Regression (LAR), mitigate these challenges by reducing complexity and improving efficiency. \cite{blatman2009} further explores alternative strategies for constructing efficient sparse PC expansions.

In the sequel, the description of our algorithm does not depend on the particular choice of the method to choose the truncation scheme and to compute the coefficients. It only assumes that given:
\begin{itemize}
    \item a dataset of input-output samples $\{(\mathbf{x}^{(k)}, y^{(k)})\}_{1 \leq k \leq N}$ in $\mathcal{D}_\mathbf{X} \times \R$;
    \item a PDF $f_\mathbf{X}$ on $\mathcal{D}_\mathbf{X}$;
    \item possibly a certain number of hyperparameters, determining for instance the order of truncation, the enumeration strategy, and estimation/regularization hyperparameters;
\end{itemize}
a PCE model 
\[
    G^\mathrm{PCE}(\mathbf{x}) = \sum_{\boldsymbol{\alpha} \in \mathcal{A}} y_{\boldsymbol{\alpha}} \Psi_{\boldsymbol{\alpha}}(\mathbf{x})
\]
can be returned, where $\mathcal{A}$ is a finite subset of $\N^d$ and $(\Psi_{\boldsymbol{\alpha}})_{\boldsymbol{\alpha} \in \mathcal{A}}$ is a family of multivariate functions, which is orthonormal in $L^2(\R^d,P_\mathbf{X})$. 

In the numerical experiments of this work, this task is carried out thanks to the open-source library for uncertainty treatment “OpenTURNS”, standing for “Open source initiative to Treat Uncertainties, Risks’N Statistics” (www.openturns.org) \citep{Baudin2017}.

\subsection{Challenge of Applying PCE to Complex Models}
\begin{figure}[H]
    \centering
    \begin{tabular}{cc}
    \subfloat[$\mathrm{TSE}$ as a function of the number of nonzero coefficients in the metamodel\label{fig:error-space}]{%
     \includegraphics[width=0.45\linewidth]{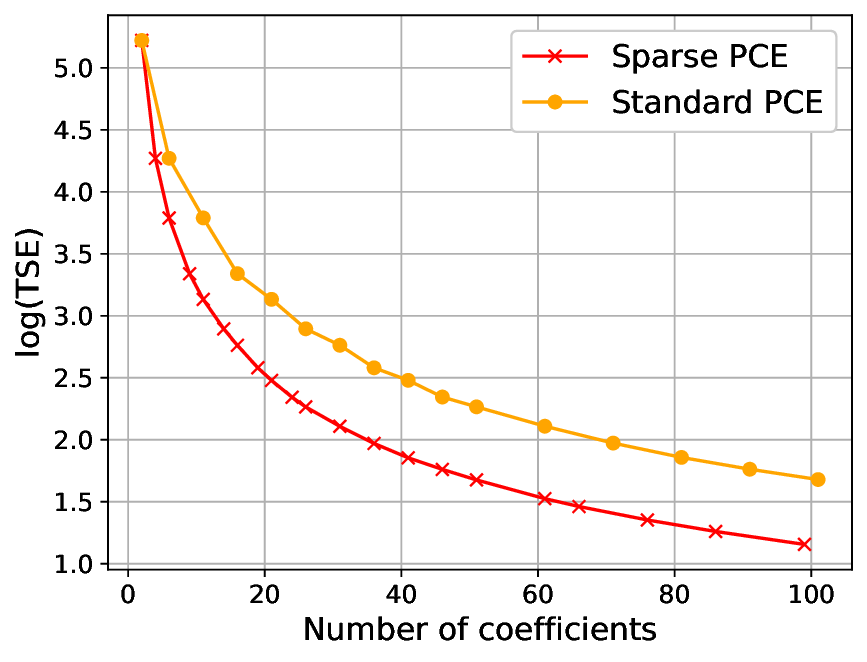}} &
    \subfloat[PCE approximation and Gibbs phenomenon\label{fig:comp-samples}]{%
    \includegraphics[width=0.45\linewidth]{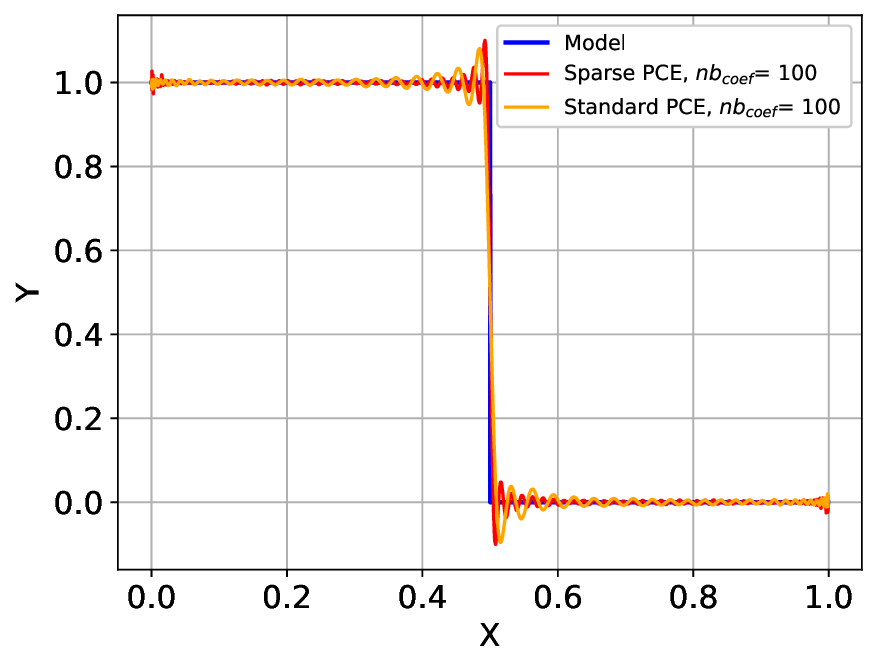}} \\
    \end{tabular}
\caption{PCE on a piecewise-constant function $Y = G(X) = \ind{X > 1/2}$: (a) evolution of the $\mathrm{TSE}$ with respect to the number of nonzero coefficients $y_\alpha$, (b) comparison between the true function and its PCE approximations.}
\label{fig:pce_limit}
\end{figure}

When applied to complex models, namely discontinuous or highly nonlinear functions $G$, the PCE method encounters significant difficulties due to its reliance on smooth polynomial functions. Specifically, it struggles to accurately represent abrupt changes in the model response, often leading to oscillatory artifacts near discontinuities, an issue known as the Gibbs phenomenon. While increasing the polynomial degree generally improves the global approximation and reduces global residual errors, it fails to suppress these oscillations. Consequently, PCE becomes less effective for models exhibiting threshold effects or piecewise-defined behaviors, where discontinuities or abrupt variations occur.
To illustrate this limitation, we consider a piecewise-constant model given by $Y = G(X) = \ind{X>1/2}$ in dimension $d=1$, with $X$ uniformly distributed on $[0,1]$, and approximate it using standard and Sparse PCE expansions with increasing polynomial degrees.  
Denoting by $G^\mathrm{PCE}$ the metamodel obtained either by standard or Sparse PCE, Figure \ref{fig:error-space} shows the evolution of the Total Squared Error (TSE)
\[
    \mathrm{TSE} = \sum_{k=1}^N \left(y^{(k)} - G^\mathrm{PCE}(x^{(k)})\right)^2
\]
as a function of the number of nonzero coefficients $y_\alpha$ of the metamodel, while Figure \ref{fig:comp-samples} illustrates the oscillatory behavior induced by high-degree polynomial approximations. Even with a polynomial with high number of nonzero coefficients for standard and Sparse PCE, significant oscillations persist, emphasizing the challenges of accurately approximating discontinuous functions using global polynomial expansions.

\section{Tree-based Polynomial Chaos Expansion (Tree-PCE)}\label{sec3} 

In this section, we present the \textbf{Tree-PCE} algorithm, inspired by regression tree methodologies \citep{Breiman1984} and designed to construct surrogate models for complex systems through a combination of local PCEs. The algorithm adaptively partitions the input space into a collection of disjoint hyper-rectangular subdomains, to which we shall simply refer as \emph{rectangle}. Over each subdomain, a local PCE is constructed to approximate the model response. This localized modeling approach aims to overcome limitations of global PCE. What sets Tree-PCE apart is its use of a partitioning strategy that directly minimizes the global approximation error. At each step, the algorithm evaluates all eligible subdomains and selects the one whose binary split leads to the greatest decrease in the global $\mathrm{TSE}$. Among the possible splits along different input directions, the one yielding the largest improvement is applied. This process continues recursively, producing a hierarchical structure in the form of a binary tree. Each leaf node of the tree corresponds to a subdomain equipped with its own local PCE. These local surrogates are then assembled into a global approximation that adapts to the complexity of the model across different regions of the input space.

The basic Tree-PCE algorithm is presented in Section~\ref{ss:describ-TreePCE}. The construction of the underlying tree structure, and its use in the metamodel evaluation, are presented in Section~\ref{ss:tree-structure}. Last, stopping criteria allowing to achieve a balance between high accuracy and low complexity are discussed in Section~\ref{ss:stop-crit}.

\subsection{Description of the Algorithm}\label{ss:describ-TreePCE}

The overall input of the algorithm is:
\begin{itemize}
    \item a dataset of input-output samples $\{(\mathbf{x}^{(k)}, y^{(k)})\}_{1 \leq k \leq N}$ in $\mathcal{D}_\mathbf{X} \times \R$;
    \item a PDF $f_\mathbf{X}$ on $\mathcal{D}_\mathbf{X}$;
    \item hyperparameters related to the construction of local PCE models.
\end{itemize}
It returns a metamodel of the form
\[
    G^\mathrm{Tree-PCE}(\mathbf{x}) = \sum_{r=1}^R \ind{\mathbf{x} \in \mathcal{R}^r} G^\mathrm{PCE}_{\mathcal{R}^r}(\mathbf{x}), \qquad G^\mathrm{PCE}_{\mathcal{R}^r}(\mathbf{x}) = \sum_{\boldsymbol{\alpha} \in \mathcal{A}^r} y_{\boldsymbol{\alpha}}^r \Psi_{\boldsymbol{\alpha}}^r(\mathbf{x}),
\]
where $(\mathcal{R}^r)_{1 \leq r \leq R}$ is a family of rectangles, indexed by a binary tree structure, which forms a partition\footnote{Throughout this work, our use of the word `partition' is slighlty abusive as, in fact, the intersection between two neighbouring rectangles is not empty but reduced to their common face. This abuse of terminology does not affect the meaning of our results.} of $\mathcal{D}_\mathbf{X}$ and, for each $r \in \{1, \ldots, R\}$, $G^\mathrm{PCE}_{\mathcal{R}^r}$ is a PCE model for the restriction of $G$ and $f_\mathbf{X}$ to the domain $\mathcal{R}^r$. 

\begin{figure}[H] 
    \centering
    \begin{tabular}{cc}
    \subfloat[Partition of the input domain $\mathcal{D}_\mathbf{X}$ induced by Tree-PCE of the 16 first classes]{%
     \includegraphics[width=0.4\linewidth]{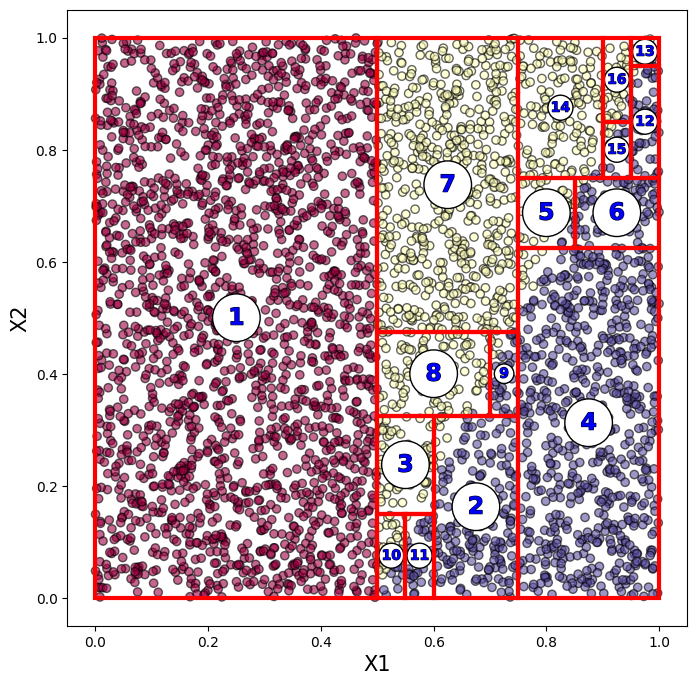}%
     \label{fig:treepce-partition}
    } &
    \subfloat[Corresponding decision tree: the 16 leaf nodes represent each rectangle]{%
     \includegraphics[width=0.6\linewidth]{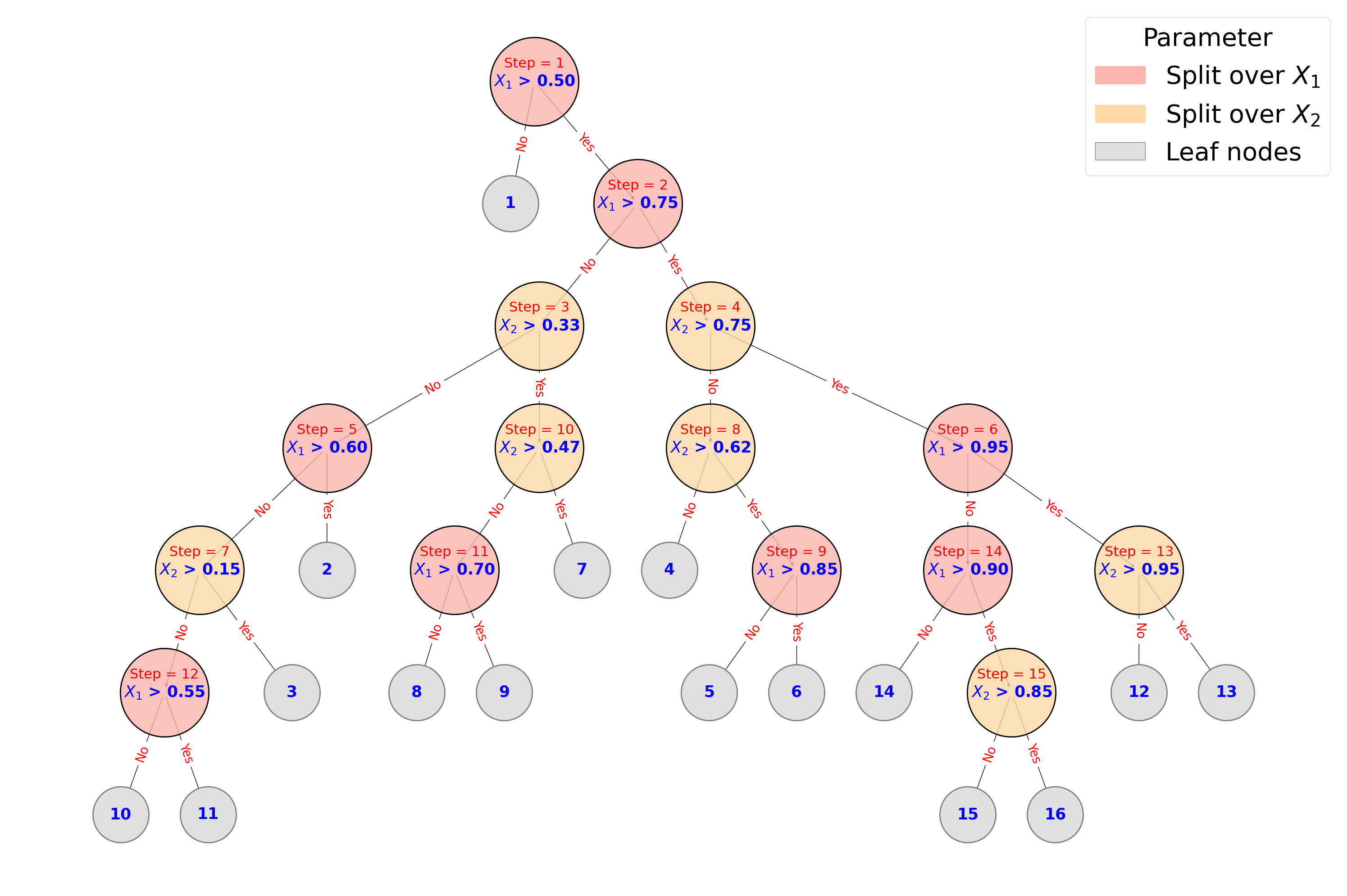}%
     \label{fig:treepce-tree}
    } \\
    \end{tabular}
    \caption{Tree-PCE applied to the example~\eqref{simple-2d-model}
     defined on $\mathcal{D}_\mathbf{X} = [0,1] \times [0,1]$.
    (a) shows how the unit square is partitioned; (b) shows the corresponding binary tree of the 16 first classes. Each colored node corresponds to a split along a single input parameter: the 'Yes' branch (right) leads to the region where the input is above the split point, and the 'No' branch (left) to the region below.  
    Leaf nodes, shown in gray, correspond to the final rectangular regions depicted in (a).
    See Section~\ref{ss:tree-structure} for more details about tree construction.  }\label{fig:diagonal2D}
    \label{fig:treepce-2d-example}
\end{figure}

Figure~\ref{fig:treepce-2d-example} illustrates the output of the Tree-PCE algorithm on the following simple two-dimensional example, with a diagonal discontinuity:
\begin{equation}\label{simple-2d-model}
    Y= G(\mathbf{X}) = \begin{cases}
        0 & \text{if $X_1 < 0.5$,}\\
        1 & \text{if $X_1 \geq 0.5$ and $X_2>2X_1-1$,}\\
        2 & \text{if $X_1 \geq 0.5$ and $X_2 \leq 2X_1-1$,}
    \end{cases}
\end{equation}
and $\mathbf{X}=(X_1,X_2)$ is uniformly distributed on $[0,1]^2$. The algorithm is stopped after that 16 classes have been created. 
Subfigure (a) shows how the input domain \(\mathcal{D}_{\mathbf{X}} = [0,1]^2\) is partitioned.
The algorithm begins by isolating the region where \(X_1 < 0.5\), corresponding to \(Y=0\), and then recursively subdivides the remaining domain by adaptively refining the partition near the diagonal discontinuity.

Throughout the description of the algorithm, we will use the Total Squared Error of a local model $G^\mathrm{PCE}_\mathcal{R}$ defined on a subset $\mathcal{R}$ of $\mathcal{D}_\mathbf{X}$, given by
\begin{equation}\label{eq:TSE-R}
    \mathrm{TSE}_\mathcal{R} = \sum_{k=1}^N \ind{\mathbf{x}^{(k)} \in \mathcal{R}} \left(y^{(k)} - G^\mathrm{PCE}_\mathcal{R}(\mathbf{x}^{(k)})\right)^2.
\end{equation}

\subsubsection{Mesh Generation of Test Split points}

In the initial step, a set of threshold values is defined over the input domain \(\mathcal{D}_{\mathbf{X}}\) to guide the partitioning process. For each input component \( X_i \), where \( i \in \{1, \dots, d\} \), a finite set of candidate threshold values is considered, denoted by  
\[
\mathcal{T}_i = \{t_{i,1} < \cdots < t_{i,n_i} \} \subset \mathcal{D}_{X_i}.
\]
This set consists of \( n_{i} \) candidate split values for each \( X_i \), either specified by the user or automatically generated based on a predefined selection strategy. If the domain \(\mathcal{D}_{X_i}\) is bounded, the candidate thresholds are typically chosen as uniformly spaced points within the input range excluding the boundaries. Alternatively, in an adaptive approach, threshold values can be selected according to the distribution of observed data points, for example by taking empirical quantiles.

\subsubsection{Splitting Procedure}\label{subsec_split}

We present the splitting procedure, which takes as inputs a rectangle $\mathcal{R} = \prod_{i=1}^d [t_{i, \underline{j_{i}}}, t_{i,\overline{j_{i}}}]$, with $1 \leq \underline{j_{i}} < \overline{j_{i}} \leq n_i$ and a PCE model $G^\mathrm{PCE}_\mathcal{R}$ of $G$ restricted to $\mathcal{R}$. This procedure either determines that the rectangle is ineligible for splitting or proceeds with the partitioning, returning the two resulting sub-rectangles along with their corresponding PCEs. The primary stages of the procedure are outlined below:

\begin{enumerate}
\item \textbf{Minimum Data Points Verification:}  
 The algorithm first verifies that the rectangle \(\mathcal{R}\) contains at least a user-specified minimum number $n_\text{min}$ of data points to justify a PCE. This criterion ensures that the partitioning remains statistically meaningful and mitigates the risk of overfitting due to insufficient sample sizes. If this condition is not met, the algorithm does not proceed with the splitting operation and instead returns this information. In~\citep{blatman2009}, the criterion \(n_{\min} = 3 \binom{p_\mathrm{loc}+d}{p_\mathrm{loc}}\) is proposed, where $p_\mathrm{loc}$ is the maximal degree of the local PCE.
\item \textbf{Threshold Evaluation and Splitting on each Direction:} For each direction \( i \in \{1, \dots, d\} \):
\begin{enumerate}
\item If $\overline{j_i} = \underline{j_i}+1$, the rectangle $\mathcal{R}$ cannot be split in the direction $i$.
\item If $\overline{j_i} > \underline{j_i}+1$, then for all $j_i \in \{\underline{j_i}+1, \ldots, \overline{j_i}-1\}$, the algorithm sets
\[
    \mathcal{R}_1(t_{i,j_i}) = \{\mathbf{x} \in \mathcal{R} | x_i \leq t_{i,j_i}\} \quad \text{and} \quad \mathcal{R}_2(t_{i,j_i}) = \{\mathbf{x} \in \mathcal{R} | t_{i,j_i} \leq x_i\},
\]
and applies PCE separately to each subdomain, using the input distribution restricted to that subdomain, namely
\[
    f_{\mathbf{X}|\mathcal{R}_1(t_{i,j_i})}(\mathbf{x}) = \frac{\ind{\mathbf{x} \in \mathcal{R}_1(t_{i,j_i})}f_\mathbf{X}(\mathbf{x})}{\int_{\mathcal{R}_1(t_{i,j_i})} f_\mathbf{X}(\mathbf{x}')d\mathbf{x}'}, \qquad f_{\mathbf{X}|\mathcal{R}_2(t_{i,j_i})}(\mathbf{x}) = \frac{\ind{\mathbf{x} \in \mathcal{R}_2(t_{i,j_i})}f_\mathbf{X}(\mathbf{x})}{\int_{\mathcal{R}_2(t_{i,j_i})} f_\mathbf{X}(\mathbf{x}')d\mathbf{x}'}.
\]
This generates metamodels $G^\mathrm{PCE}_{\mathcal{R}_1(t_{i,j_i})}$ and $G^\mathrm{PCE}_{\mathcal{R}_2(t_{i,j_i})}$ to approximate the system's behavior within each subdomain. 

The residual error of the combined local models is given by $\mathrm{TSE}_{\mathcal{R}_1(t_{i,j_i})} + \mathrm{TSE}_{\mathcal{R}_2(t_{i,j_i})}$, where we recall the definition~\eqref{eq:TSE-R} of the Total Squared Error. The algorithm then selects the threshold $t_{i,j_i^*}$ that maximizes the error gain: 
\[
        j_i^* = \arg \min_{\underline{j_i} < j_i < \overline{j_i}} \left(\mathrm{TSE}_{\mathcal{R}_1(t_{i,j_i})} + \mathrm{TSE}_{\mathcal{R}_2(t_{i,j_i})} \right).
\]
\end{enumerate}

\item \textbf{Optimal Split Across All Directions}: If, for all $i \in \{1, \ldots, d\}$, $\overline{j_i} = \underline{j_i}+1$, so the rectangle $\mathcal{R}$ cannot be split in any direction, the algorithm returns this information. Otherwise, let
    \[
        i^* = \arg \min_{i} \left(\mathrm{TSE}_{\mathcal{R}_1(t_{i,j^*_i})} + \mathrm{TSE}_{\mathcal{R}_2(t_{i,j^*_i})} 
        \right),
    \]
   where the $\arg\min$ is taken over directions in which the rectangle $\mathcal{R}$ can be split. The algorithm then returns the pairs $(\mathcal{R}_1(t_{i^*,j^*_{i^*}}), G^\mathrm{PCE}_{\mathcal{R}_1(t_{i^*,j^*_{i^*}})})$ and $(\mathcal{R}_2(t_{i^*,j^*_{i^*}}), G^\mathrm{PCE}_{\mathcal{R}_2(t_{i^*,j^*_{i^*}})})$.
\end{enumerate} 
\begin{rk}
\label{rq:errormeasure}
    In the literature, the quality of a metamodel $G^\mathrm{meta}$ is often quantified using the coefficient of determination\footnote{Or its Leave-One-Out version $Q^2$, which we shall not discuss here.} $R^2$, defined by
    \begin{equation*}
        R^2 = 1 - \frac{\sum_{k=1}^N (y^{(k)}-G^\mathrm{meta}(\mathbf{x}^{(k)}))^2}{\sum_{k=1}^N (y^{(k)}-\overline{y})^2},
    \end{equation*}
    where $\overline{y}$ is the sample mean of $\{y^{(k)}\}_{1 \leq k \leq N}$. So it would make sense to decide to split a rectangle $\mathcal{R}$ if a combination of the coefficients $R^2_{\mathcal{R}_1}$ and $R^2_{\mathcal{R}_2}$, corresponding to the PCEs on $\mathcal{R}_1$ and $\mathcal{R}_2$, shows a `better' accuracy than the coefficient $R^2_\mathcal{R}$ corresponding to the PCE on $\mathcal{R}$. In Appendix~\ref{app:A}, we exhibit a simple example where the mere knowledge of the values of the coefficients $R^2_{\mathcal{R}_1}$ and $R^2_{\mathcal{R}_2}$ does not allow to determine an optimal splitting point, thereby evidencing the necessity to consider a criterion based on the TSE over the whole rectangle $\mathcal{R}$.
\end{rk}
 
\subsubsection{Tree-PCE Algorithm Steps}\label{sss:tree-pce-steps}

We are now in position to describe the overall Tree-PCE algorithm.

    \begin{enumerate}
        \item \textbf{Initialization}:
        \begin{enumerate}
            \item Apply a global PCE across the entire domain $\mathcal{R}= \mathcal{D}_{\mathbf{X}}$ to get a metamodel $G^\mathrm{PCE}_\mathcal{R}$.
            \item Apply the splitting procedure described in Subsection~\ref{subsec_split} to $(\mathcal{R},G^\mathrm{PCE}_\mathcal{R})$. If the domain $\mathcal{R}$ can be split, let $(\mathcal{R}_1,G^\mathrm{PCE}_{\mathcal{R}_1})$, $(\mathcal{R}_2,G^\mathrm{PCE}_{\mathcal{R}_2})$ be given by the splitting procedure, and initialize a list $L$ which contains as a single element the tuple
            \begin{equation}\label{eq:tupleL}
                \{(\mathcal{R},G^\mathrm{PCE}_\mathcal{R},\mathrm{TSE}_\mathcal{R}); (\mathcal{R}_1,G^\mathrm{PCE}_{\mathcal{R}_1},\mathrm{TSE}_{\mathcal{R}_1}); (\mathcal{R}_2,G^\mathrm{PCE}_{\mathcal{R}_2},\mathrm{TSE}_{\mathcal{R}_2})\}.
            \end{equation}
            If the domain $\mathcal{R}$ cannot be split, initialize the list $L$ with the tuple
            \begin{equation}\label{eq:tupleL-empty}
                \{(\mathcal{R},G^\mathrm{PCE}_\mathcal{R},\mathrm{TSE}_\mathcal{R}); \emptyset; \emptyset\}
            \end{equation}
            as single element.

        \end{enumerate}
At the end of each step of the algorithm, the list \( L \) will contain tuples of the form~\eqref{eq:tupleL} or~\eqref{eq:tupleL-empty}, whose first components $(\mathcal{R},G^\mathrm{PCE}_\mathcal{R},\mathrm{TSE}_\mathcal{R})$ describe the current partition of $\mathcal{D}_\mathbf{X}$, with the associated metamodels. In particular, unless the construction of the metamodel fails for the whole domain at the step~(a) above, the list $L$ will never be empty. Moreover, to determine which rectangle must be split in priority at the next step, the list $L$ will be kept sorted in descending order based on the gain in TSE owed to splitting, which is defined by
\begin{equation}\label{eq:TSEgain}
    \Delta \mathrm{TSE}_\mathcal{R} =\mathrm{TSE}_\mathcal{R} - \left(\mathrm{TSE}_{\mathcal{R}_1} + \mathrm{TSE}_{\mathcal{R}_2}\right)
\end{equation}
for tuples of the type~\eqref{eq:tupleL}, and by
\begin{equation}\label{eq:TSEgain-empty}
    \Delta \mathrm{TSE}_\mathcal{R} = 0
\end{equation}
for tuples of the type~\eqref{eq:tupleL-empty}. This is ensured by the following Recursive Step.

      \item \textbf{Recursive Step}:
\begin{enumerate}
   \item Let $\Delta \mathrm{TSE}_\mathcal{R}$ be the TSE gain associated with the first element of $L$, given by~\eqref{eq:TSEgain}--\eqref{eq:TSEgain-empty}. If $\Delta \mathrm{TSE}_\mathcal{R} \leq 0$, then the algorithm stops\footnote{The stopping criterion will be further discussed in Subsection~\ref{ss:stop-crit}.} and returns the list $L$.
    \item Otherwise, extract the first element
    \[
        \{(\mathcal{R},G^\mathrm{PCE}_\mathcal{R},\mathrm{TSE}_\mathcal{R}); (\mathcal{R}_1,G^\mathrm{PCE}_{\mathcal{R}_1},\mathrm{TSE}_{\mathcal{R}_1}); (\mathcal{R}_2,G^\mathrm{PCE}_{\mathcal{R}_2},\mathrm{TSE}_{\mathcal{R}_2})\}
    \]
    from \( L \), corresponding to the rectangle with highest current value of $\Delta \mathrm{TSE}_\mathcal{R}$ (which is necessarily positive), and remove it from $L$. 
    \item For $l=1,2$, apply the splitting procedure to $(\mathcal{R}_l,G^\mathrm{PCE}_{\mathcal{R}_l})$, and insert the resulting tuple, either of the type~\eqref{eq:tupleL} if the rectangle is split, or of the type~\eqref{eq:tupleL-empty} otherwise, in the list $L$, while maintaining the order based on $\Delta \mathrm{TSE}_{\mathcal{R}_l}$.
\end{enumerate}
\end{enumerate}

The output of the algorithm is the list $L$, which contains (in this order) tuples of the type~\eqref{eq:tupleL-empty}, whose first component $(\mathcal{R},G^\mathrm{PCE}_\mathcal{R},\mathrm{TSE}_\mathcal{R})$ can no longer be split, and then tuples of the type~\eqref{eq:tupleL}, whose first component splitting would increase the TSE. Therefore, denoting by $R$ the cardinality of $L$ and by $(\mathcal{R}^r,G^\mathrm{PCE}_{\mathcal{R}^r},\mathrm{TSE}_{\mathcal{R}^r})$, $r \in \{1, \ldots, R\}$ the first components of all tuples in $L$, the final metamodel constructed by the algorithm reads

\[
    G^\mathrm{Tree-PCE}(\mathbf{x}) = \sum_{r=1}^R \ind{\mathbf{x} \in \mathcal{R}^r} G^\mathrm{PCE}_{\mathcal{R}^r}(\mathbf{x}).
\]

This construction, however, gives rise to the following two observations: 

\begin{itemize}
    \item since rectangles which are split are dropped from the list $L$, the tree structure underlying the construction of the partition $(\mathcal{R}^r)_{1 \leq r \leq R}$ is seemingly lost, making in particular the evaluation of the metamodel $G^\mathrm{Tree-PCE}$ computationally suboptimal;
    \item the only stopping criterion being that the TSE gain stops being positive, almost useless splits of already acceptable rectangles may occur, leading to a metamodel with high number of classes, and therefore high complexity.
\end{itemize}

These remarks are examined in the subsequent two subsections.

\subsection{Tree structure and model evaluation}\label{ss:tree-structure}

Given $\mathbf{x} \in \mathcal{D}_\mathbf{X}$, to evaluate the metamodel $G^\mathrm{Tree-PCE}$ computed above at $\mathbf{x}$, one needs to look over the list $L$ and check, at each element $r$, if $\mathbf{x}$ belongs to $\mathcal{R}^r$. The cost of this operation is proportional to the number of classes $R$, and each test requires in principle $2d$ comparisons. It is however possible to reduce this computational cost by keeping track of the tree structure underlying the algorithm, and constructing a decision tree together with the local metamodels, as follows.

\begin{enumerate}
    \item At initialization, if the PCE is successfully constructed, define a tree with a single node (the root), which is associated with $(\mathcal{D}_\mathbf{X}, G^\mathrm{PCE}_{\mathcal{D}_\mathbf{X}})$.
    \item At each step, the first components of the tuples in the list $L$ are associated with the leaves of the tree. When the first element of the list is extracted, the corresponding leaf, associated with the rectangle $\mathcal{R}$, is appended two children (which become leaves), associated with $\mathcal{R}_1$ and $\mathcal{R}_2$. Moreover, the parent node is assigned the pair $(i^*,t_{i^*,j^*_{i^*}})$ as a label.  
\end{enumerate}

Using this binary tree, to determine the rectangle $\mathcal{R}$ in which a given $\mathbf{x} \in \mathcal{D}_\mathbf{X}$ lies, it suffices to browse the tree, starting from the root, and, at each internal node with label $(i^*,t_{i^*,j^*_{i^*}})$, to select the left- or right-child depending on whether $x_{i^*} < t_{i^*,j^*_{i^*}}$ or not. This requires to browse a number of rectangles which is at most the height of the tree (which may be of the order $\log R$ instead of $R$ \citep{Bentley1975}), and to perform a single comparison at each step. An example of such a tree is shown in Figure~\ref{fig:treepce-tree}.

\subsection{Stopping criterion}\label{ss:stop-crit}

As it is described above, the algorithm only stops when rectangles in the list can no longer be split, or their splitting would increase the TSE. It is however likely that in the last steps of the algorithm, splitting rectangles only bring a small decrease of TSE and that stopping the algorithm earlier would provide a metamodel with significantly fewer classes, and thus much lower complexity, without losing too much on accuracy. In order to tune this accuracy/complexity tradeoff, more restrictive stopping criteria can be introduced in Step~(a) of the recursive step of the algorithm. 

For instance, one may consider fixing the total number of classes, or the height of the tree, a priori. This may easily be implemented in the algorithm by defining a counter keeping track of the current number of classes, or current height of the tree. This allows to control the complexity of the resulting metamodel, but not its accuracy.

\begin{figure}[H]
    \centering
    \begin{tabular}{ccc}
    \subfloat[\(\log(\mathrm{TSE})\) as a function of the number of classes\label{fig:TSE_k}]{%
     \includegraphics[width=0.45\linewidth]{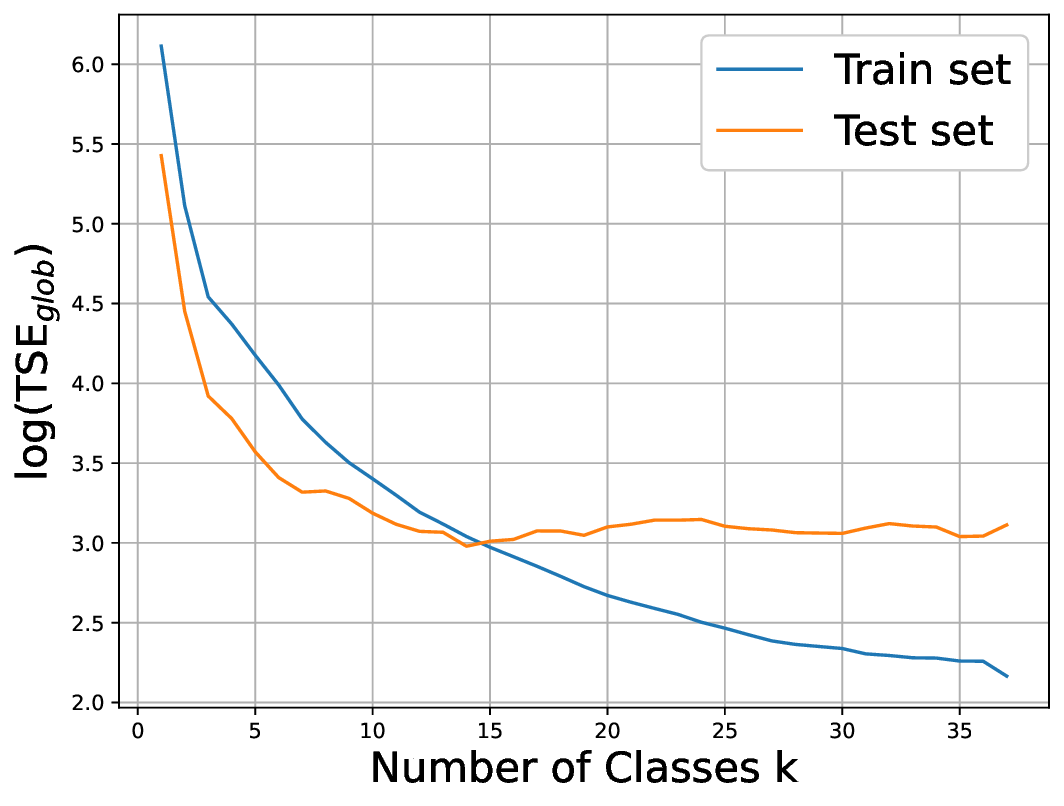}} &
      \subfloat[ number of classes as a function of  \(\epsilon\) in the stopping criterion~\eqref{eq:stopping_criterion}\label{fig:k_eps}]{%
     \includegraphics[width=0.45\linewidth]{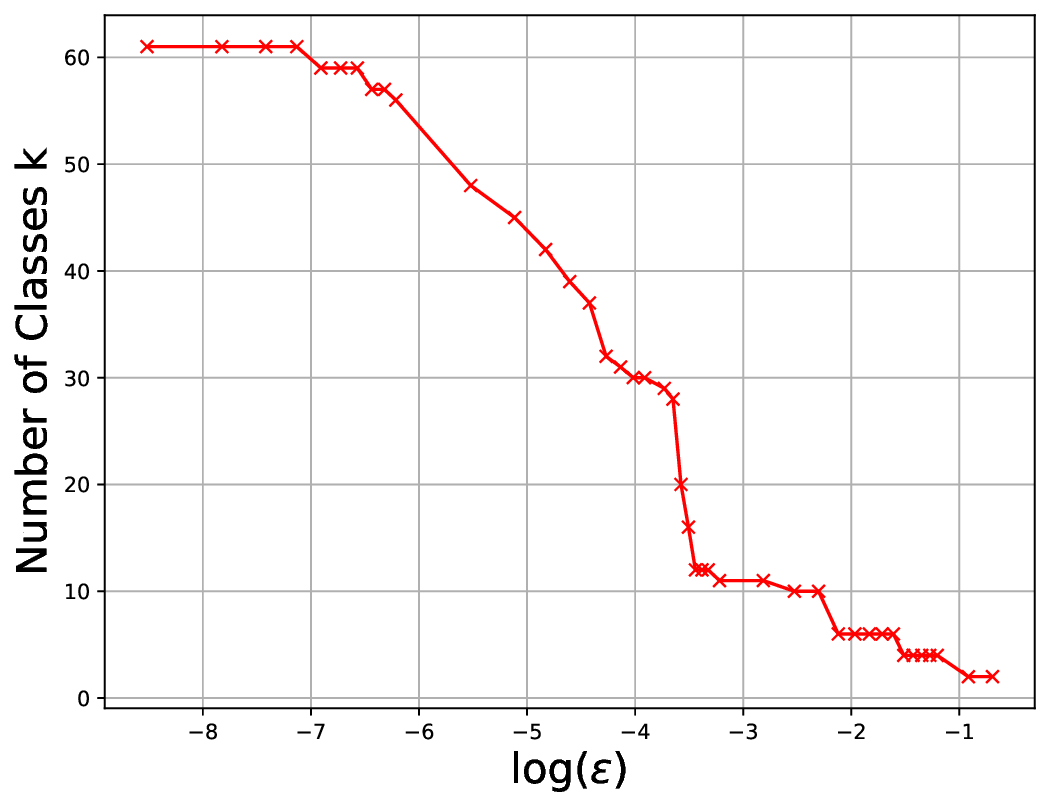}}\\
    \end{tabular}
    \caption{Evolution of global error and number of classes in Tree-PCE algorithm applied to the 2D example of Figure~\ref{fig:diagonal2D}  
    (a) Evolution of the total squared error (in logarithmic scale) as function of the number of classes.  
    (b) Evolution of the Tree-PCE generated number of classes as a function of \(\epsilon\) (in logarithmic scale).}
    \label{fig:TSE_vs_nb_classes}
\end{figure}

Another approach may consist in deciding to split a rectangle only if the gain in TSE exceeds a certain threshold, determined a priori. Since TSE is a quantity with a physical dimension, the threshold must be relative and not absolute. We suggest here to decide to split a rectangle \emph{only if it brings a significative decrease in the TSE of the current metamodel on the overall domain $\mathcal{D}_\mathbf{X}$}: in practice, we fix $\epsilon>0$ and a rectangle $\mathcal{R}$ is split only if
\begin{equation}\label{eq:stopping_criterion}
    (1+\epsilon)\left(\mathrm{TSE}_\mathrm{glob} - \Delta \mathrm{TSE}_\mathcal{R}\right) < \mathrm{TSE}_\mathrm{glob},
\end{equation}
where $\mathrm{TSE}_\mathrm{glob}$ is the TSE of the current metamodel. In other words, a rectangle is split only if the decrease in global TSE is more than $\frac{\epsilon}{1+\epsilon}$ times its current value. This quantity can be computed from scratch at each step by summing the terms $\mathrm{TSE}_\mathcal{R}$ of all first components of tuples in $L$, but it is more efficient to keep track of it by introducing a variable $\mathrm{TSE}_\mathrm{glob}$ which is initialized to $\mathrm{TSE}_{\mathcal{D}_\mathbf{X}}$ and updated by letting $\mathrm{TSE}_\mathrm{glob} \leftarrow \mathrm{TSE}_\mathrm{glob} - \Delta \mathrm{TSE}_\mathcal{R}$ at each step. In this approach, the smaller $\epsilon$ and the higher the number of classes in the final metamodel and consequently, the more accurate the approximation as illustrated in Figures ~\ref{fig:TSE_k} and ~\ref{fig:k_eps}.

\subsection{Comparison with Stochastic Spectral Embedding}

The Tree-PCE algorithm bears close similarities with Stochastic Spectral Embedding (SSE), introduced in~\citep{Marelli2021}. Indeed, both approaches rely on successive partitions of the input space according to splits into sub-rectangles, and construct local polynomial approximations. This allows to trade global regularity for higher local accuracy, at the cost of a potential increase in complexity. 

The construction of SSE may be summarized as follows. The algorithm first constructs a global metamodel $G^\mathrm{PCE}_{\mathcal{D}_\mathbf{X}}$ of $G$; then proceeds to split $\mathcal{R}^{0,1} := \mathcal{D}_\mathbf{X}$ into two rectangles $\mathcal{R}^{1,1}$ and $\mathcal{R}^{1,2}$; and then reapplies iteratively the procedure to the \emph{residual} $G-G^\mathrm{PCE}_{\mathcal{D}_\mathbf{X}}$ separately on $\mathcal{R}^{1,1}$ and $\mathcal{R}^{1,2}$. The output metamodel therefore naturally writes as a multilevel expansion
\begin{equation*}
    G^\mathrm{SSE}(\mathbf{x}) = \sum_{\ell=0}^L \sum_{p=1}^{P_\ell} \ind{\mathbf{x} \in \mathcal{R}^{\ell,p}} G^\mathrm{residual\ PCE}_{\mathcal{R}^{\ell,p}}(\mathbf{x}),
\end{equation*}
where $\ell=0, \ldots, L$ are the different \emph{levels} of approximation, $\mathcal{R}^{\ell,1}, \ldots, \mathcal{R}^{\ell,P_\ell}$ are the different rectangles within the level $\ell$, and $G^\mathrm{residual\ PCE}_{\mathcal{R}^{\ell,p}}$ is the local metamodel constructed for the residual in the associated rectangle. If there is a sufficient number of samples in each rectangle, we typically have $P_\ell=2^\ell$ and  $G^\mathrm{SSE}(\mathbf{x})$ is then, at each point $\mathbf{x}$, a sum of $L+1$ PCE metamodels computed on a decreasing sequence of $L+1$ rectangles. With the underlying binary tree representation of the partition of the input space, the levels $\ell$ correspond to the height of nodes, so in contrast with Tree-PCE, which only retains local PCE models on the leaves of the tree (the `terminal rectangles'), SSE provides a superposition of local models on all nodes of the tree. Of course, this decomposition can be \emph{flattened} by summing the contributions of metamodels on all levels for each terminal rectangle, but recovering local orthogonal decompositions requires further computation. This is the first main difference with Tree-PCE, which only returns local PCE models of the whole model $G$ on terminal rectangles.

A second notable difference lies in the splitting strategy. While Tree-PCE scans several splitting points $t_{i,j_i} \in \mathcal{T}_i$ in each direction $i$, and then selects the pair $(i^*,t_{i^*,j^*_{i^*}})$ which leads to the largest gain in TSE, the default presentation of SSE only considers one split per direction, located at the median of the sample. This certainly makes a single step of the algorithm significantly faster, but may in turn lead to an increase in the number of steps if discontinuities of the mapping $\mathbf{x} \mapsto G(\mathbf{x})$ are not well aligned with these medians.

Last, it is worth mentioning that while the presentation of SSE in~\citep{Marelli2021} does not elaborate much on the stopping criterion, the implementation of the method described in~\citep[Section~1.3.1]{UQdoc_21_118} shows the same ordering strategy for the choice of the next rectangle to split as Tree-PCE, however based on a leave-one-out cross-validation error estimator of the residual expansion rather than the relative gain in global TSE. This leads to a stopping criterion analogous to the one introduced in Subsection~\ref{ss:stop-crit}.

\section{Experimental comparison of Tree-PCE with other metamodels}\label{sec4}

The motivation for the introduction of the Tree-PCE algorithm is the hope that for complex models, low-degree local approximations result in a better accuracy than high-degree global approximations. The goal of this section is thus to assess, on a simple test model, how Tree-PCE compares, in terms of complexity and accuracy, with other polynomial-based metamodels, either global (standard and sparse PCE) or local (SSE). The different metamodels that will be compared, together with the criterion selected to quantify their complexity, are presented in Subsection~\ref{ss:perf-metamodel}. The test model is then introduced in Subsection~\ref{ss:perf-testcase}, and the experimental results are reported and discussed in Subsection~\ref{ss:perf-results}.

\subsection{Metamodels and Criterion to Quantify their Complexity}\label{ss:perf-metamodel}

To assess the performance of Tree-PCE relative to standard/sparse PCE and SSE, we compare the global approximation error of the resulting metamodels as a function of the number of coefficients required to represent them. For each method, the number of coefficients is given by:

\begin{itemize}
    \item \textbf{Standard PCE}:  If using linear truncation strategy, the metamodel is constructed using a full polynomial basis up to degree \( p \) in \( d \) dimensions. The number of coefficients required is given by:
    \begin{equation}\label{nb_coef_standard}
        N_{\text{coeff, standard}} =|\mathcal{A}_{d,p}|= \frac{(d+p)!}{d! p!} 
    \end{equation}
    \item \textbf{Sparse PCE}: Instead of using the full polynomial basis, sparse PCE selects a subset of the most relevant basis functions, reducing the number of coefficients. The number of coefficients is expressed as:
    \begin{equation}\label{nb_coef_sparse}
        N_{\text{coeff, sparse}} = |\mathcal{A}|,
    \end{equation}
    where \( |\mathcal{A}| \) denotes the number of selected basis functions. 
    \item \textbf{Stochastic Spectral Embedding (SSE)}: SSE builds a hierarchy of local PCE models, where each local model approximates the residual left by previous levels, leading to a progressive refinement of the global metamodel. The overall complexity is:
\begin{equation*}
    N_{\text{coeff, SSE}} = \sum_{l=0}^{L} \sum_{k=1}^{P_l} |\mathcal{A}^{l,k}|,
\end{equation*}
where \( L \) is the maximum refinement level, \( P_l \) is the number of partitions at level \( l \), and \( |\mathcal{A}^{l,k}| \) is the number of selected basis functions in partition \( p \) at level \( k \) computed as $\frac{(d + p_\mathrm{loc})!}{d!p_\mathrm{loc}!}$ if using full polynomial basis of local degree $p_\mathrm{loc}$.
\item \textbf{Sparse SSE:}
When using the sparse option, $\bigl|\mathcal{A}^{l,k}\bigr|= \bigl|\mathcal{A}^{l,k}_{\mathrm{sparse}}\bigr|$ is simply the number of basis functions selected by the sparse‐PCE algorithm.

\end{itemize}

We shall compare standard PCE, sparse PCE and SSE with the Tree-PCE algorithm as it is described in Section~\ref{sec3}, with local polynomial approximations using the linear truncation scheme, and with its sparse version, where local polynomial approximations are performed using the sparse PCE method. The numbers of coefficients needed to represent these metamodels are counted as follows:

\begin{itemize}
    \item \textbf{Tree-PCE}: The Tree-PCE method partitions the input space into \( R \) subrectangles and constructs local PCE metamodels within each subdomain. The number of coefficients of Tree-PCE is given by:
    \begin{equation*}
        N_{\text{coeff, Tree-PCE}} = R \times  \frac{(d+p_{\text{loc}})!}{d! p_{\text{loc}}!} ,
    \end{equation*}
    where
     \( R \) is the number of subrectangles created by the Tree-PCE algorithm and
        \( p_{\text{loc}} \) is the polynomial degree of the local PCE models within each subrectangle.

    \item  \textbf{Tree-PCE sparse:} Tree-PCE can also be ran using Sparse PCE to approximate local behaviors. In this case, the metamodel number of coefficients is given by:
    \begin{equation*} 
        N_{\text{coeff, Tree-PCE sparse}} = \sum_{r=1}^R  |\mathcal{A}^r|, 
    \end{equation*}
    where $|\mathcal{A}^r|$  represents the number of selected basis functions for the metamodel in the rectangle $\mathcal{R}^r$.
\end{itemize}

\subsection{Multi-Dimensional Test Problems}\label{ss:perf-testcase}

We consider a benchmark problem that combines high dimensionality, discontinuities, and local oscillatory behavior, testing the capabilities of different methods. The model is defined as:
\begin{equation}\label{model_multid}
Y = G(\mathbf{X}) = \sum_{i=1}^d a_i \left[c + \sin\left(k\pi \left(X_i - \tfrac{1}{3}\right)\right)\right] \ind{X_i > \tfrac{1}{3}},    
\end{equation}
where the variables $X_i$ are independent and uniformly distribued on $[0,1]$, the numbers \( a_i \in \R \) quantify the effect of the \(i^{\text{th}}\) parameter on the response,  with a higher value indicating a greater effect, and \( c \in \R \) controls the amplitude of the discontinuity. The parameter $k$ modulates the frequency of oscillations in the function: lower values of  $k$
yield a smoother profile, while higher values induce increasingly rapid oscillatory behavior.

In these experiments, we set \( a_i = i \) for $1 \leq i \leq d$ and \( c = 1 \). 
We conduct tests for different values of \(d\) and \(k\) to investigate the impact of both dimensionality and oscillation frequency on Tree-PCE. We begin with a low-dimensional case (\(d = 4\)) and mild oscillations (\(k = 1\)) in Subsection~\ref{sec:d-4_k-1}. In this setting, the function exhibits smooth variations, and a low-degree polynomial provides a reasonable approximation of the behavior.
Next, we increase the oscillation frequency by setting \(k = 4\) in Subsection~\ref{sec:d-4_k-4}, which introduces faster oscillations in the function. This change challenges the expansion, requiring a higher-degree polynomial to accurately capture the more rapid fluctuations while still providing a good approximation of the overall behavior.
Finally, in Subsection~\ref{sec:d-10_k-1-c-1}, we test a higher-dimensional case (\(d = 10, k=1\)). These tests allow us to analyze the scalability and performance of Tree-PCE under varying levels of complexity and compare its performance to other methods.

\subsection{Numerical Experiments}\label{ss:perf-results}

For the three cases of interest: $d=4$, $k=1$; $d=4$, $k=4$; $d=10$, $k=1$; we first apply the Tree-PCE algorithm on a mesh given by $\mathcal{T}_i = \{1/9, \ldots, 8/9\}$ in each direction, so that at most $10^d$ classes may be generated before the algorithm stops. We always take for $n_\mathrm{min}$, the minimal number of samples required to be in a rectangle to start the splitting procedure, the value $3\binom{p_{\text{loc}}+d}{ p_{\text{loc}}}$, with $p_{\text{loc}}$ the degree of local PCE. For the low-dimensional, highly oscillatory case $d=4$, $k=4$, we actually apply Tree-PCE with two different choices of degree for local polynomial approximations in order to better capture the high-frequency oscillations.

\begin{figure}[H]
    \centering
    \includegraphics[width=0.5\linewidth]{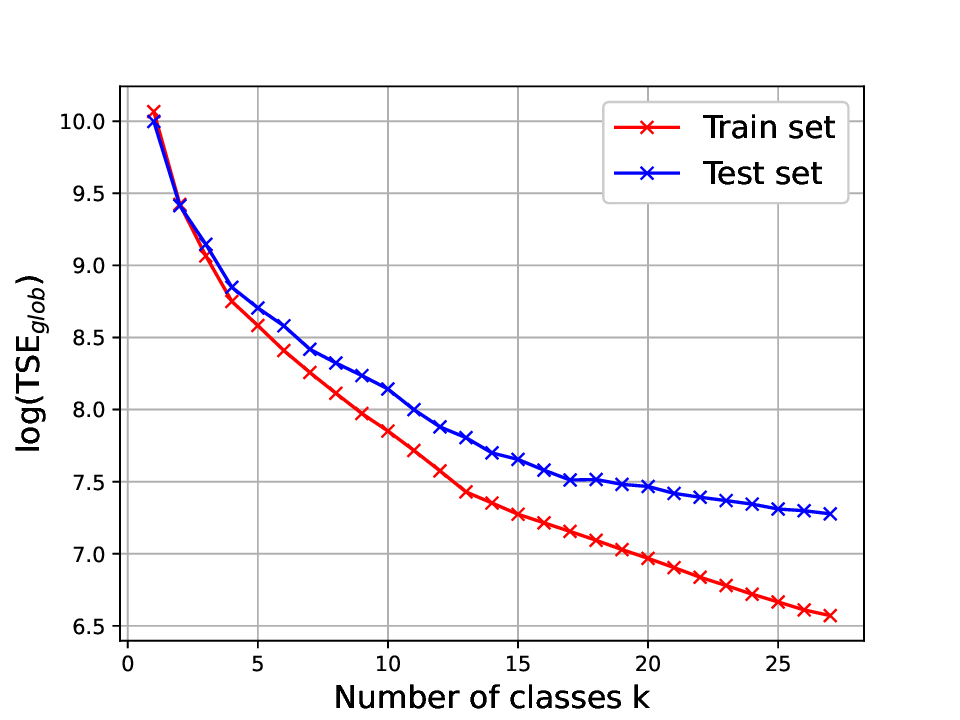}
    \caption{Global $\mathrm{TSE}$ evolution in function of classes for the output~\eqref{model_multid} with \(d = 4\), \(k = 4\) and \(p_\mathrm{loc} = 3\)}
    \label{fig:TSE-d4-k4-p3}
\end{figure}

\begin{figure}[H]
    \centering
    \includegraphics[width=0.5\linewidth]{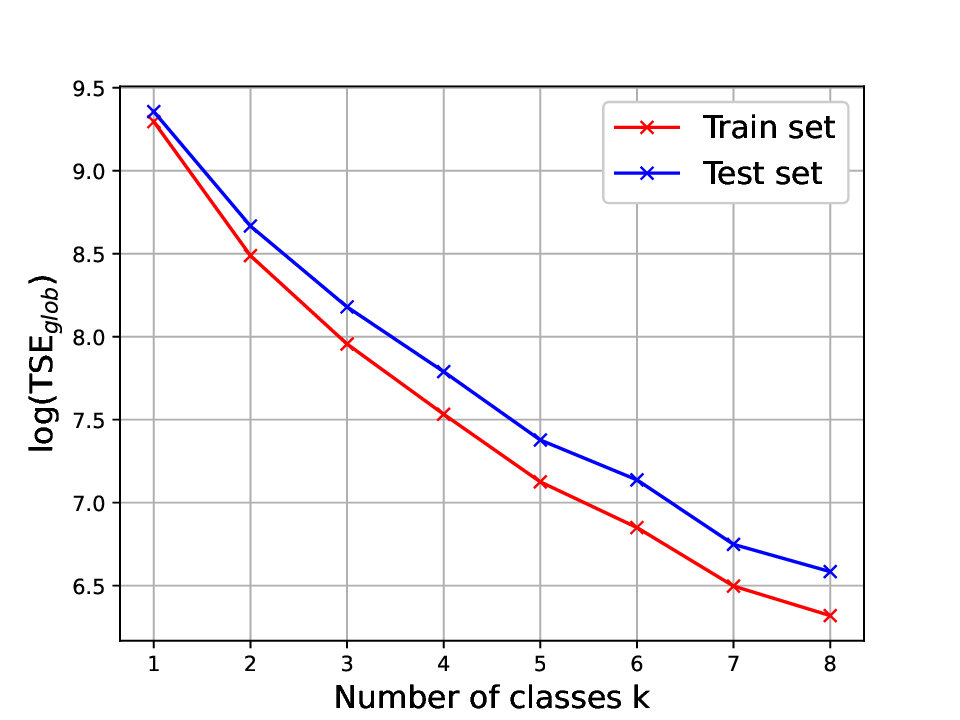}
    \caption{Global $\mathrm{TSE}$ evolution in function of classes for the output~\eqref{model_multid} with \(d = 4\), \(k = 4\) and \(p_\mathrm{loc} = 5\)}
    \label{fig:TSE-d4-k4-p5}
\end{figure}

\begin{figure}[H]
    \centering
    \includegraphics[width=0.5\linewidth]{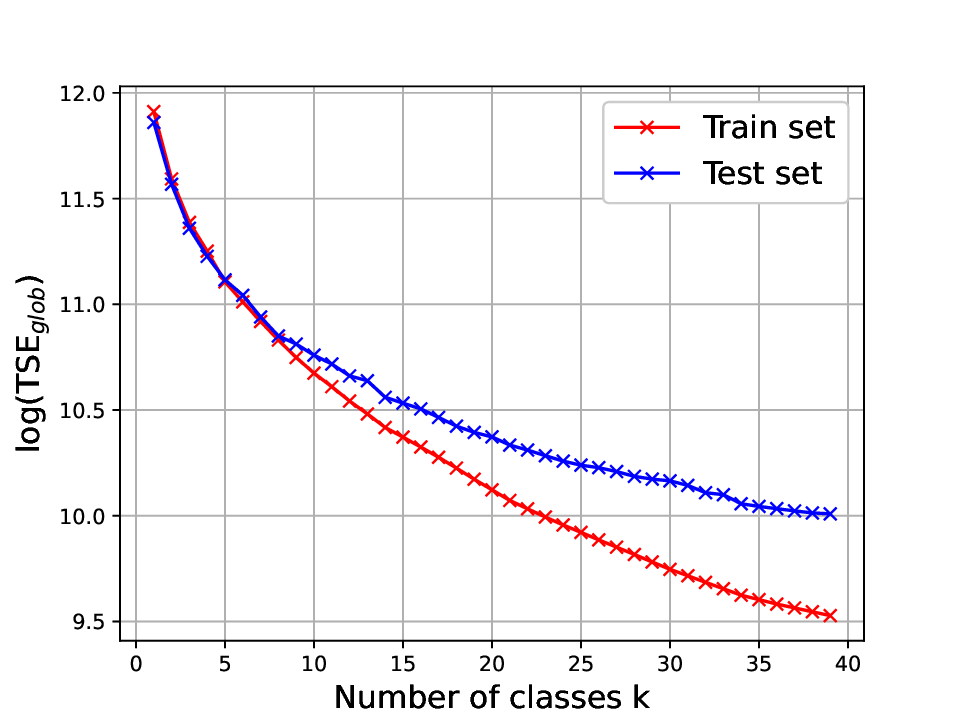}
    \caption{Global $\mathrm{TSE}$ evolution in function of classes for the output~\eqref{model_multid} with \(d = 10\) and \(k = 1\)}
    \label{fig:TSE-d10-k1}
\end{figure}

In order to observe the behavior of the algorithm, we shall plot:
\begin{itemize}
    \item the decision tree returned by the algorithm (stopped at a sufficiently large number of classes), in Figures~\ref{fig:tree-d4-k1}, \ref{fig:tree-d4-k4-p3}, \ref{fig:tree-d4-k4-p5} and~\ref{fig:tree-d10-k1-c1};
    \item the evolution of the TSE of the metamodel returned by the algorithm as a function of the number of classes, both on the training set and on an independent test set with the same size, in Figures~\ref{fig:TSE-d4-k1}, \ref{fig:TSE-d4-k4-p3}, \ref{fig:TSE-d4-k4-p5} and~\ref{fig:TSE-d10-k1}.
\end{itemize}

We shall then present performance comparisons between Tree-PCE and other metamodels, in terms of complexity (quantified as is explained in Subsection~\ref{ss:perf-metamodel}) and accuracy (measured by the TSE). More precisely, in Figures~\ref{fig:Comparison-4dk1}, \ref{fig:Comparison-4dk4} and~\ref{fig:Comparison-10dk1c1}, we shall plot for each test case the TSE as a function of the number of coefficients for:
\begin{itemize}
    \item standard PCE with varying number of coefficients;
    \item sparse PCE with varying number of coefficients;
    \item Sparse SSE with fixed degree of local polynomials and varying number of classes; as well as with fixed number of classes and varying number of coefficients;
    \item Tree-PCE and sparse Tree-PCE with fixed degree of local polynomials and varying number of classes; as well as with fixed number of classes and varying number of coefficients.
\end{itemize}
These computations are carried out both on the training set and on the test set. In all figures, Tree-PCE metamodels are represented with a solid line while other metamodels are presented with dashed lines.

\subsubsection{Low-Dimensional Smooth Case ($d=4$ and $k=1$)}\label{sec:d-4_k-1}

\paragraph{Tree-PCE metamodel} The Tree-PCE algorithm is applied with local polynomials of order \( p_{\text{loc}} = 2 \), on a training set of size $N_{\text{train}} = 2000$.

\begin{landscape}       
\begin{figure}[p]      
  \centering
  \includegraphics[width=\linewidth]{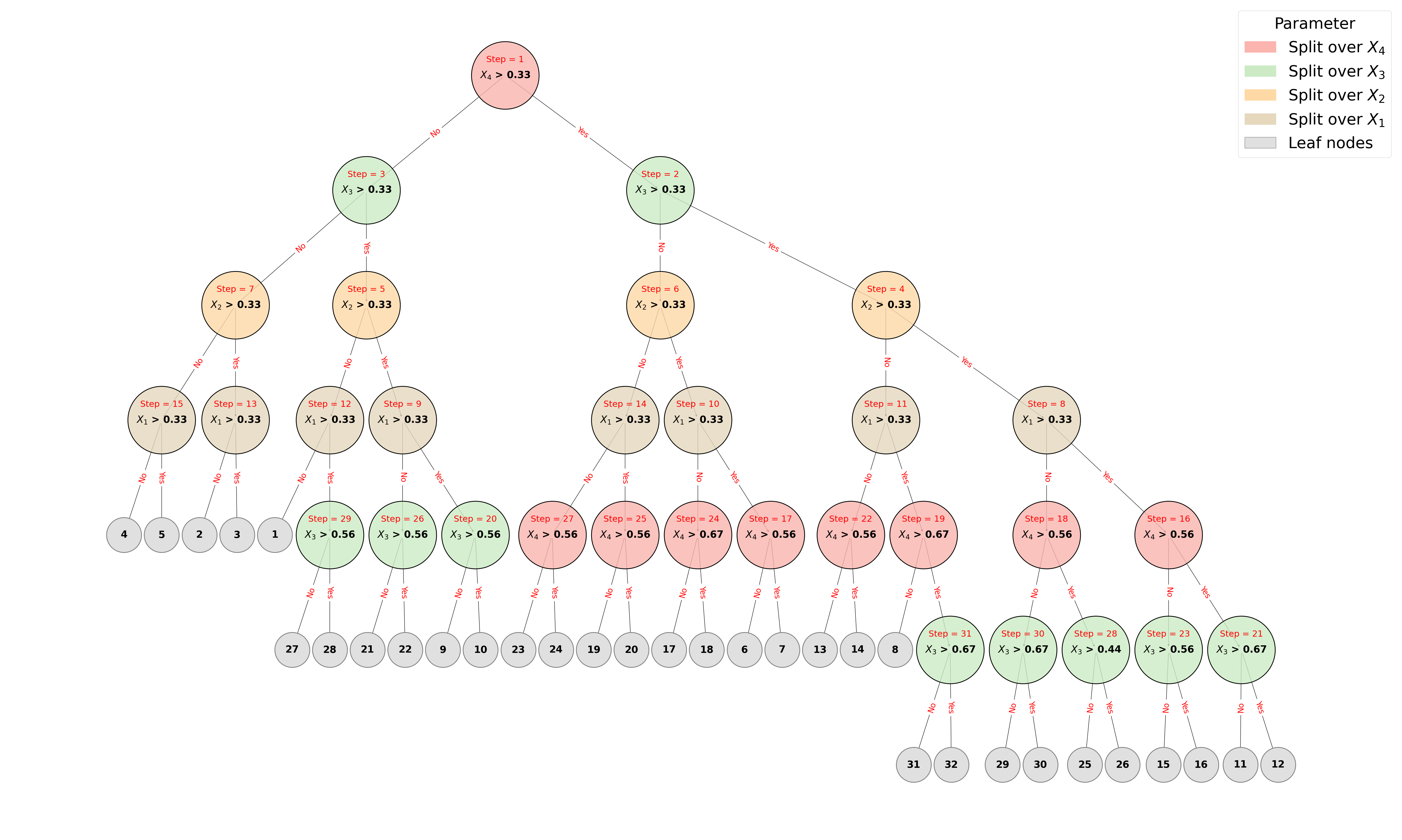}
  \caption{Tree structure for the output~\eqref{model_multid} with \(d = 4\) and \(k = 1\)}
  \label{fig:tree-d4-k1}
\end{figure}
\end{landscape}

The associated decision tree, stopped at $32$ classes, is represented on Figure~\ref{fig:tree-d4-k1}. As is expected, this figure shows that the first $4$ steps of the algorithm produce splits at the discontinuity points of the function $a_i \left[c + \sin\left(k\pi \left(x_i - \tfrac{1}{3}\right)\right)\right] \ind{x_i > \tfrac{1}{3}}$, in decreasing order of the value of $a_i$. After $16$ steps, all discontinuities of the model have been identified, in the sense that $G$ is continuous on each of the $16$ rectangles. Note that the discontinuities are aligned with the mesh grid here, which makes the identification of these discontinuities by the Tree-PCE algorithm possible; otherwise, a similar approximating behavior to the example of Figure~\ref{fig:diagonal2D} would have been observed.

\begin{figure}[H]
    \centering
    \includegraphics[width=0.5\linewidth]{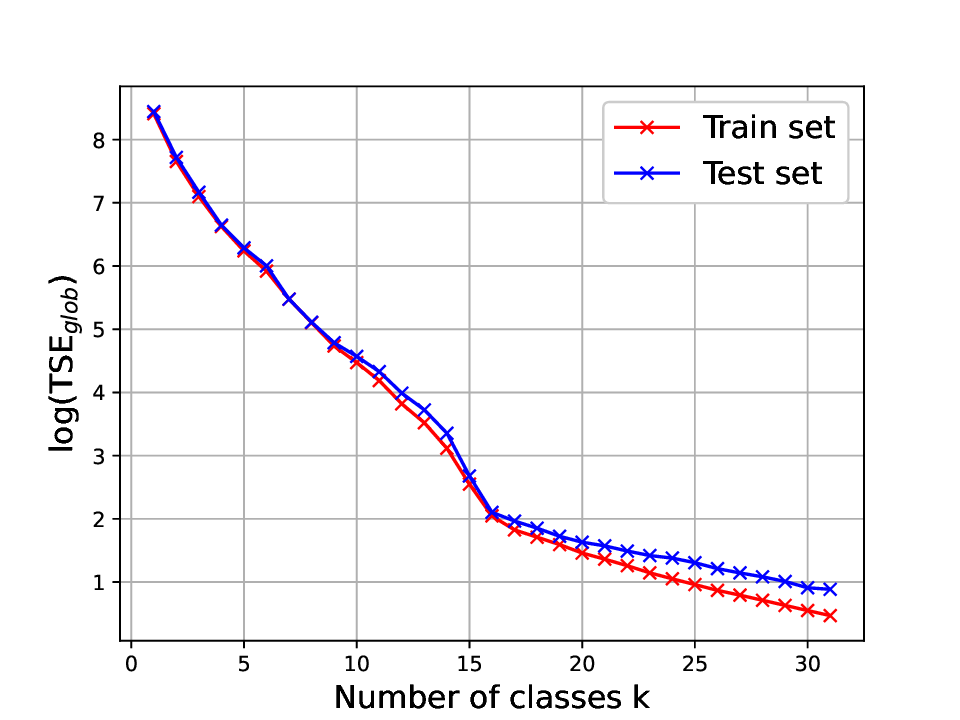}
    \caption{Global $\mathrm{TSE}$ evolution in function of classes for the output~\eqref{model_multid} with \(d = 4\) and \(k = 1\)}
    \label{fig:TSE-d4-k1}
\end{figure}

This threshold is also observed on Figure~\ref{fig:TSE-d4-k1}, which shows the TSE of the metamodel, both on the training set and on the test set, as a function of the number of classes. A change of slope is clearly visible at $16$ classes, showing that the marginal gain in TSE, albeit still positive, is lower once discontinuities have been sorted out.

\begin{figure}[H]
    \centering
    \includegraphics[width=0.45\linewidth]{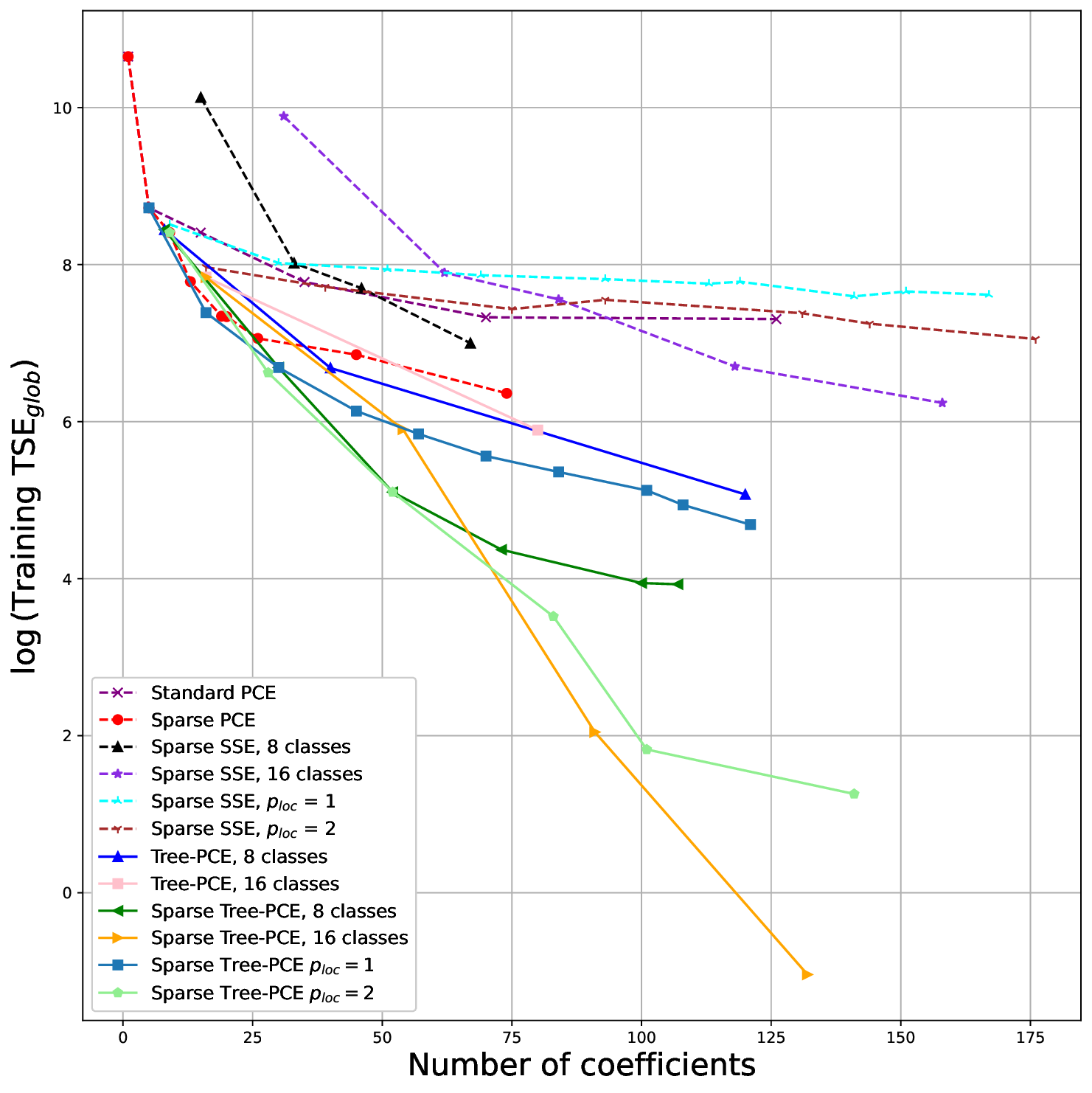}
    \includegraphics[width=0.45\linewidth]{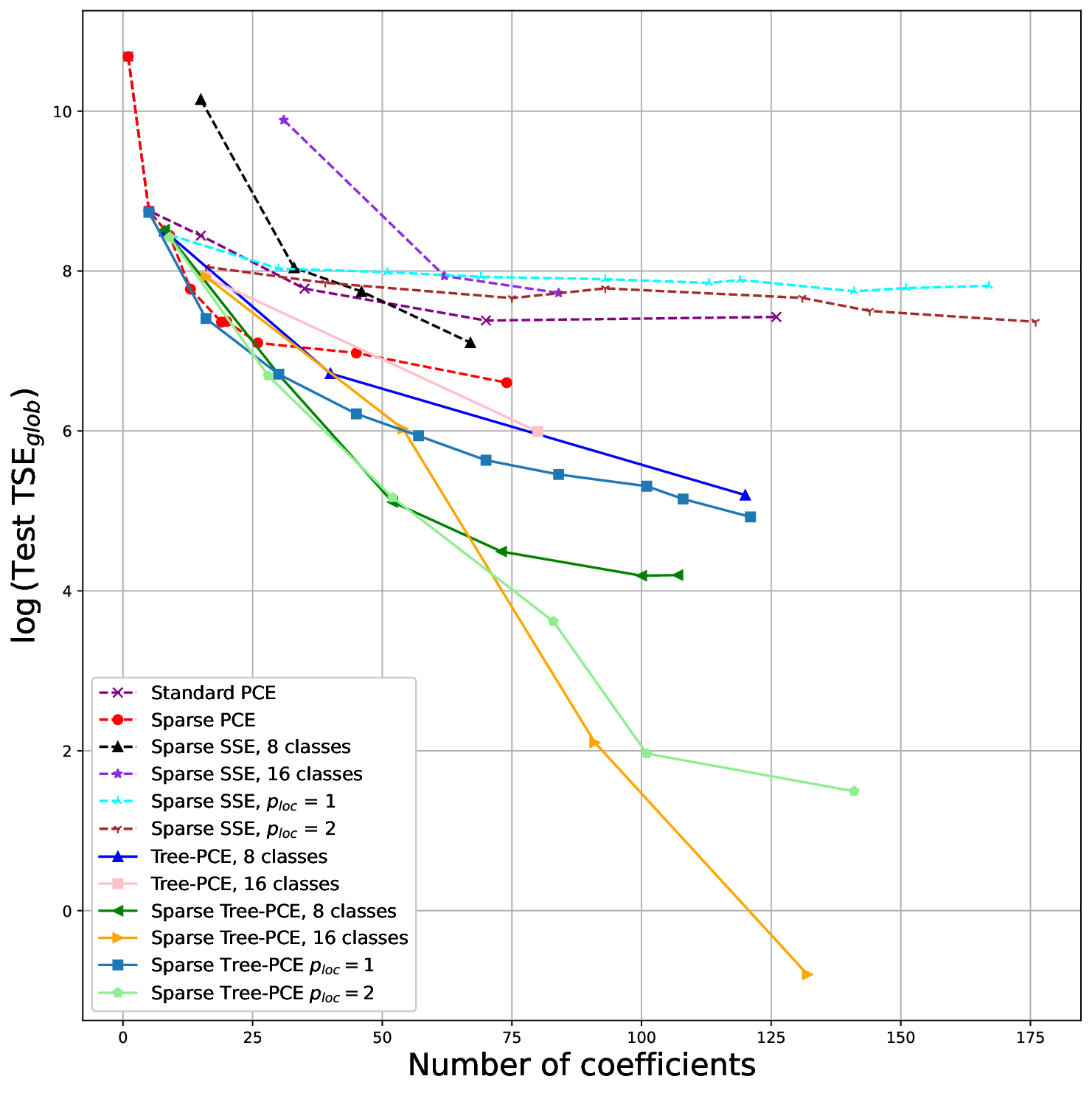}
    \caption{Methods comparison for the output~\eqref{model_multid} with \(d = 4\) and \(k = 1\)}
    \label{fig:Comparison-4dk1}
\end{figure}

\paragraph{Performance comparison} The performance, in terms of complexity and accuracy, of standard PCE, sparse PCE, SSE, Tree-PCE and sparse Tree-PCE is presented on Figure~\ref{fig:Comparison-4dk1}.

The results show that the Tree-PCE and Sparse Tree-PCE methods achieve lower errors with a smaller number of non-zero coefficients. 
The figures further reveal that, for a fixed number of classes, increasing the local polynomial degree $p_\mathrm{loc}$
consistently improves the accuracy. This is because higher-degree polynomials better approximate complex local behaviors within each subdomain. Conversely, when the local degree is held constant, increasing the number of classes leads to an even more substantial reduction in error. This improvement stems from the metamodel’s enhanced capacity to localize and represent variations more precisely. However, this gain in accuracy comes at the cost of model complexity, as the total number of coefficients grows with finer partitions.
In contrast, the Sparse SSE method exhibits the least favorable behavior. It struggles to reduce the error effectively, particularly when using a low number of classes or low-degree polynomials. This poor performance is due to the method’s inability to directly detect and adapt to discontinuity points in the model. As a result, it requires either more classes or higher local polynomial degrees to reach an acceptable level of accuracy.

\subsubsection{Low-Dimensional Oscillatory Case ($d=4$ and $k=4$)}
\label{sec:d-4_k-4}
 
\paragraph{Tree-PCE metamodel} The Tree-PCE algorithm is applied with local polynomials of order \( p_{\text{loc}} = 3 \) and then \( p_{\text{loc}} = 5 \), on a training set of size $N_{\text{train}} = 2000$.

\begin{figure}[H]
    \centering
   \includegraphics[width=\linewidth]{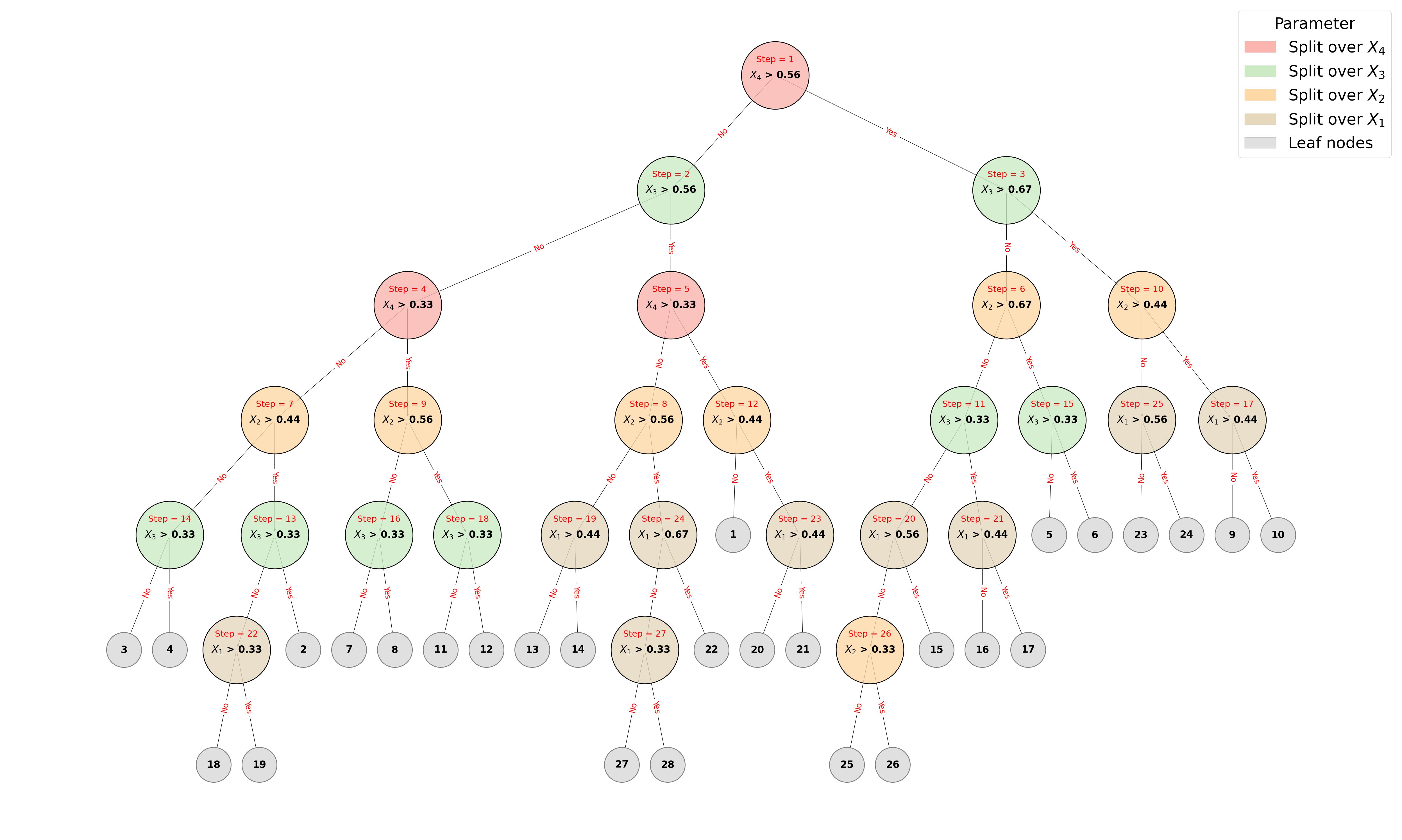}
    \caption{Tree structure of the output~\eqref{model_multid} with \(d = 4\), \(k = 4\) and \(p_\mathrm{loc} = 3\)}
    \label{fig:tree-d4-k4-p3}
\end{figure}

\begin{figure}[H]
    \centering
   \includegraphics[width=\linewidth]{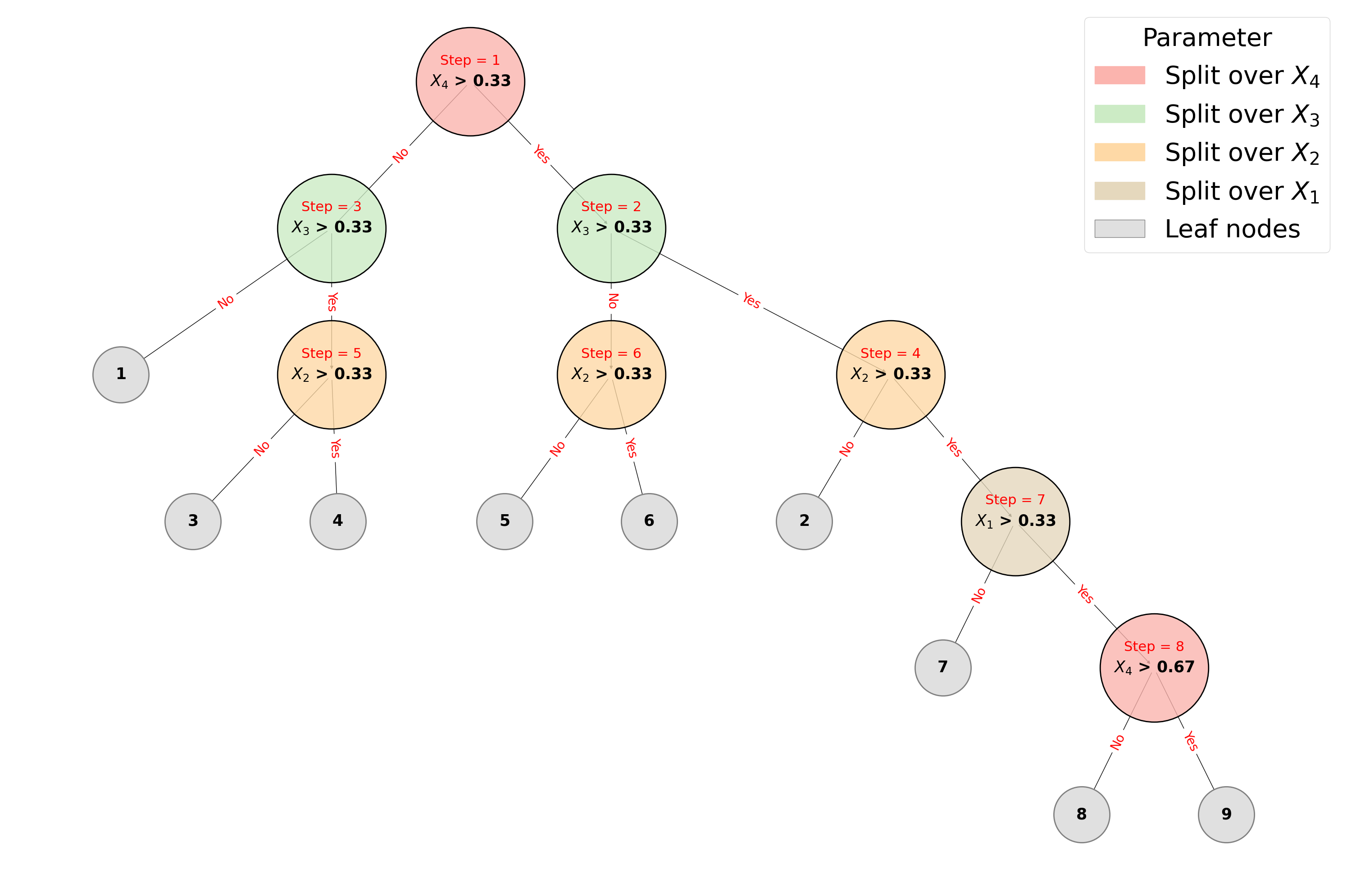}
    \caption{Tree structure of the output~\eqref{model_multid} with \(d = 4\), \(k = 4\) and \(p_\mathrm{loc} = 5\)}
    \label{fig:tree-d4-k4-p5}
\end{figure}

The function exhibits strong local oscillations within each subdomain, which significantly increases the complexity of the response. In this context, using a local polynomial degree of 3 is not sufficient to capture both the oscillatory behavior and the underlying discontinuity at \(x_i = \tfrac{1}{3}\). This limitation is illustrated in the tree structure presented in Figure~\ref{fig:tree-d4-k4-p3}, where Tree-PCE performs splits along the most influential parameters but fails to detect the true discontinuity. The resulting partitioning does not align with the structural break, which leads to suboptimal localization of the response.
When the local degree is increased to 5 (Figure~\ref{fig:tree-d4-k4-p5}), the improved approximation capacity allows the Tree-PCE algorithm to better represent local variations and successfully identify the discontinuity point. In this case, the tree structure shows a split that aligns with the location of the discontinuity, demonstrating the importance of having a sufficiently rich local polynomial basis in highly oscillatory settings.
However, the refinement of the partition is limited by the available data. In this experiment, the algorithm stops after producing 9 classes, as the remaining samples in the candidate leaves are insufficient to support further splits according to the minimum node size constraint $n_\mathrm{min} = 3 \binom{p_\mathrm{loc}+d}{p_\mathrm{loc}}$, which increases fast with $p_\mathrm{loc}$. This highlights a trade-off between model complexity and data availability: while a higher local degree improves accuracy and structural detection, it also reduces the number of admissible partitions due to increased sample requirements.


\begin{figure}[H]
    \centering
    \includegraphics[width=0.45\linewidth]{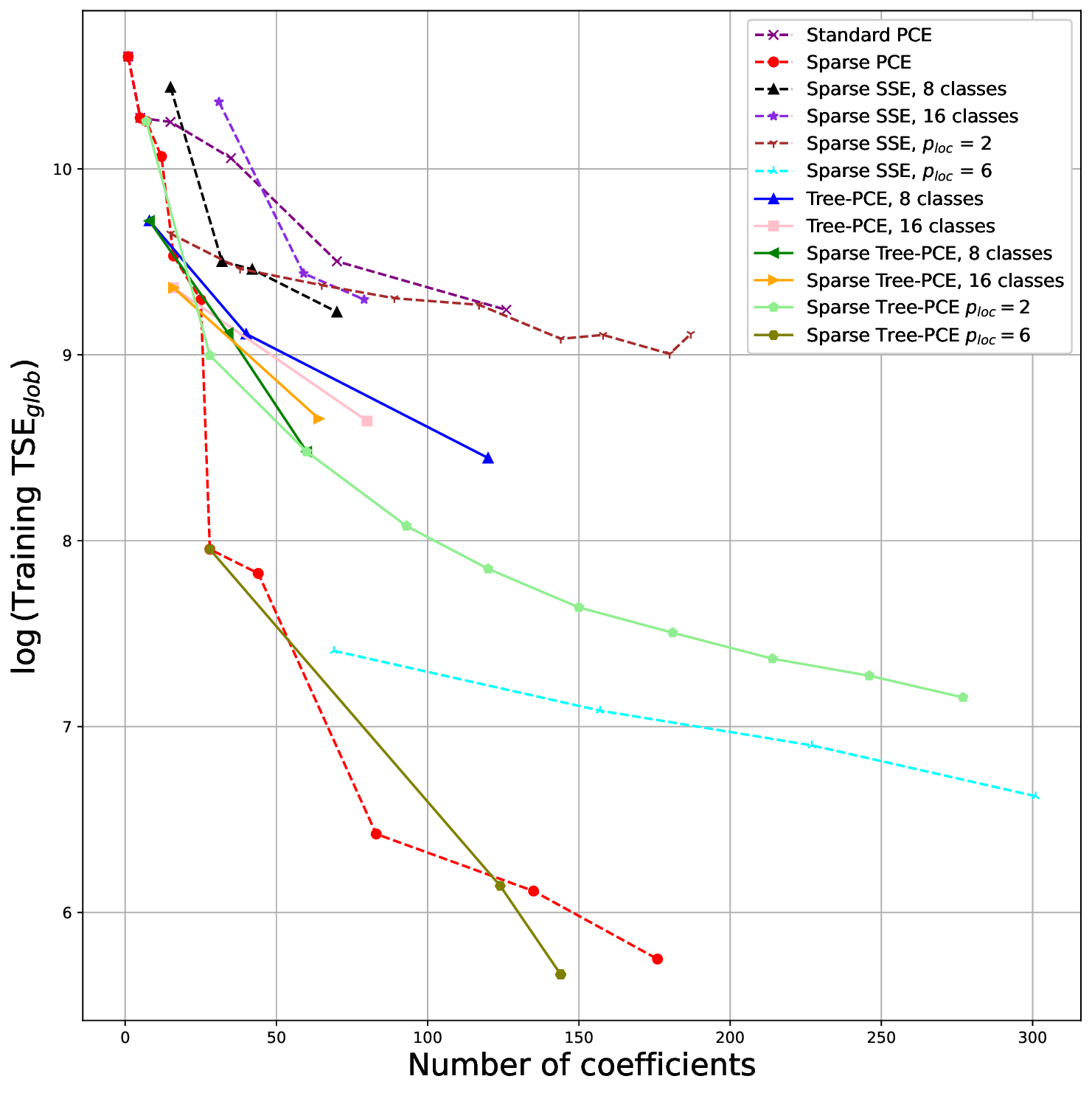}
    \includegraphics[width=0.45\linewidth]{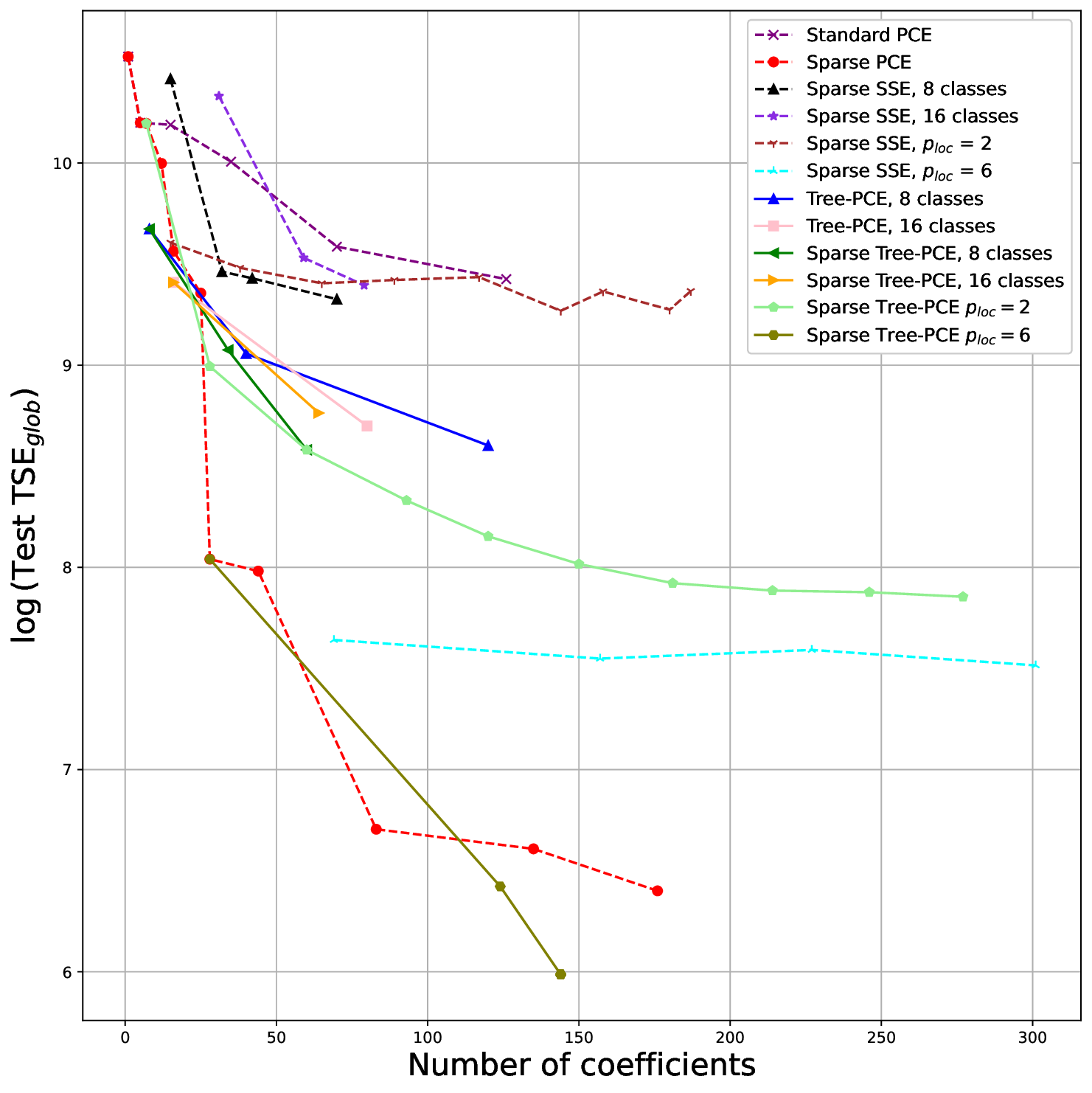}
    \caption{Methods comparison for the output~\eqref{model_multid} with \(d = 4\) and \(k = 4\)}
    \label{fig:Comparison-4dk4}
\end{figure}

\paragraph{Performance comparison}\  We see in Figure~\ref{fig:Comparison-4dk4} that sparse Tree-PCE with \(p_{\text{loc}} = 6\) achieves the best performance in terms of error, thanks to its adaptive partitioning and the expressiveness of high-degree local polynomials. However, this accuracy comes with a high complexity, due to the large number of coefficients.
Sparse PCE remains competitive in terms of accuracy but also requires many coefficients.
Sparse SSE improves significantly with \(p_{\text{loc}} = 6\), but still performs slightly worse, with a comparable level of complexity.
Variants with low polynomial degree or a small number of classes perform less effectively. The choice of method thus depends on the desired trade-off between accuracy and complexity.

\subsubsection{High-Dimensional Smooth Case ($d=10$ and $k=1$)}\label{sec:d-10_k-1-c-1}

\paragraph{Tree-PCE metamodel} The Tree-PCE algorithm is applied with local polynomials of order \( p_{\text{loc}} = 2 \), on a training set of size $N_{\text{train}} = 5000$. 

\begin{figure}[H]
    \centering
    \includegraphics[width=1\linewidth]{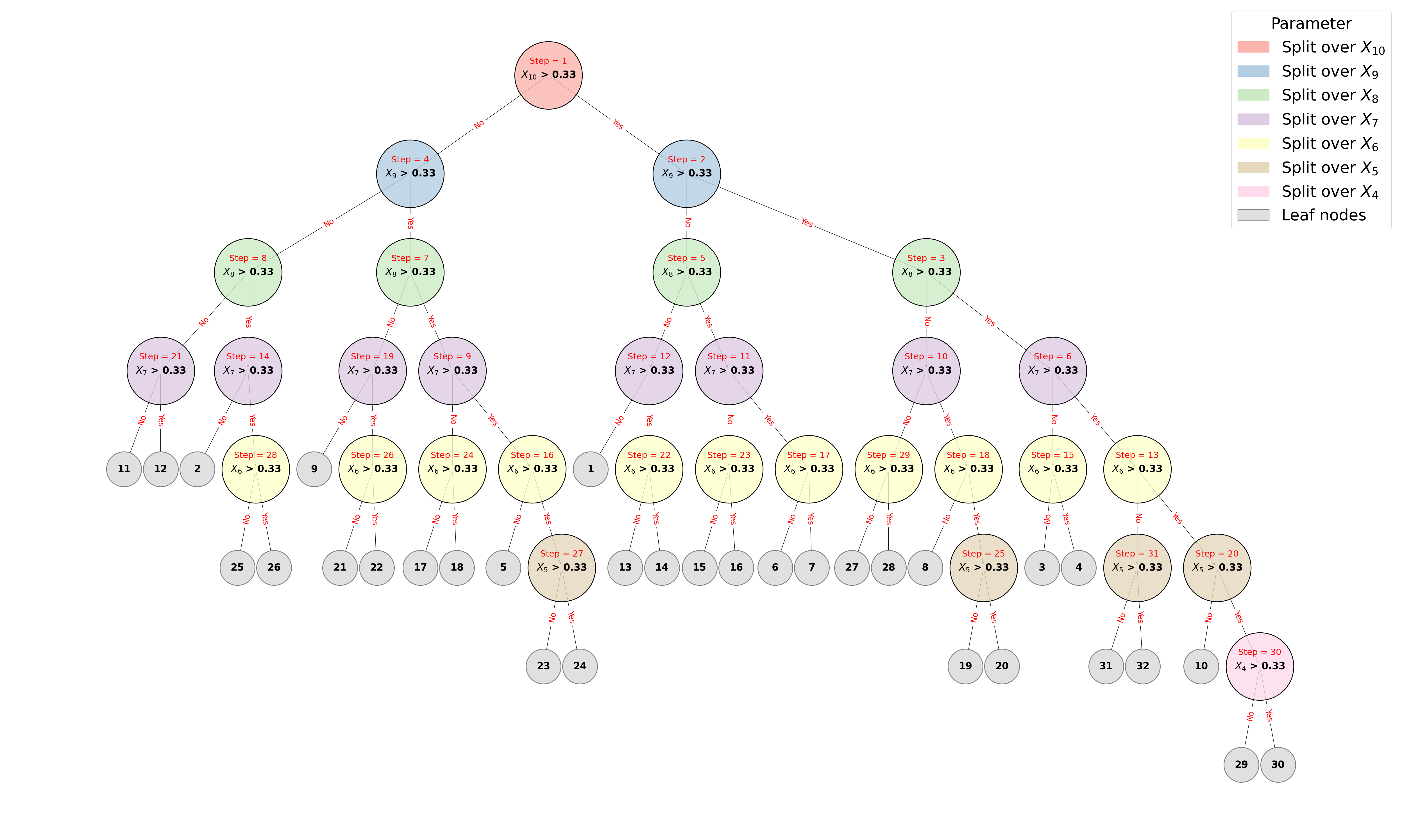}
    \caption{Tree structure for the case with \(d = 10\) and \(k = 1\)}
    \label{fig:tree-d10-k1-c1}
\end{figure}

In this example, the function exhibits smooth behavior, but the complexity arises from the high dimensionality. From the tree structure (Figure~\ref{fig:tree-d10-k1-c1}), we observe that the Tree-PCE succeeds in detecting discontinuity points by first splitting on the most influential parameters. The global $\mathrm{TSE}$ decreases as the number of classes increases, but we do not observe a distinct peak beyond which further improvements become negligible; such behavior only becomes noticeable when a sufficiently large number of classes is allowed. In theory, resolving all possible combinations of binary discontinuities would require up to $2^{10} = 1024$ classes, which would imply a huge model complexity and would require an even larger training dataset.

\paragraph{Performance comparison}

\begin{figure}[H]
    \centering
    \includegraphics[width=0.45\linewidth]{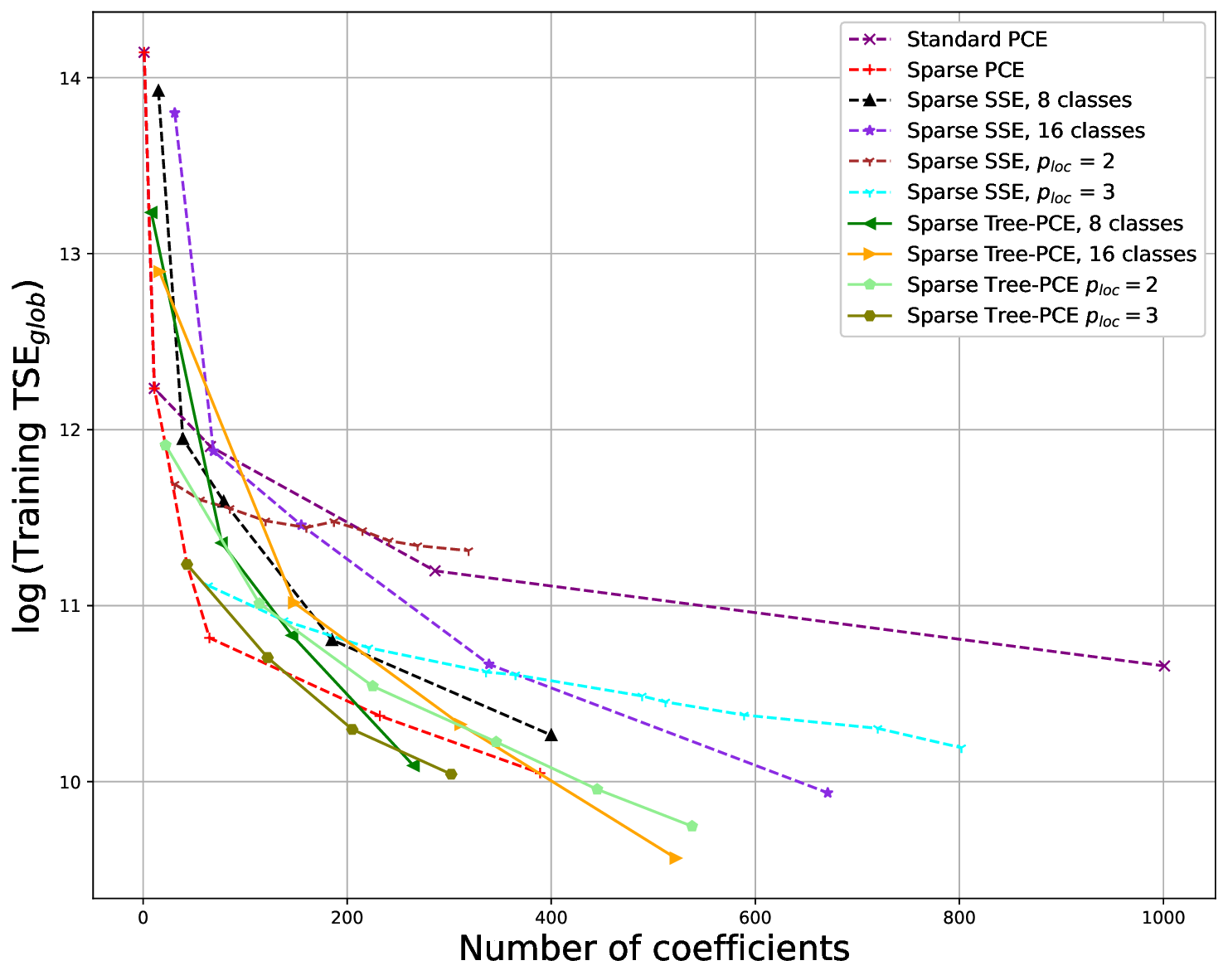}
    \includegraphics[width=0.45\linewidth]{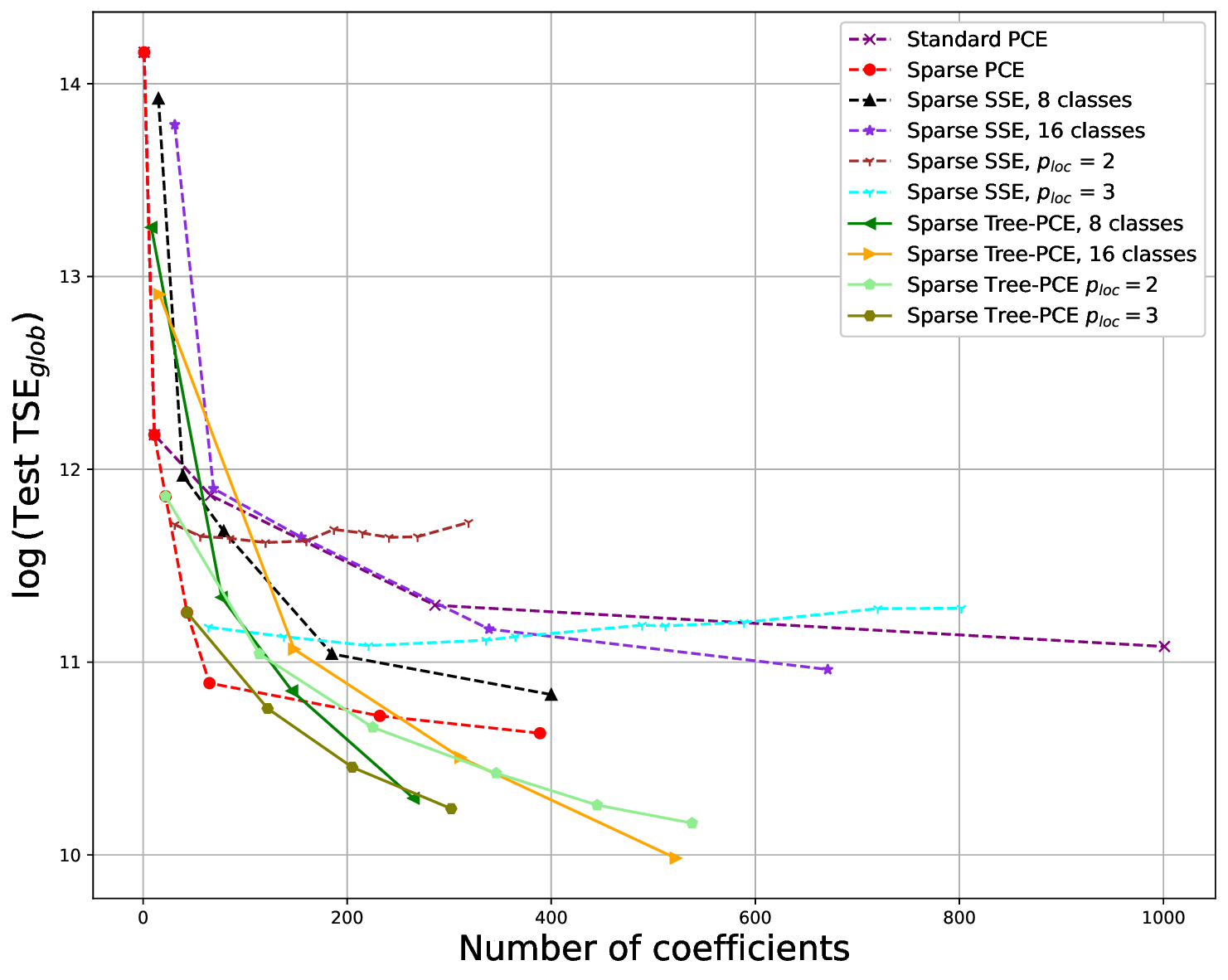}
    \caption{Methods comparison for the output~\ref{model_multid} with \(d = 10\) and \(k = 1\)}
    \label{fig:Comparison-10dk1c1}
\end{figure}
Figure~\ref{fig:Comparison-10dk1c1} compares the performance of the different methods. In this example, we focus on Tree-PCE with the sparse option, as it was found to be more efficient in the previous comparison. Both sparse Tree-PCE and sparse PCE deliver the best results, with Tree-PCE achieving slightly superior performance by consistently yielding lower global $\mathrm{TSE}$ values than standard PCE, sparse PCE, and SSE. While sparse PCE remains competitive thanks to its reduced number of terms and improved accuracy, it is still a global method and may fail to capture localized features. Conversely, SSE continues to struggle, especially when the local polynomial degree is fixed at 2 or 3, as these degrees are insufficient to capture the complex behavior of the model. Furthermore, because its splitting strategy always occurs at the median, SSE cannot directly detect discontinuities, limiting its ability to represent the function complex behavior.
\subsubsection{Summary of Results}

Across all the tested cases, the Tree-PCE method demonstrates strong performance, consistently achieving lower errors with fewer coefficients compared to standard PCE, sparse PCE, and SSE. Its adaptive partitioning and local polynomial expansions allow it to effectively capture both smooth variations and localized features, including discontinuities. Sparse PCE remains competitive, particularly in non-smooth, oscillatory problems, but its global nature limits its ability to handle localized phenomena as efficiently as Tree-PCE. SSE benefits from local modeling; however, its strategy of always splitting at the median reduces its effectiveness in detecting discontinuities. Overall, Tree-PCE offers the best balance between accuracy and model complexity, making it a promising approach for high-dimensional and complex functions.
Notably, other tests performed with 
a higher value of the discontinuity amplitude $c$ show that the performance of sparse PCE degrades when handling large discontinuities, while Tree-PCE remains insenstitive to the discontinuity amplitude, provided that sufficiently many classes have been identified to remove discontinuities from local models. Tree-PCE’s adaptive splitting captures sharp transitions effectively, whereas sparse PCE struggles due to its global nature, leading to higher errors.

\section{Application to Sensitivity Analysis}\label{sec5}

When the input variables $X_1, \ldots, X_d$ are independent, PCEs are commonly used in global sensitivity analysis to compute Sobol' indices directly from expansion coefficients, with no extra computational cost. This section revisits the standard PCE-based formulas, extends them to the Tree-PCE framework for handling complex models, and introduces Tree-PCE sensitivity indices, a novel set of measures derived from the tree structure itself, offering a new view on input importance.

\subsection{Sobol' Indices}\label{subsec5_1}

Throughout this subsection, we assume that the components of the input vector $\mathbf{X}=(X_1, \dots, X_d)$ are independent. In this setting, a central mathematical framework in global sensitivity analysis is the functional decomposition of a model response into orthogonal components associated with individual input variables and their interactions. Notably, the Hoeffding decomposition \eqref{Hoeffding} \citep{Hoeff1992}and the ANOVA (ANalysis Of VAriance) decomposition \eqref{ANOVA} \citep{St1989} provide the theoretical foundation for this approach and serve as the basis for the definition and computation of Sobol' sensitivity indices.

Let $\mathcal{P}_d$ be the set of all subsets of $\{1, \ldots, d\}$. Assume that $G \in L^2\left(\R^d, P_{\mathbf{X}}\right)$, and for any $A \in \mathcal{P}_d$, let us denote by $\mathbf{X}_A$ the random vector $(X_i)_{i \in A}$. The Hoeffding decomposition is the unique decomposition of $Y=G(\mathbf{X})$ of the form 
\begin{equation}\label{Hoeffding}
    G(\mathbf{X})=\sum_{A \in \mathcal{P}_d} G_A(\mathbf{X}_A),
\end{equation}
such that the random variables $(G_A(\mathbf{X}_A)_{A \in \mathcal{P}_d}$ are orthogonal. Writing $V_A = \Var(G_A(\mathbf{X}_A))$, we deduce the ANOVA decomposition
\begin{equation}\label{ANOVA}
    \Var (G(\mathbf{X}))=\sum_{A \in \mathcal{P}_d} V_A.
\end{equation} 
From these decompositions, the first order Sobol' index associated to $\mathbf{X}_A$ is defined as 
        $$S_A = \frac{\Var\left(\Exp[G(\mathbf{X})|\mathbf{X}_A]\right)}{\Var\left( G(\mathbf{X})\right)}= \frac{\sum_{A' \subset A} V_{A'}}{V},$$
and the total Sobol' index associated to $\mathbf{X}_A$ is defined as $$S_A^T = 1- S_{\overline{A}}= \frac{\sum_{A' \cap A \ne \emptyset} V_{A'}}{V}$$
with $\overline{A}= \{1, \ldots, d\}\backslash  A$.

Among the available methods for estimating Sobol' indices, one prominent approach is using pick-freeze estimator \citep{Saltelli2002}, which relies on Monte Carlo simulations.
While this method provides reliable results with associated confidence intervals, it is computationally expensive due to the large number of simulations required to achieve accurate estimates. To mitigate this issue, PCE offers a viable alternative. PCE has the advantage of being less computationally costly, as Sobol' indices can be directly derived from the coefficients of the metamodel, eliminating the need for extensive simulations while maintaining accuracy.

\subsubsection{PCE-Based Derivation of Sobol' Sensitivity Indices}\label{sss:sobol-pce}

The derivation of Sobol' indices from PCE was first introduced by \citep{Sudret2008}. For completeness, the key elements are briefly recalled here.

Given a PCE metamodel of the form
\[
    G^\mathrm{PCE}(\mathbf{x}) = \sum_{\boldsymbol{\alpha} \in \mathcal{A}} y_{\boldsymbol{\alpha}} \Psi_{\boldsymbol{\alpha}}(\mathbf{x}),
\]
the associated Hoeffding decomposition writes
\[
    G^\mathrm{PCE}(\mathbf{X}) = \sum_{A \in \mathcal{P}_d} G^\mathrm{PCE}_A(\mathbf{X}_A),
\]
where for any $A \in \mathcal{P}_d$,
\[
    G^\mathrm{PCE}_A(\mathbf{X}_A) = \sum_{\boldsymbol{\alpha} \in \mathcal{A}_A} y_{\boldsymbol{\alpha}} \Psi_{\boldsymbol{\alpha}}(\mathbf{X}_A), \qquad \mathcal{A}_A = \{\boldsymbol{\alpha} \in \mathcal{A}: \text{$\alpha_i>0$ if and only if $i \in A$}\}.
\]

As a consequence, the Sobol' indices 
\begin{align*}
    S^\mathrm{PCE}_A &= \frac{\displaystyle\sum_{A' \in \mathcal{P}_d, A' \subset A} \Var\left(G^\mathrm{PCE}_{A'}(\mathbf{X}_{A'})\right)}{\displaystyle\sum_{A' \in \mathcal{P}_d} \Var\left(G^\mathrm{PCE}_{A'}(\mathbf{X}_{A'})\right)} = \frac{\Var\left(\Exp\left[G^\mathrm{PCE}(\mathbf{X})|\mathbf{X}_A\right]\right)}{\Var\left(G^\mathrm{PCE}(\mathbf{X})\right)},\\
    S^{\mathrm{PCE},T}_A &= \frac{\displaystyle\sum_{A' \in \mathcal{P}_d, A' \cap A \not= \emptyset} \Var\left(G^\mathrm{PCE}_{A'}(\mathbf{X}_{A'})\right)}{\displaystyle\sum_{A' \in \mathcal{P}_d} \Var\left(G^\mathrm{PCE}_{A'}(\mathbf{X}_{A'})\right)} = 1-S^\mathrm{PCE}_{\overline{A}},
\end{align*}
may be obtained directly from the coefficients of the PCE decomposition by the formulas
\[
    S^\mathrm{PCE}_A = \frac{\displaystyle \sum_{\substack{\boldsymbol{\alpha} \in \mathcal{A}, \boldsymbol{\alpha} \not= \boldsymbol{0}\\ \forall i \in \overline{A}, \alpha_i = 0}}y_{\boldsymbol{\alpha}}^2}{\displaystyle \sum_{\boldsymbol{\alpha} \in \mathcal{A}, \boldsymbol{\alpha} \not= \boldsymbol{0}}y_{\boldsymbol{\alpha}}^2}, \qquad S_A^{\mathrm{PCE},T} = \frac{\displaystyle \sum_{\substack{\boldsymbol{\alpha} \in \mathcal{A}, \boldsymbol{\alpha} \not= \boldsymbol{0}\\ \exists i \in A: \alpha_i > 0}}y_{\boldsymbol{\alpha}}^2}{\displaystyle \sum_{\boldsymbol{\alpha} \in \mathcal{A}, \boldsymbol{\alpha} \not= \boldsymbol{0}}y_{\boldsymbol{\alpha}}^2}.
\]
   
\subsubsection{Tree-PCE-Based Derivation of Sobol' Sensitivity Indices}

The Tree-PCE metamodel obtained by the algorithm described in Section~\ref{ss:describ-TreePCE} writes
\[
    G^\mathrm{Tree-PCE}(\mathbf{x}) = \sum_{r=1}^R \ind{\mathbf{x} \in \mathcal{R}^r} G^\mathrm{PCE}_{\mathcal{R}^r}(\mathbf{x}), \qquad G^\mathrm{PCE}_{\mathcal{R}^r}(\mathbf{x}) = \sum_{\boldsymbol{\alpha} \in \mathcal{A}^r} y_{\boldsymbol{\alpha}}^r \Psi_{\boldsymbol{\alpha}}^r(\mathbf{x}).
\]
The associated Sobol' indices
\begin{align*}
    S^\mathrm{Tree-PCE}_A &= \frac{\displaystyle\sum_{A' \in \mathcal{P}_d, A' \subset A} \Var\left(G^\mathrm{Tree-PCE}_{A'}(\mathbf{X}_{A'})\right)}{\displaystyle\sum_{A' \in \mathcal{P}_d} \Var\left(G^\mathrm{Tree-PCE}_{A'}(\mathbf{X}_{A'})\right)} = \frac{\Var\left(\Exp[G^\mathrm{Tree-PCE}(\mathbf{X})|\mathbf{X}_A]\right)}{\Var\left(G^\mathrm{Tree-PCE}(\mathbf{X})\right)},\\
    S^{\mathrm{Tree-PCE},T}_A &= \frac{\displaystyle\sum_{A' \in \mathcal{P}_d, A' \cap A \not= \emptyset} \Var\left(G^\mathrm{Tree-PCE}_{A'}(\mathbf{X}_{A'})\right)}{\displaystyle\sum_{A' \in \mathcal{P}_d} \Var\left(G^\mathrm{Tree-PCE}_{A'}(\mathbf{X}_{A'})\right)} = 1-S^\mathrm{Tree-PCE}_{\overline{A}},
\end{align*}
can be expressed in terms of the coefficients $y^r_{\boldsymbol{\alpha}}$ thanks to the formulas given in Proposition~\ref{prop:sobol-tree} below. In this statement, we assume that for any $r \in \{1, \ldots, R\}$, the local multivariate basis $(\Psi^r_{\boldsymbol{\alpha}})_{\boldsymbol{\alpha} \in \N^d}$ is given by the tensor product of univariate polynomials, namely
\[
    \Psi^r_{\boldsymbol{\alpha}}(\mathbf{x}) = \prod_{i=1}^d \phi_{\alpha_i}^{(i),r}(x_i), \qquad \boldsymbol{\alpha} = (\alpha_1, \ldots, \alpha_d),
\]
as in Section~\ref{ss:construction-multivariate-basis}. Moreover, we write each rectangle $\mathcal{R}^r$ as the Cartesian product
\[
    \mathcal{R}^r = \prod_{i=1}^d \mathcal{I}^{(i),r}
\]
of intervals $\mathcal{I}^{(i),r}$.

\begin{prop}[Analytical formulas for Sobol' indices]\label{prop:sobol-tree}
    We have
    \[
        \Var\left(G^\mathrm{Tree-PCE}(\mathbf{X})\right) = \sum_{r=1}^R \left( \sum_{\boldsymbol{\alpha} \in \mathcal{A}^r} (y_{\boldsymbol{\alpha}}^r)^2\right) \Pr(\mathbf{X} \in \mathcal{R}^r) - \left(\sum_{r=1}^R y^r_{\boldsymbol{0}} \Pr(\mathbf{X} \in \mathcal{R}^r)\right)^2,
    \]
    with the convention that $y^r_{\boldsymbol{0}}=0$ if $\boldsymbol{0} \not\in \mathcal{A}^r$.
    
    Besides, for any $A \in \mathcal{P}_d$,
    \begin{align}    &\Var\left(\Exp\left[G^\mathrm{Tree-PCE}(\mathbf{X})|\mathbf{X}_A\right]\right) \label{eq:Sobol_formula}\\
        &= \sum_{r,r'=1}^R \left(\sum_{\substack{\boldsymbol{\alpha}, \boldsymbol{\alpha}' \in \mathcal{A}^r\\ \forall i \in \overline{A}, \alpha_i=\alpha_i'=0}} y_{\boldsymbol{\alpha}}^r y_{\boldsymbol{\alpha}'}^{r'} \prod_{i \in A} J^{(i),r,r'}_{\alpha_i,\alpha_i'}\right)\Pr\left(\mathbf{X}_{\overline{A}} \in \mathcal{R}^r_{\overline{A}}\right)\Pr\left(\mathbf{X}_{\overline{A}} \in \mathcal{R}^{r'}_{\overline{A}}\right) - \left(\sum_{r=1}^R y^r_{\boldsymbol{0}} \Pr(\mathbf{X} \in \mathcal{R}^r)\right)^2,\notag
    \end{align}
    where
    \[
        \mathcal{R}^r_{\overline{A}} = \prod_{i \in \overline{A}} \mathcal{I}^{(i),r},
    \]
    and
    \begin{equation}\label{def_Js}
        J^{(i),r,r'}_{\alpha_i,\alpha_i'} = \Exp\left[\ind{X_i \in \mathcal{I}^{(i),r} \cap \mathcal{I}^{(i),r'}} \phi_{\alpha_i}^{(i),r}(X_i) \phi_{\alpha_i'}^{(i),r'}(X_i)\right].
    \end{equation}
        
\end{prop}

Similar formulas for SSE are described in~\cite[Appendix~A]{Marelli2021}. In contrast with the formulas given in Section~\ref{sss:sobol-pce} for $S^\mathrm{PCE}_A$ and $S^{\mathrm{PCE},T}_A$, the numerical evaluation of $S^\mathrm{Tree-PCE}_A$ and $S^{\mathrm{Tree-PCE},T}_A$ using the formulas from Proposition~\ref{prop:sobol-tree} requires the computation of supplementary terms beyond the coefficients $y^r_{\boldsymbol{\alpha}}$, namely terms of the form $\Pr(\mathbf{X} \in \mathcal{R}^r)$, $\Pr(\mathbf{X}_{\overline{A}} \in \mathcal{R}^r_{\overline{A}})$ and $J^{(i),r,r'}_{\alpha_i,\alpha_i'}$. All these terms can be decomposed as products of one-dimensional integrals of terms of the form $x^k f_{X_i}(x)$, so in principle their computation can be done with very good accuracy as described in Appendix~\ref{app:D}. However, the number of such terms to compute may become huge as soon as $R$ or $d$ are large, so in practice the approximation of Sobol' indices based on these analytical formulas may turn out to be intractable. Still, in the situation where the evaluation of the actual model $G$ is costly, the Tree-PCE metamodel $G^\mathrm{Tree-PCE}$ may be employed as a surrogate to $G$ to generate samples from $(\mathbf{X}, G^\mathrm{Tree-PCE}(\mathbf{X}))$, which may then be used to estimate Sobol' indices by the pick-freeze method.

\begin{rk}
    Let us consider the case $\mathcal{A}^r=\{ \alpha \in \N^d, |\alpha|\le p \}$ with $p\in \N$, for all $r$.  To calculate first order (resp. total)  Sobol' indices with $A=\{i\}$ (resp. $\bar{A}=\{i\}$), the number of terms to evaluate in the first sum of~\eqref{eq:Sobol_formula} is $R^2p^2$ (resp. $R^2 \frac{(d-1+p)!}{(d-1)!p!})$, and the evaluation of each term requires to calculate one value (resp. $d-1$ values)  of~\eqref{def_Js}. Thus, these formulas are computationally efficient to calculate Sobol' indices when~$R$ is not too large, otherwise the pick-freeze method has to be preferred. 
\end{rk}

\subsection{Innovative sensitivity indices based on Tree-PCE}

The tree structure produced by the Tree-PCE algorithm gives a detailed information on the sensitivity. For example, in Figure~\ref{fig:Comparison-4dk1}, we see that the first split is on $X_4$, then on $X_3$, etc. This gives a hierarchy of the inputs' influence in the different regions of the domain. However, the information given by the tree is very rich, and we would like to summarize the main facts by indices. 
In addition to the well-established Sobol' sensitivity measures, we introduce a novel importance metric derived directly from the Tree-PCE algorithm: the Tree-PCE sensitivity index. This index remains applicable in the presence of dependent input variables and quantifies the influence of each input by evaluating its contribution to the surrogate model's construction. 

The Tree-PCE sensitivity index associated with an input variable $X_i$ is defined from the cumulative reduction in residual error achieved by all binary splits performed along the \(i\)th direction during the recursive partitioning of the input space. Since Tree-PCE prioritizes splits that most effectively reduce the global approximation error, a large cumulative contribution from \( X_i \) indicates that the function \( X_i \mapsto G(\mathbf{X}) \) cannot be well-approximated by a low-degree polynomial across the entire domain. In contrast, a low contribution suggests that either \( G \) depends weakly on \( X_i \), or that the dependence is smooth and easily captured by low-degree polynomials. This index therefore reflects both the global importance of \( X_i \) and the local complexity of its effect on the output.

To define the Tree-PCE sensitivity index, we first introduce some notation. We denote by $\mathrm{TSE}_0 := \mathrm{TSE}_{\mathcal{D}_\mathbf{X}}$ the TSE of the global PCE metamodel constructed at the initialization of the algorithm (see Subsubsection~\ref{sss:tree-pce-steps}), and for any $k \geq 1$, by $\mathrm{TSE}_k$ the TSE of the global Tree-PCE metamodel obtained after the $k$-th recursive step. By construction,
\[
    \Delta \mathrm{TSE}_k := \mathrm{TSE}_{k-1} - \mathrm{TSE}_k > 0.
\]
We also denote by $i^*_k \in \{1, \ldots, d\}$ the index of the direction in which the split has been done at the $k$-th step, and by $K \geq 0$ the total number of steps of the algorithm. Notice that since $K$ denotes the number of internal nodes of the binary tree associated with the metamodel, and $R$ denotes the number of leaves, we always have $K=R-1$.

\begin{defi}[Tree-PCE sensitivity index]
Let $K\ge 1$ and $i^*_k$, $1\le k\le K$ defined as above. The Tree-PCE sensitivity index associated with the input $X_i$ is defined by
\[
    \text{Tree-PCE}_i := \frac{\displaystyle \sum_{k=1}^K \ind{i^*_k=i} \Delta \mathrm{TSE}_k}{\mathrm{TSE}_0}.
\]
\end{defi}
\noindent It is worth to stress that these indices can be calculated with no additional computational cost, since the quantities $\mathrm{TSE}_k$ are already calculated by the Tree-PCE Algorithm.

By construction, \(\text{Tree-PCE}_i \in [0,1]\) for all \(i \in \{1,\dots,d\}\). Moreover, the indices satistfy
\[
    \sum_{i=1}^d \text{Tree-PCE}_i \leq 1,
\]
and the residual term \(1 - \sum_{i=1}^d \text{Tree-PCE}_i\) corresponds to the portion of the initial error that could not be reduced by the algorithm, due to stopping criteria or the ineligibility of further splits.

\begin{rk}
    Tree-PCE indices depend on the local polynomial degree $p_\mathrm{loc}$: higher degrees enable more flexible approximations, capturing complex model behaviors and potentially reducing residual error more effectively. For example, the output $Y=\ind{X_1>1/2}(X_1^2+X_2^2)$ will lead to $\text{Tree-PCE}_1=1$ and $\text{Tree-PCE}_2=0$ if $p_\mathrm{loc}=2$ and $1/2 \in \mathcal{T}_1$ (i.e. $1/2$ is in the thresholds grid of the first component). Indeed, the output is reached exactly by the Tree-PCE after one iteration. If we take instead $p_\mathrm{loc}=1$, Tree-PCE algorithm will first identify $X_1>1/2$ as above, but will then go on splitting the rectangle  $[1/2,1]\times[0,1]$ in both directions to approximate $X_1^2+X_2^2$ by piecewise affine function. We therefore get $\text{Tree-PCE}_1<1$ and $0<\text{Tree-PCE}_2$ in this case, as well as $\text{Tree-PCE}_1+\text{Tree-PCE}_2<1$ since the Tree-PCE model cannot coincide exactly with the output~$Y$.
\end{rk}

Roughly speaking, $\text{Tree-PCE}_i$ indicates how splitting the domain with respect to the coordinate $X_i$ has been useful to reduce the TSE of the metamodel for a given polynomial basis.  

These sensitivity indices provide an alternative means of assessing input importance, useful for ranking and screening. Geometrically, the Tree-PCE index reflects the distance between the true model response and the space of polynomials of degree at most  \( p_\mathrm{loc} \). In each subdomain, the algorithm minimizes the \( L^2\) residual error, representing this distance under the input distribution. Inputs that significantly reduce this error through splitting are deemed important, as they enhance the surrogate model's accuracy.

\subsection{Application example}

To benchmark the described metamodel-based sensitivity analysis methods, we select the low dimensional oscillatory case ($d=4$, $c=1$ and $k=4$) previously presented in Section \ref{sec:d-4_k-4}. First, Sobol' indices are computed using both Tree-PCE and Sparse PCE, and the results are compared with analytical reference values derived from the exact decomposition of the model output variance:

\[S_j = S_j^T = \frac{ a_j^2}{\sum_{i=1}^d a_i^2}.\]

\begin{figure}[H]
    \centering
    \includegraphics[width=0.5\linewidth]{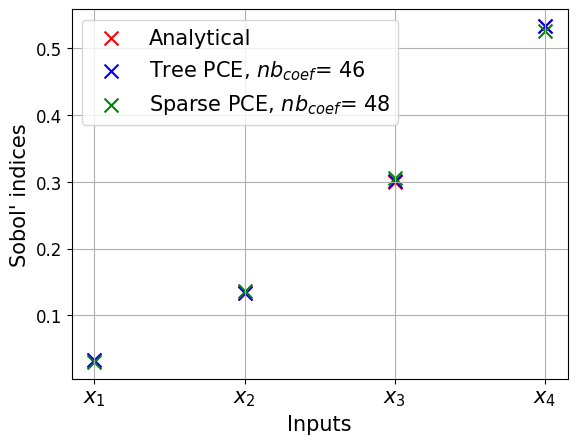}
    \caption{Sobol' indices for the output~\eqref{model_multid} with $d=4$ and $k=4$.}
    \label{fig:Sobol_indices_d4_k4}
\end{figure}

For Tree-PCE, we use $N = 2000, p_\mathrm{loc} = 2, nb_{classes} = 10$, \( \mathcal{T}_i = \left(\frac{k}{9}\right)_{0 < k < 9} \) for all $1\le i\le d$. For sparse PCE, we use $N=2000$, $p=8$. Figure~\ref{fig:Sobol_indices_d4_k4} shows that both Tree-PCE and sparse PCE, with a similar number of coefficients, successfully estimate the analytical Sobol' indices.

\begin{figure}[H]
    \centering
 \includegraphics[width=0.6\textwidth]{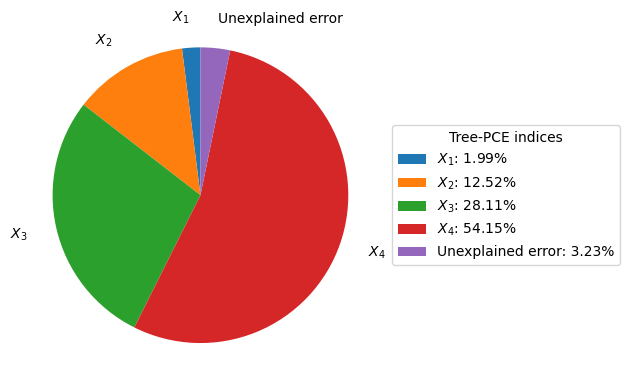}
    \caption{Tree-PCE indices the output~\eqref{model_multid} with $d=4$, $k=4$, and \( p_{\text{loc}} = 2 \). }
    \label{fig:Treepceindices}
\end{figure}

The pie chart in Figure~\ref{fig:Treepceindices} presents the Tree-PCE indices computed using \( p_{\text{loc}} = 2 \). The algorithm stops after 65 classes. We observe that the variable ranking matches that of the Sobol' indices. Only 3.23\% of the $\mathrm{TSE}$ remains unexplained, indicating an excellent fit of the metamodel. 

\section{Application to 2D morphodynamic model}\label{sec6}

In this section, we demonstrate the relevance of the proposed method for a representative morphodynamic engineering example exhibiting discontinuous response. This study examines an experiment proposed by \citep{Yen1995} involving sediment transport and bed evolution in a 180° curved channel. The setup features a laboratory flume with controlled flow conditions and sediment feed, enabling detailed observation of scour and deposition patterns under controlled inflow and sediment properties.

\subsection{Presentation of the channel bend study case}

The experimental setup consisted of a 180° channel bend with a constant radius curvature $r_c = 4$ m and bed width \( B = 1 \) m (Figure \ref{fig:Point_channel}). A 0.2 m thick uniform sand layer, with a median sediment diameter of \( d_{50} = 1 \) mm covered the channel bend. The experiment started with a flat bed and a constant longitudinal slope set at 0.002. The inlet discharge was unsteady, starting from a base flow of  0.002 m$^3$/s (corresponding to an initial water depth of 5.44 cm) and peaking at 0.0530 m$^3$/s approximately one-third into the 300 min duration.

To simulate the experiment, coupled simulations were used aligning the hydrodynamic 2D model, TELEMAC-2D \citep{Hervouet2007}, and the sediment transport and bed evolution model GAIA \citep{Gaia2023}, which are part of the openTELEMAC numerical platform (\url{www.opentelemac.org}). These module are used to solve respectively the Shallow Water and Exner equations complemented by boundary conditions. The sediment transport is assumed to be induced by bedload. For this purpose, the Meyer-Peter and Müller formula \citep{Meyer_1948} is considered in the following statements. In practical sediment transport modeling, empirical corrections are often applied to account for the magnitude and direction of sediment bedload transport. These formulations often rely on empirical coefficients that are not directly measurable and must be calibrated for each specific case, introducing uncertainty and limiting their general applicability. To address this issue, the following analysis focuses on key morphodynamic parameters such as porosity ($\lambda$), median sediment diameter ($d_{50}$), transport coefficients ($\alpha_{MPM}$), the critical Shields number ($\theta_{cr}$), and correction factors for bedload magnitude and direction ($\alpha_{ks}$, $\alpha_{sc}$, $\Phi$, and $\beta_2$) which are essential for improving the accuracy and physical consistency of sediment transport modeling. Table \ref{tab:uncertain_params} outlines the parameter variation ranges, derived either from established literature or from quantified measurement uncertainties.

\renewcommand{\arraystretch}{1.2}
\begin{table}[h!]
\centering
\scriptsize
\begin{tabular}{|c|c|c|c|}
\hline
\textbf{Definition} & \textbf{Parameter} & \textbf{Law} & \textbf{Reference} \\ \hline
Median sediment diameter (mm) & \( d_{50} \) & \( \mathcal{U}([1, 2]) \) & Measurement errors \\ \hline
Meyer-Peter \& Müller coefficient & \( \alpha_{MPM} \) & \( \mathcal{U}([3, 18]) \) & \cite{WongParker2006} \\
 & & & \cite{Zech2007}  \\ \hline
Critical Shields number & \( \theta_{cr} \) & \( \mathcal{U}([0.03, 0.06]) \) & \cite{bosboom2013coastaldynamics}\\ \hline
Bed material porosity & \( \lambda \) & \( \mathcal{U}([0.22, 0.42]) \) &\cite{Frings2011} \\ \hline
Skin friction parameter & \( \alpha_{k_s} \) & \( \mathcal{U}([1, 3]) \) &\cite{mendoza2016effect}\\ \hline
% Slope effect factor & \( \beta \) & \( \mathcal{U}([0.1, 5.0]) \) & Koch \& Flokstra (1981) (=5?) \\ \hline
Deviation angle & \( \beta_2 \) & \( \mathcal{U}([0.35, 1.5]) \) & \cite{talmon1995laboratory} \\ \hline
Secondary currents coefficient & \( \alpha_{sc} \) & \( \mathcal{U}([0.75, 1.0]) \) & GAIA User Manual \citep{Gaia2023} \\ \hline
Sediment angle of repose & \( \Phi \) & \( \mathcal{U}([25^\circ, 45^\circ]) \) & \cite{beakawi2021review}  \\\hline
\end{tabular}
\caption{The 8 uncertain input parameters, their probabilistic laws and sources. These parameters are assumed to be independent.}
\label{tab:uncertain_params}
\end{table}

\subsection{Tree-PCE metamodel}

\begin{figure}
    \centering
    \includegraphics[width=0.5\linewidth]{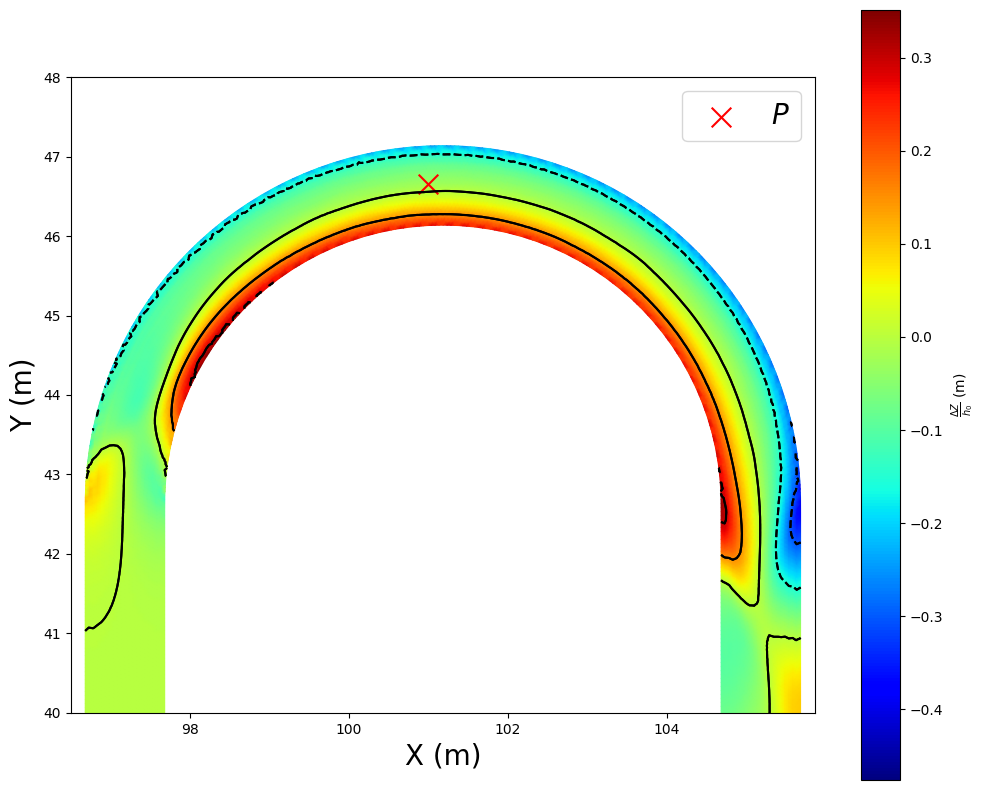}
    \caption{Illustration of the selected point P in the channel of the Yen experiment.  The color map indicates the bed elevation (red) or lowering (blue).}
    \label{fig:Point_channel}
\end{figure}

To handle dynamic system behavior under parameter uncertainty, a set of sample configurations is generated using random sampling. In the sampling procedure, the parameter uncertainties are taken in a uniform distribution whose limits are defined in able \ref{tab:uncertain_params}. The hydro-morphodynamic TELEMAC-2D/GAIA model ensures the relationship between a configuration vector of model uncertain inputs and the output quantity of interest, namely the channel bed elevation change. Hereafter, the results are presented on a selected point at the channel center, located at the transition between deposition and erosion zones at the end of the simulation (\( t = 400 \) min) (Figure~\ref{fig:Point_channel}). This location is characterized by complex sediment dynamics, where small input perturbations can induce significant bed elevation changes. Analyzing this sensitive zone provides critical insights into the impact of input uncertainties on morphodynamic responses.

\begin{figure}[H]
    \centering
    \includegraphics[width=0.5\linewidth]{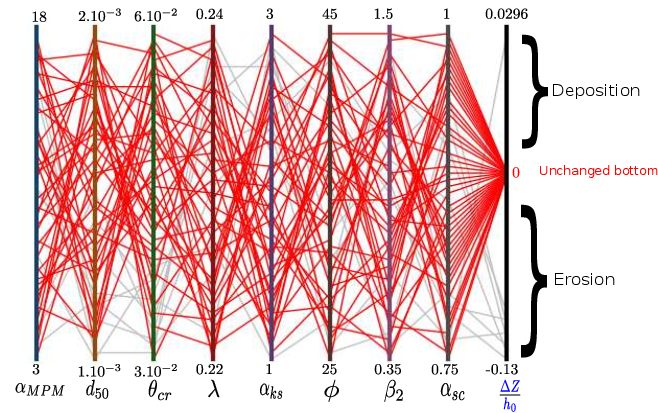}
    \caption{Parallel coordinates graph illustrating the relationship between bottom variation and the input parameters.}
    \label{fig:cob}
\end{figure}

\begin{figure}[H]
    \centering
    \includegraphics[width=1\linewidth]{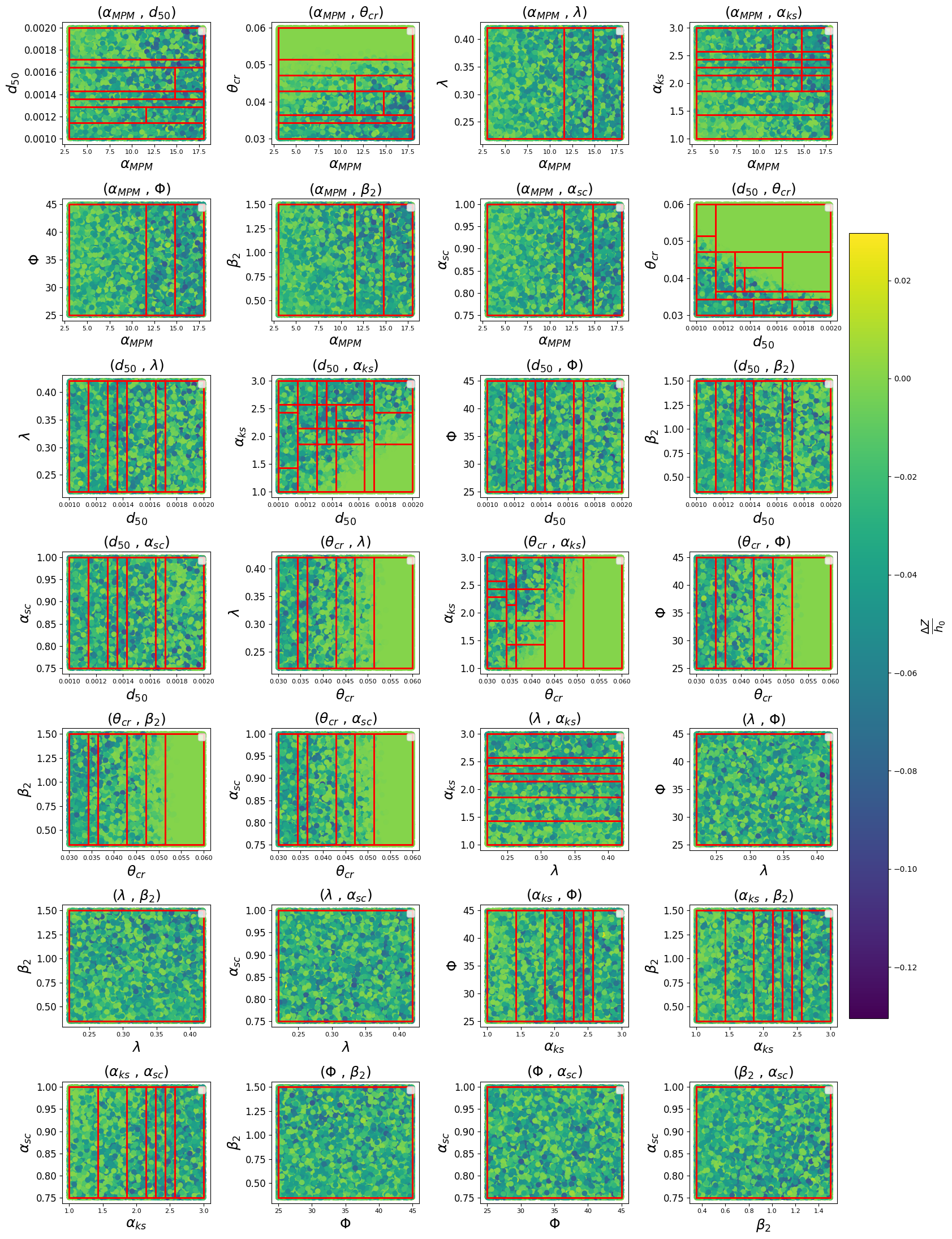}
    \caption{The $28=\binom{8}{2}$ 2d projections of the Tree-PCE partition. The color map indicates the bed elevation (yellow) or lowering (blue).}
    \label{fig:Yen-exp-Tree-PCE-decomp}
\end{figure}

The parallel coordinates plot (Figure~\ref{fig:cob}) reveals how specific input parameter combinations influence bed elevation, with red lines indicating configurations where no morphological change occurs. Complementarily, the 2D output sample plot (Figure~\ref{fig:Yen-exp-Tree-PCE-decomp}) illustrates model responses to input pairs: light green areas denote stable bed elevation, while multicolored regions highlight active sediment transport. These visualizations collectively expose the model’s sensitivity to input variations and delineate regions of the parameter space associated with morphological activity. Notably, the output exhibits discontinuities with respect to \( (d_{50}, \theta_{cr}, \alpha_{ks} )\), suggesting a threshold behavior beyond which input variations significantly affect bed elevation.

\begin{figure}[H]
    \centering
    \includegraphics[width=0.4\linewidth]{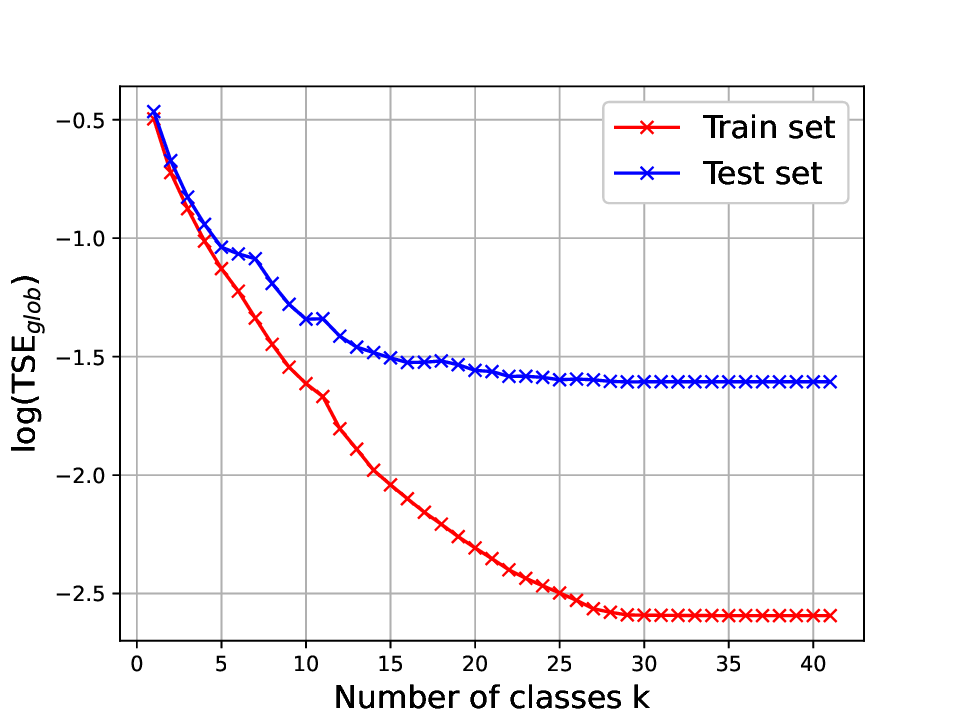}
    \caption{Global error evolution over steps.}
    \label{fig:Yen-exp-globalerror}
\end{figure}

\begin{figure}[H]
    \centering
    \includegraphics[width=1.1\linewidth]{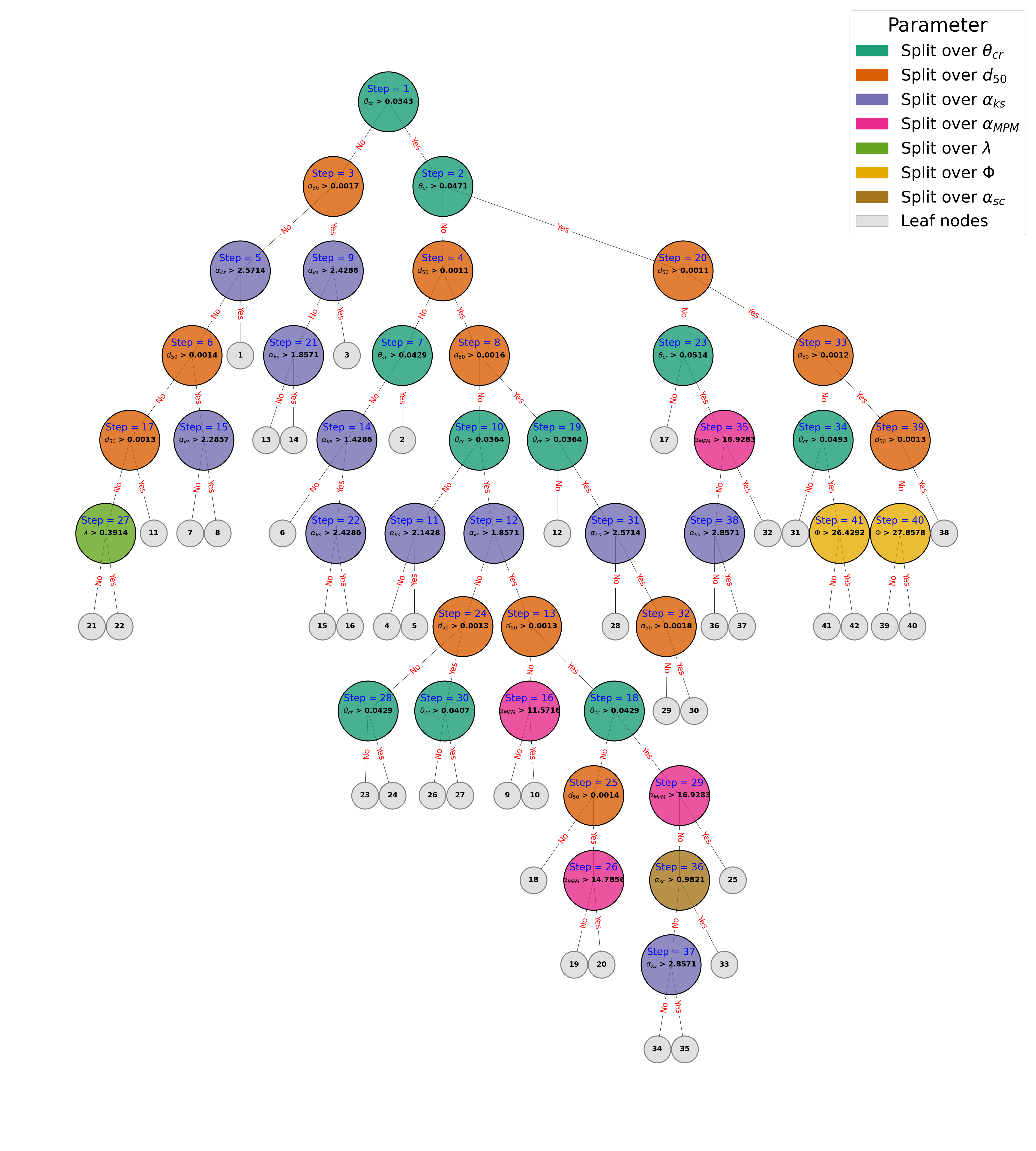}
    \caption{Binary Tree (obtained with $\epsilon =0$ in~\eqref{eq:stopping_criterion}).}
    \label{fig:Yeneps0}
\end{figure}

The Tree-PCE method was applied with \( N = 5000 \), a local polynomial degree \( p_{\text{loc}} = 2 \), and a tensorized grid defined as \(\mathcal{T} = \prod_{i=1}^d \mathcal{T}_{i} \) where \(\mathcal{T}_i = \left( x_{i,\min} + \frac{j}{14} (x_{i,\max} - x_{i,\min}) \right)_{0 < j < 14}\). The algorithm was run without a stopping criterion (\( \epsilon = 0 \)), allowing up to \( k_{\max} = 100 \) partitions, with sparsity enforced. Figure~\ref{fig:Yen-exp-globalerror} shows the evolution of the global TSE error, which stabilizes for \( k \geq 27 \), indicating diminishing improvement beyond this point. Figure~\ref{fig:Yeneps0} illustrates the hierarchical subdivision process, where all the partitions after step 27 are quite insignificant for the improvement of the Tree-PCE metamodel. The initial splits prioritize the most influential parameters, as later confirmed by Sobol' indices. Finally, Figure~\ref{fig:Yen-exp-Tree-PCE-decomp} presents the domain decomposition after the first 27 significant partitions, revealing the method’s ability to distinguish regions of constant bottom elevation from those with morphological variation, thereby demonstrating its adaptive refinement capabilities.


\begin{figure}[H]
    \centering
    \includegraphics[width=0.45\linewidth]{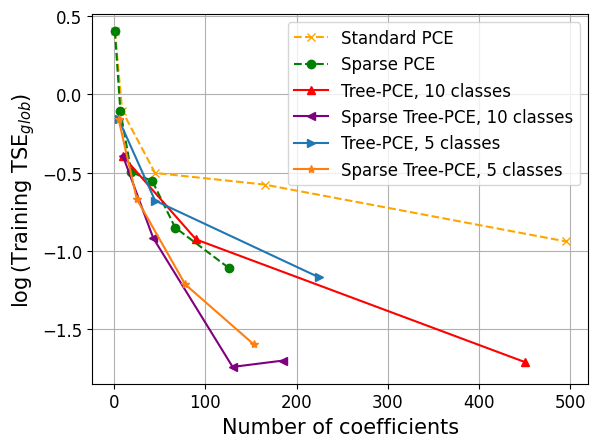}
    \includegraphics[width=0.45\linewidth]{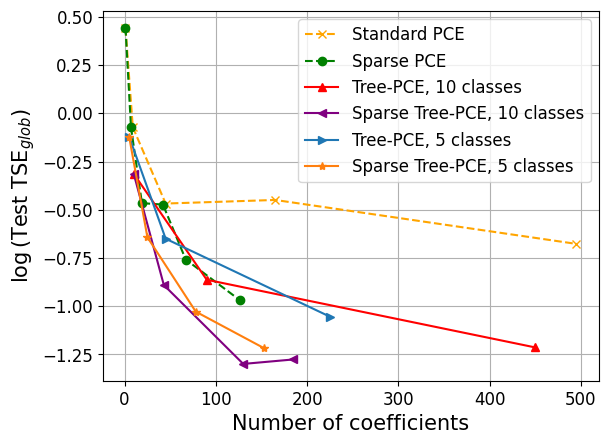}
    \caption{Complexity Comparison of the metamodels for the Yen experience}
    \label{fig:CoplexityYen}
\end{figure}

To assess computational complexity, Tree-PCE was run with fixed class counts of 5 and 10. For each configuration, various polynomial degrees were tested, and both global errors (on training and test sets) and the number of coefficients were evaluated. As shown in Figure~\ref{fig:CoplexityYen}, Tree-PCE consistently achieves higher accuracy than standard and sparse PCE for a given model complexity. Additionally, increasing the number of classes from 5 to 10 further reduces global error, as expected.

\subsection{Sensitivity analysis}

This section explores Tree-PCE’s effectiveness for global sensitivity analysis in the morphodynamic case study. Tree-PCE is well-suited for this task, offering both analytical Sobol' index computation from local coefficients and tree-based sensitivity metrics that capture the surrogate model’s hierarchical structure. Together, these methods provide a comprehensive understanding of how input uncertainties influence model behavior.

\subsubsection{Variance-based Sensitivity analysis}

\begin{figure}[H]
        \centering
        \includegraphics[width=0.45\linewidth]{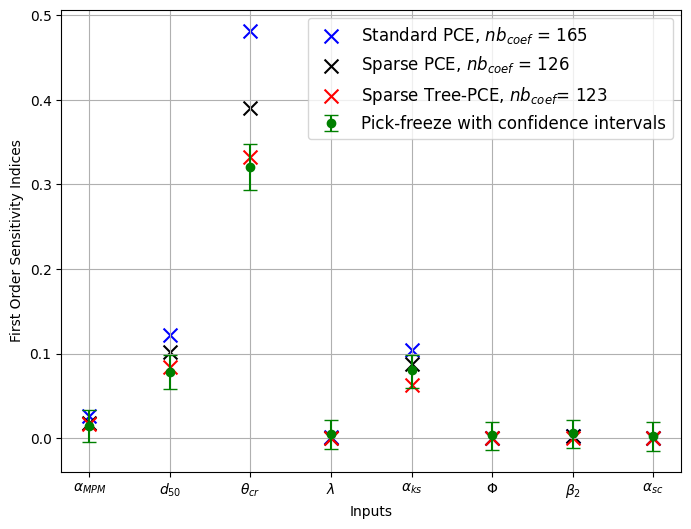}
        \includegraphics[width=0.45\linewidth]{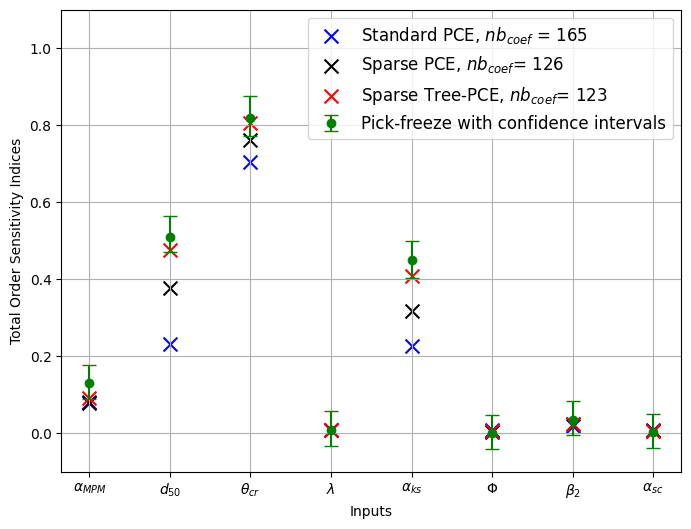}
       \caption{Sobol' indices, (left) First order, (right) Total order.}
        \label{fig:Sobolindices}
    \end{figure}

Sobol' indices were computed using three different methods to evaluate parameter sensitivity. For standard/sparse/Tree-PCE, we used $N=5000$, while the Pick-freeze method was applied with $N=200000$. For Tree-PCE, we used a local polynomial degree \( p_{\text{loc}} = 2 \), a maximum of 9 classes, and 10 evaluation points per input dimension. As shown in Figure~\ref{fig:Sobolindices}, the indices obtained with Tree-PCE closely match those from Pick-freeze’s method, while requiring a significantly less complex metamodel than sparse PCE. 

The sensitivity analysis reveals that the critical Shields parameter \( \theta_{cr} \) is the most influential input, exhibiting the highest first-order and total-order effects. It is followed by the median grain diameter \( d_{50} \) and and the skin friction correction factor \(\alpha_{ks} \), both of which contribute significantly through direct effects and interactions. The transport coefficient \( \alpha_{MPM} \) shows limited direct influence but plays a more substantial role through interactions. In contrast, the deviation angle coefficient \( \beta_2 \) has minimal impact, with negligible direct and interaction effects. Finally, the sediment angle of repose \( \Phi \), the secondary currents coefficient 
\( \alpha_{sc} \), and bed porosity \( \lambda \)  exhibit no influence, as indicated by their null Sobol' indices.

From a physical standpoint, the dominant role of the critical Shields parameter \( \theta_{cr} \)  underscores the importance of the threshold shear stress in controlling sediment mobilization within the curved channel. The influence of the median grain diameter \( d_{50} \) highlights the role of particle size in determining sediment transportability under varying flow conditions.  The significance of the skin friction parameter \(\alpha_{ks}\) reflects the impact of shear stress partitioning between grain-scale roughness and larger-scale flow structures, particularly relevant in curved geometries. Although, the transport coefficient \( \alpha_{MPM} \) has limited direct sensitivity, its role in calibrating empirical transport formulas makes it influential in combination with other parameters.  The negligible effect of the deviation angle coefficient \( \beta_2 \) suggests that sediment transport remains largely aligned with the main flow direction, despite channel curvature. Finally, the negligible influence of parameters like the sediment angle of repose \( \Phi \), the secondary currents coefficient \( \alpha_{sc} \), and bed porosity \( \lambda \) indicates that slope stability, secondary flows, and sediment packing have limited relevance under the tested conditions.

\subsubsection{Tree-PCE sensitivity measures}

\begin{figure}[H]
    \centering
    \includegraphics[width=0.6\linewidth]{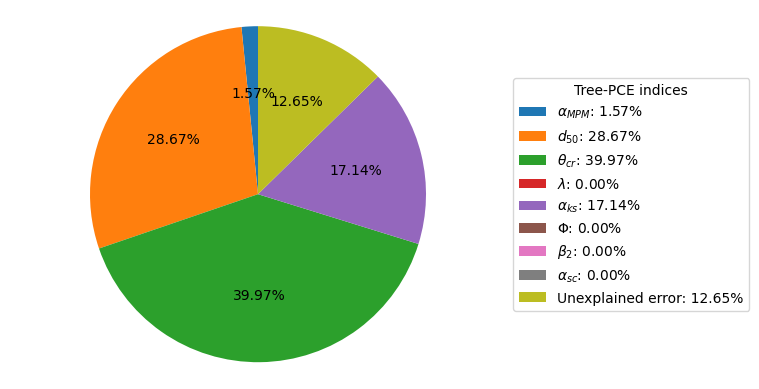}
    \caption{Tree PCE indices for the Yen experience. }
    \label{fig:enter-label}
\end{figure}

When computing the Tree-PCE indices with 27 classes, the same dominant input parameters are identified, namely the critical Shields parameter  \( \theta_{cr} \), followed by the median grain diameter  \( d_{50} \) and the skin friction correction factor \(\alpha_{ks} \). However, the differences between the three Tree-PCE sensitivity measure indices are less important than for Sobol' indices, indicating that the splits in these three directions contribute quite equally to reduce the metamodel error. Furthermore, the pie chart presented in Figure \ref{fig:enter-label} indicates that 12.65\% of the output variance remains unexplained by the Tree-PCE model. It is interesting to note that the Tree-PCE indices put in evidence the same 4 inputs as Sobol' indices, with the same ranking between them (\( \theta_{cr}, \ d_{50}, \ \alpha_{ks} \) and then $\alpha_{MPM}$). Besides, Tree-PCE indices are a byproduct of the method and do not require extra computation.

\section{Conclusion and Perspectives}

In this study, we introduced the Tree-PCE algorithm, an extension of Polynomial Chaos Expansion (PCE) specifically designed to handle the complexities of nonlinear models. By adaptively partitioning the input space and applying local PCE approximations, Tree-PCE improves the accuracy of global approximations while maintaining computational efficiency by allowing low-degree polynomial approximations in each subdomain, whereas high-degree polynomial models would be needed for global approximations. 

Tree-PCE was then argued to be well-suited to perform global sensitivity analysis. On the one hand, the orthogonality properties of local metamodels allow for the derivation of analytical formula for Sobol' indices based on the coefficients of the Tree-PCE model. The computational cost to evaluate these formulas may however reveal impractical when the dimension, the degree of local polynomial approximations or the number of classes increase. In this case, the Tree-PCE metamodel can still be used as a surrogate for the estimation of Sobol' indices by the pick-freeze method. On the other hand, the construction of the Tree-PCE metamodel provides, without any supplementary computational cost, sensitivity indices which indicate the contribution of each input variable in the variations of the output, and may be usefully employed for input ranking and screening.

The applicability of Tree-PCE was demonstrated on both a simple test model and an actual hydromorphodynamic test case, and it was demonstrated to be competitive with global models (standard and sparse PCE) as well as other local models (Stochastic Spectral Embedding).

Designing an efficient Tree-PCE metamodel requires balancing accuracy and computational cost by tuning key hyperparameters such as the number of candidate splits, splitting tolerance $\epsilon$, maximum number of subdomains, local polynomial degree, and sparsity settings. Increasing the number of subdomains enhances local adaptability while higher polynomial degrees improve approximation quality, both at the expense of computational efficiency. This trade-off is analogous to the classical balance between mesh refinement and numerical order in numerical methods. Ideally, both the number of subdomains and the polynomial degree should be minimized simultaneously to reduce complexity while maintaining accuracy. In this study, we explored various combinations of these parameters and observed that performance gains are regime-dependent. However, a systematic optimization of this trade-off remains an open perspective. To this end, optimization algorithms such as the Tree-structured Parzen Estimator (TPE), available in the open-source Python library Optuna \citep{Akiba_2019}, offer a promising avenue for automating hyperparameter tuning and identifying optimal configurations.

\subsection*{Acknowledgement}

This research is partially funded by Association Nationale de Recherche et de la Technologie (ANRT) Grant No. [CIFRE 2022/1501]. The authors gratefully acknowledge contributions from the open-source community, especially that of OpenTURNS (Open source initiative for the Treatment of Uncertainties, Risks’N Statistics). The authors also would like to address special thanks to Kamal El Kadi Abderrezzak for his support in this work.

\appendix

\section{Error normalization in the splitting procedure}
\label{app:A}
The goal of this section is to present a simple example where, in the splitting procedure described in Section~\ref{subsec_split}, using a criterion based on the coefficients of determination $R^2$ on $\mathcal{R}_1$ and $\mathcal{R}_2$ does not allow to determine the optimal splitting point (see Remark~\ref{rq:errormeasure} in particular). We consider the model $Y = G(X) = \ind{X>1/2}$, where $X$ is uniformly distributed on $[0,1]$. We take as admissible metamodels the set of models of the form
\[
    G^\mathrm{meta}(x) = a_1 \ind{x<y} + a_2 \ind{x \geq y},
\]
with 3 parameters $a_1, a_2 \in \R$ and $y \in [0,1]$ to be chosen. The global Mean Squared Error between the model and the metamodel is defined by
\[
    \Exp\left[(Y-G^\mathrm{meta}(X))^2\right] = \begin{cases}
        \displaystyle\int_0^y (0-a_1)^2 dx + \int_y^{1/2} (0-a_2)^2 dx + \int_{1/2}^1 (1-a_2)^2 dx & \text{if $y \leq 1/2$,}\\
        \displaystyle\int_0^{1/2} (0-a_1)^2 dx + \int_{1/2}^y (1-a_1)^2 dx + \int_y^1 (1-a_2)^2 dx & \text{if $y > 1/2$,}
    \end{cases}
\]
and this error admits as unique minimizer the solution $a_1=0$, $a_2=1$, and $y=1/2$. 

On the other hand, assuming for simplicity that $y > 1/2$, the $R^2$ of the metamodel
\[
    G^\mathrm{meta}_1(x) = a_1, \qquad x < y,
\]
is defined by
\[
    R^2_1 = 1 - \frac{\displaystyle \Exp\left[(Y-G^\mathrm{meta}_1(X))^2|X<y\right]}{\displaystyle \Var(Y|X<y)},
\]
which rewrites
\[
    R^2_1 = 1 - \frac{\displaystyle \Exp\left[(Y-a_1)^2|X<y\right]}{\displaystyle \Exp\left[(Y-m_y)^2|X<y\right]} = 1-\frac{\displaystyle\int_0^{1/2} (0-a_1)^2 dx + \int_{1/2}^y (1-a_1)^2 dx}{\displaystyle\int_0^{1/2} (0-m_y)^2 dx + \int_{1/2}^y (1-m_y)^2 dx},
\]
where
\[
    m_y = \Exp[Y|X<y] = 1 - \frac{1}{2y}.
\]
Similarly, the $R^2$ of the metamodel
\[
    G^\mathrm{meta}_2(x) = a_2, \qquad x \geq y,
\]
is in principle given by
\[
    R^2_2 = 1 - \frac{\displaystyle \Exp\left[(Y-G^\mathrm{meta}_2(X))^2|X \geq y\right]}{\displaystyle \Var(Y|X \geq y)},
\]
but the latter quantity is not well-defined since $\Var(Y|X \geq y)=0$. It may however be taken to be $1$ by convention. 

The point here is that, for $y$ fixed, the value of $a_1$ which maximizes $R^2_1$ is $a_1 = m_y$, for which $R^2_1 = 0$. Therefore, whatever the value of $y > 1/2$, the maximal value of $R^2_1$ over $a_1$ will always be $0$, independently from $y$. Thus, the mere knowledge of the coefficient of determination of the optimal metamodel on each subdomain $[0,y]$ and $[y,1]$ does not allow to recover the optimal value of $y$.

\section{Proof of Proposition \ref{prop:sobol-tree}}\label{app:C}
\begin{proof}
    We start with the computation of
    \[
        \Var\left(G^\mathrm{Tree-PCE}(\mathbf{X})\right) = \Exp\left[G^\mathrm{Tree-PCE}(\mathbf{X})^2\right] - \Exp\left[G^\mathrm{Tree-PCE}(\mathbf{X})\right]^2.
    \]
    On the one hand,
    \[
        \Exp\left[G^\mathrm{Tree-PCE}(\mathbf{X})\right] = \sum_{r=1}^R \sum_{\boldsymbol{\alpha} \in \mathcal{A}^r} y_{\boldsymbol{\alpha}}^r \Exp\left[\ind{\mathbf{X} \in \mathcal{R}^r} \Psi_{\boldsymbol{\alpha}}^r(\mathbf{X})\right].
    \]
    By othonormality of $(\Psi_{\boldsymbol{\alpha}}^r)_{\boldsymbol{\alpha} \in \N^d}$ in $L^2(\mathcal{R}^r,P_\mathbf{X}(\cdot|\mathcal{R}^r))$, for any $r \in \{1, \ldots, R\}$ and $\boldsymbol{\alpha} \in \mathcal{A}^r$, 
    \[
        \Exp\left[\ind{\mathbf{X} \in \mathcal{R}^r} \Psi_{\boldsymbol{\alpha}}^r(\mathbf{X})\right] = \begin{cases}
            \Pr(\mathbf{X} \in \mathcal{R}^r) & \text{if $\boldsymbol{\alpha}=\boldsymbol{0}$,}\\
            0 & \text{otherwise.}
        \end{cases}
    \]
    Therefore,
    \[
        \Exp\left[G^\mathrm{Tree-PCE}(\mathbf{X})\right] = \sum_{r=1}^R y^r_{\boldsymbol{0}} \Pr(\mathbf{X} \in \mathcal{R}^r),
    \]
    with the convention that $y^r_{\boldsymbol{0}}=0$ if $\boldsymbol{0} \not\in \mathcal{A}^r$. On the other hand,
    \begin{align*}
        \Exp\left[G^\mathrm{Tree-PCE}(\mathbf{X})^2\right] &= \sum_{r,r'=1}^R \sum_{\boldsymbol{\alpha} \in \mathcal{A}^r}\sum_{\boldsymbol{\alpha}' \in \mathcal{A}^{r'}} y_{\boldsymbol{\alpha}}^ry_{\boldsymbol{\alpha}'}^{r'} \Exp\left[\ind{\mathbf{X} \in \mathcal{R}^r \cap \mathcal{R}^{r'}} \Psi_{\boldsymbol{\alpha}}^r(\mathbf{X})\Psi_{\boldsymbol{\alpha}'}^{r'}(\mathbf{X})\right]\\
        &= \sum_{r=1}^R \sum_{\boldsymbol{\alpha} \in \mathcal{A}^r} (y_{\boldsymbol{\alpha}}^r)^2 \Pr(\mathbf{X} \in \mathcal{R}^r),
    \end{align*}
    where we have used the fact that, if $r\not=r'$, then $\Pr(\mathbf{X} \in \mathcal{R}^r \cap \mathcal{R}^{r'})=0$, and then the othonormality of $(\Psi_{\boldsymbol{\alpha}}^r)_{\boldsymbol{\alpha} \in \N^d}$ in $L^2(\mathcal{R}^r,P_\mathbf{X}(\cdot|\mathcal{R}^r))$ again.

    Let us now fix $A \in \mathcal{P}_d$ and compute
    \[
        \Var\left(\Exp\left[G^\mathrm{Tree-PCE}(\mathbf{X})|\mathbf{X}_A\right]\right) = \Exp\left[\Exp\left[G^\mathrm{Tree-PCE}(\mathbf{X})|\mathbf{X}_A\right]^2\right] - \Exp\left[G^\mathrm{Tree-PCE}(\mathbf{X})\right]^2.
    \]
    We first write
    \[
        \Exp\left[G^\mathrm{Tree-PCE}(\mathbf{X})|\mathbf{X}_A\right] = \sum_{r=1}^R \sum_{\boldsymbol{\alpha} \in \mathcal{A}^r} y_{\boldsymbol{\alpha}}^r \Exp\left[\ind{\mathbf{X} \in \mathcal{R}^r} \Psi_{\boldsymbol{\alpha}}^r(\mathbf{X})|\mathbf{X}_A\right].
    \]
    We have
    \begin{align*}
        \Exp\left[\ind{\mathbf{X} \in \mathcal{R}^r} \Psi_{\boldsymbol{\alpha}}^r(\mathbf{X})|\mathbf{X}_A\right] &= \Exp\left[\prod_{i=1}^d \ind{X_i \in \mathcal{I}^{(i),r}} \phi_{\alpha_i}^{(i),r}(X_i)|\mathbf{X}_A\right]\\
        &= \prod_{i \in A} \ind{X_i \in \mathcal{I}^{(i),r}} \phi_{\alpha_i}^{(i),r}(X_i)\prod_{i \in \overline{A}} \Exp\left[\ind{X_i \in \mathcal{I}^{(i),r}} \phi_{\alpha_i}^{(i),r}(X_i)\right],
    \end{align*}
    and
    \[
        \Exp\left[\ind{X_i \in \mathcal{I}^{(i),r}} \phi_{\alpha_i}^{(i),r}(X_i)\right] = \begin{cases}
            \Pr(X_i \in \mathcal{I}^{(i),r}) & \text{if $\alpha_i=0$,}\\
            0 & \text{otherwise.}
        \end{cases}
    \]
    As a consequence,
    \[
        \Exp\left[G^\mathrm{Tree-PCE}(\mathbf{X})|\mathbf{X}_A\right] = \sum_{r=1}^R \left(\sum_{\substack{\boldsymbol{\alpha} \in \mathcal{A}^r\\ \forall i \in \overline{A}, \alpha_i=0}} y_{\boldsymbol{\alpha}}^r \prod_{i \in A} \ind{X_i \in \mathcal{I}^{(i),r}} \phi_{\alpha_i}^{(i),r}(X_i) \right)\Pr\left(\mathbf{X}_{\overline{A}} \in \mathcal{R}^r_{\overline{A}}\right).
    \]
    Taking the square and then the expectation, we get
    \begin{align*}
        &\Exp\left[\Exp\left[G^\mathrm{Tree-PCE}(\mathbf{X})|\mathbf{X}_A\right]^2\right]\\
        &= \sum_{r,r'=1}^R \left(\sum_{\substack{\boldsymbol{\alpha}, \boldsymbol{\alpha}' \in \mathcal{A}^r\\ \forall i \in \overline{A}, \alpha_i=\alpha_i'=0}} y_{\boldsymbol{\alpha}}^r y_{\boldsymbol{\alpha}'}^{r'} \prod_{i \in A} J^{(i),r,r'}_{\alpha_i,\alpha_i'}\right)\Pr\left(\mathbf{X}_{\overline{A}} \in \mathcal{R}^r_{\overline{A}}\right)\Pr\left(\mathbf{X}_{\overline{A}} \in \mathcal{R}^{r'}_{\overline{A}}\right),
    \end{align*}
    which finally yields the claimed formula.
\end{proof}

\section{Calculation of $J^{(i),r,r'}_{\alpha_i,\alpha_i'}$}\label{app:D}

In this section, we provide more detail on the numerical evaluation of the quantities $J^{(i),r,r'}_{\alpha_i,\alpha_i'}$ defined in~\eqref{def_Js}, which write
\[
   J^{(i),r,r'}_{\alpha_i,\alpha_i'} = \Exp\left[\Phi^{(i),r}_{\alpha_i}(X_i)\Phi^{(i),r'}_{\alpha'_i}(X_i)\right] = \int_{\mathcal{I}^{(i),r} \cap \mathcal{I}^{(i),r'}} \Phi^{(i),r}_{\alpha_i}(x_i)\Phi^{(i),r'}_{\alpha'_i}(x_i) f_{X_i}(x_i) d x_i.
\]
We clearly have $J^{(i),r,r'}_{\alpha_i,\alpha_i'}=0$ when $\mathcal{I}^{(i),r} \cap \mathcal{I}^{(i),r'}$ has zero Lebesgue measure and $J^{(i),r,r'}_{\alpha_i,\alpha_i'}=\delta_{\alpha_i,\alpha'_i}$ when  $\mathcal{I}^{(i),r}= \mathcal{I}^{(i),r'}$ by using the orthonormality of the basis. However, it may happen that $\mathcal{I}^{(i),r} \not = \mathcal{I}^{(i),r'}$ and $\mathcal{I}^{(i),r} \cap \mathcal{I}^{(i),r'}$ has a positive Lebesgue measure, see for example Figure~\ref{fig:diagonal2D} where 
the left rectangle shares its right edge with many other rectangles. 
Then, to compute this quantity, we may proceed as follows. For each $i \in \{1, \ldots, d\}$, let us introduce
\[
   a^{(i)}_1 < \cdots < a^{(i)}_{L^{(i)}}
\]    
the boundaries of all the intervals $\mathcal{I}^{(i),r}$, $r=1, \ldots, R$. Given $r,r'$ and $\alpha_i, \alpha_i'$, the integral $J^{(i),r,r'}_{\alpha_i,\alpha_i'}$ can be computed as follows. First, either $\mathcal{I}^{(i),r} \cap \mathcal{I}^{(i),r'}=\emptyset$, in which case $J^{(i),r,r'}_{\alpha_i,\alpha_i'}=0$, or there exist $\ell_- < \ell_+$ such that $\mathcal{I}^{(i),r} \cap \mathcal{I}^{(i),r'} = [a^{(i)}_{\ell_-}, a^{(i)}_{\ell_+}]$, and thus
\[
   J^{(i),r,r'}_{\alpha_i,\alpha_i'} = \sum_{\ell=\ell_-}^{\ell_+-1} \int_{a^{(i)}_\ell}^{a^{(i)}_{\ell+1}} \Phi^{(i),r}_{\alpha_i}(x_i)\Phi^{(i),r'}_{\alpha'_i}(x_i) f_{X_i}(x_i) d x_i.
\]
Then, let us write
\[
   \Phi^{(i),r}_{\alpha_i}(x_i) = \sum_{\rho=0}^p \beta^{(i),r}_{\alpha_i,\rho} x_i^\rho, \qquad \Phi^{(i),r'}_{\alpha_i'}(x_i) = \sum_{\rho=0}^p \beta^{(i),r'}_{\alpha_i',\rho} x_i^\rho,
\]
so that
\[
   \Phi^{(i),r}_{\alpha_i}(x_i)\Phi^{(i),r'}_{\alpha_i'}(x_i) = \sum_{\rho=0}^{2p} B^{(i),r,r'}_{\alpha_i,\alpha_i'} x_i^\rho, \qquad B^{(i),r,r'}_{\alpha_i,\alpha_i'} = \sum_{l=0}^\rho \beta^{(i),r}_{\alpha_i,l}\beta^{(i),r'}_{\alpha_i',\rho-l}.
\]
Then, $J^{(i),r,r'}_{\alpha_i,\alpha_i'}$ rewrites
\begin{equation*}
   J^{(i),r,r'}_{\alpha_i,\alpha_i'} = \sum_{\ell=\ell_-}^{\ell_+-1} \sum_{\rho=0}^{2p} B^{(i),r,r'}_{\alpha_i,\alpha_i'} e^{(i)}_{\rho,\ell},
\end{equation*}
where, for each $\rho \in \{0, \ldots, 2p\}$ and $\ell \in \{1, \ldots, L^{(i)}-1\}$, 
\[
   e^{(i)}_{\rho,\ell} := \int_{a^{(i)}_\ell}^{a^{(i)}_{\ell+1}} x_i^\rho f_{X_i}(x_i) d x_i.
\]

To accurately estimate the integrals \(e^{(i)}_{\rho,\ell}\), one may either use the dataset $(\mathbf{X}^{(k)})_{1 \leq k \leq N}$ and introduce the estimator
\[
   \hat{e}^{(i),N}_{\rho,\ell} := \frac{1}{N}\sum_{k=1}^N (X^{(k)}_i)^\rho \ind{X^{(k)}_i \in [a^{(i)}_\ell, a^{(i)}_{\ell+1}]},
\]
in which case the $(2p+1) \times L^{(i)}$ matrix with coordinates $\hat{e}^{(i),N}_{\rho,\ell}$ can be computed with one single scan of the sample, or resort to any numerical quadrature method, since these integrals are one-dimensional and, in general, the PDF $f_{X_i}$ is assumed to be known analytically.

\bibliographystyle{plainnat}
\bibliography{references.bib}  

\begin{thebibliography}{47}
\providecommand{\natexlab}[1]{#1}
\providecommand{\url}[1]{\texttt{#1}}
\expandafter\ifx\csname urlstyle\endcsname\relax
  \providecommand{\doi}[1]{doi: #1}\else
  \providecommand{\doi}{doi: \begingroup \urlstyle{rm}\Url}\fi

\bibitem[Akiba et~al.(2019)Akiba, Sano, Yanase, Ohta, and Koyama]{Akiba_2019}
T.~Akiba, S.~Sano, T.~Yanase, T.~Ohta, and M.~Koyama.
\newblock Optuna: A {N}ext-generation {H}yperparameter {O}ptimization
  {F}ramework, 2019.
\newblock Preprint at \url{https://doi.org/10.48550/arXiv.1907.10902}.

\bibitem[Baudin et~al.(2017)Baudin, Dutfoy, Iooss, and Popelin]{Baudin2017}
Micha{\"e}l Baudin, Anne Dutfoy, Bertrand Iooss, and Anne-Laure Popelin.
\newblock {OpenTURNS: An industrial software for uncertainty quantification in
  simulation}.
\newblock \emph{Industrial \& Engineering Chemistry Research}, 56\penalty0
  (13):\penalty0 3681--3688, 2017.
\newblock \doi{10.1021/acs.iecr.6b03211}.

\bibitem[Beakawi Al-Hashemi and Al-Amoud(2021)]{beakawi2021review}
H.~M. Beakawi Al-Hashemi and O.~S.~B. Al-Amoud.
\newblock Review on granular materials, 2021.
\newblock Review on granular materials.

\bibitem[Bentley(1975)]{Bentley1975}
Jon~Louis Bentley.
\newblock Multidimensional binary search trees used for associative searching.
\newblock \emph{Communications of the ACM}, 18\penalty0 (9):\penalty0 509--517,
  1975.
\newblock \doi{10.1145/361002.361007}.

\bibitem[Blatman and Sudret(2010)]{Baltman2010}
G.~Blatman and B.~Sudret.
\newblock An adaptive algorithm to build up sparse polynomial chaos expansions
  for stochastic finite element analysis.
\newblock \emph{Probabilistic Engineering Mechanics}, 25\penalty0 (2):\penalty0
  183--197, 2010.

\bibitem[Blatman and Sudret(2011)]{blatman2009}
G.~Blatman and B.~Sudret.
\newblock Adaptive sparse polynomial chaos expansion based on least angle
  regression.
\newblock \emph{Journal of Computational Physics}, 230\penalty0 (6):\penalty0
  2345--2367, 2011.

\bibitem[Bosboom and Stive(2013)]{bosboom2013coastaldynamics}
J.~Bosboom and M.~J.~F. Stive.
\newblock Coastal dynamics: Part 1 (version 2012-0.3 and 2013-0.4).
\newblock Technical report, Delft University of Technology, 2013.

\bibitem[Breiman et~al.(1984)Breiman, Friedman, Olshen, and Stone]{Breiman1984}
Leo Breiman, Jerome Friedman, Richard Olshen, and Charles Stone.
\newblock \emph{Classification and Regression Trees}.
\newblock Wadsworth International Group, 1984.

\bibitem[Crestaux et~al.(2009)Crestaux, Le~Ma{\^\i}tre, and
  Martinez]{Crestaux2009}
Thierry Crestaux, Olivier Le~Ma{\^\i}tre, and J.-M. Martinez.
\newblock Polynomial chaos expansion for sensitivity analysis.
\newblock \emph{Reliability Engineering \& System Safety}, 94\penalty0
  (7):\penalty0 1161--1172, 2009.

\bibitem[Damblin and Ghione(2021)]{Damblin2021}
Guillaume Damblin and Alberto Ghione.
\newblock Adaptive use of replicated latin hypercube designs for computing
  sobol' sensitivity indices.
\newblock \emph{Reliability Engineering \& System Safety}, 212:\penalty0
  107507, 2021.

\bibitem[Dréau et~al.(2023)Dréau, Magnain, and Batailly]{Dreau2023}
J.~Dréau, B.~Magnain, and A.~Batailly.
\newblock Multi-element polynomial chaos expansion based on automatic
  discontinuity detection for nonlinear systems.
\newblock \emph{Journal of Sound and Vibration}, 567:\penalty0 117920, 2023.

\bibitem[Dutfoy and Lebrun(2009{\natexlab{a}})]{DutfoyLebrun2009a}
A.~Dutfoy and R.~Lebrun.
\newblock An innovating analysis of the nataf transformation from the copula
  viewpoint.
\newblock \emph{Probabilistic Engineering Mechanics}, 24:\penalty0 312--320,
  2009{\natexlab{a}}.

\bibitem[Dutfoy and Lebrun(2009{\natexlab{b}})]{DutfoyLebrun2009b}
A.~Dutfoy and R.~Lebrun.
\newblock A generalization of the nataf transformation to distributions with
  elliptical copula.
\newblock \emph{Probabilistic Engineering Mechanics}, 24 (2):\penalty0
  172--178, 2009{\natexlab{b}}.

\bibitem[El~Garroussi et~al.(2020)El~Garroussi, Ricci, De~Lozzo, Goutal, and
  Lucor]{Elgarnoussi2020}
S.~El~Garroussi, S.~Ricci, M.~De~Lozzo, N.~Goutal, and D.~Lucor.
\newblock Assessing uncertainties in flood forecasts using a mixture of
  generalized polynomial chaos expansions.
\newblock In \emph{2020 TELEMAC-MASCARET User Conference}, 2020.

\bibitem[Feinberg et~al.(2018)Feinberg, Eck, and Langtangen]{Feinberg_2018}
Jonathan Feinberg, Vinzenz~Gregor Eck, and Hans~Petter Langtangen.
\newblock Multivariate polynomial chaos expansions with dependent variables.
\newblock \emph{SIAM Journal on Scientific Computing}, 40\penalty0
  (1):\penalty0 A199--A223, 2018.
\newblock \doi{10.1137/15M1020447}.
\newblock URL \url{https://doi.org/10.1137/15M1020447}.

\bibitem[Foo et~al.(2008)Foo, Wan, and Karniadakis]{Foo2008}
Jasmine Foo, Xiaoliang Wan, and George~Em Karniadakis.
\newblock The multi-element probabilistic collocation method (me-pcm): Error
  analysis and applications.
\newblock \emph{Journal of Computational Physics}, 227\penalty0 (22):\penalty0
  9572--9595, 2008.

\bibitem[Frings et~al.(2011)Frings, Schüttrumpf, and Vollmer]{Frings2011}
R.~M. Frings, H.~Schüttrumpf, and S.~Vollmer.
\newblock Verification of porosity predictors for fluvial sand‐gravel
  deposits.
\newblock \emph{Water Resources Research}, 47\penalty0 (7), 2011.

\bibitem[Ghanem and Spanos(1991)]{Ghanem1991}
R.~Ghanem and P.~Spanos.
\newblock \emph{Stochastic finite elements – A spectral approach}.
\newblock Springer Verlag, New York, 1991.

\bibitem[Hervouet(2007)]{Hervouet2007}
J.~M. Hervouet.
\newblock \emph{Hydrodynamics of Free Surface Flows: Modelling with the Finite
  Element Method}.
\newblock John Wiley \& Sons, 2007.

\bibitem[Hoeffding(1992)]{Hoeff1992}
W.~Hoeffding.
\newblock A class of statistics with asymptotically normal distribution.
\newblock \emph{Breakthroughs in statistics: Foundations and basic theory},
  pages 308--334, 1992.

\bibitem[Kröker et~al.(2023)Kröker, Oladyshkin, and Rybak]{Kroker2023}
I.~Kröker, S.~Oladyshkin, and I.~Rybak.
\newblock Global sensitivity analysis using multi-resolution polynomial chaos
  expansion for coupled stokes–darcy flow problems.
\newblock \emph{Computational Geosciences}, 27\penalty0 (5):\penalty0 805--827,
  2023.

\bibitem[Le~Maître(2001)]{LeMaitre2001}
O.~P. Le~Maître.
\newblock Wavelet-based polynomial chaos expansions using haar basis.
\newblock \emph{SIAM Journal on Scientific Computing}, 2001.

\bibitem[Li and Mahadevan(2016)]{Li2016}
Chenzhao Li and Sankaran Mahadevan.
\newblock An efficient modularized sample-based method to estimate the
  first-order sobol' index.
\newblock \emph{Reliability Engineering \& System Safety}, 153:\penalty0
  110--121, 2016.

\bibitem[Marelli et~al.(2021)Marelli, Wagner, Lataniotis, and
  Sudret]{Marelli2021}
Simone Marelli, Patrick~R. Wagner, Christos Lataniotis, and Bruno Sudret.
\newblock Stochastic spectral embedding.
\newblock \emph{International Journal for Uncertainty Quantification},
  11\penalty0 (2):\penalty0 1--20, 2021.

\bibitem[Mendoza et~al.(2016)Mendoza, Abad, Langendoen, Wang, Tassi, and
  El~Kadi~Abderrezzak]{mendoza2016effect}
Alejandro Mendoza, Jorge~D. Abad, Eddy~J. Langendoen, Dongchen Wang, Pablo
  Tassi, and Kamal El~Kadi~Abderrezzak.
\newblock Effect of sediment transport boundary conditions on the numerical
  modeling of bed morphodynamics.
\newblock \emph{Journal of Hydraulic Engineering}, 143\penalty0 (4):\penalty0
  04016099, 2016.
\newblock \doi{10.1061/(ASCE)HY.1943-7900.0001208}.

\bibitem[Meng and Li(2021)]{Meng2021}
J.~Meng and H.~Li.
\newblock Machine learning for uq in subsurface flow.
\newblock \emph{Journal of Computational Physics}, 2021.

\bibitem[Meyer-Peter and M\"uller(1948)]{Meyer_1948}
E.~Meyer-Peter and R.~M\"uller.
\newblock Formulas for {B}ed-{L}oad transport, 1948.
\newblock Paper presented at the 2nd Meeting, IAHR, Stockholm, Sweden, 1948.

\bibitem[Nataf(1962)]{Nataf1962}
André Nataf.
\newblock Détermination des distributions dont les marges sont données.
\newblock \emph{Comptes Rendus de l'Académie des Sciences de Paris},
  225:\penalty0 42--43, 1962.

\bibitem[Oliveira et~al.(2021)Oliveira, Ballio, and Maia]{Oliveira_2021}
B.~Oliveira, F.~Ballio, and R.~Maia.
\newblock Numerical modelling-based sensitivity analysis of fluvial
  morphodynamics.
\newblock \emph{Environmental Modelling \& Software}, 135:\penalty0 104903,
  2021.
\newblock ISSN 1364-8152.
\newblock \doi{https://doi.org/10.1016/j.envsoft.2020.104903}.

\bibitem[Poëtte and Lucor(2012)]{PoetteLucor2012}
G.~Poëtte and D.~Lucor.
\newblock Non-intrusive iterative stochastic spectral representation with
  application to compressible gas dynamics.
\newblock \emph{Journal of Computational Physics}, 231\penalty0 (9):\penalty0
  3587--3609, 2012.

\bibitem[Rosenblatt(1952)]{Rosenblatt1952}
Murray Rosenblatt.
\newblock Remarks on a multivariate transformation.
\newblock \emph{Annals of Mathematical Statistics}, 23\penalty0 (3):\penalty0
  470--472, 1952.
\newblock \doi{10.1214/aoms/1177729394}.

\bibitem[Saltelli(2002)]{Saltelli2002}
A.~Saltelli.
\newblock Making best use of model evaluations to compute sensitivity indices.
\newblock \emph{Computer Physics Communications}, 145:\penalty0 280--297, 2002.

\bibitem[Soize and Ghanem(2004)]{Soize2004}
C.~Soize and R.~Ghanem.
\newblock Physical systems with random uncertainties: chaos representations
  with arbitrary probability measure.
\newblock \emph{SIAM J. Sci. Comput}, 26\penalty0 (2):\penalty0 395–410,
  2004.

\bibitem[St. and Wold(1989)]{St1989}
L.~St. and S.~Wold.
\newblock Analysis of variance (anova).
\newblock \emph{Chemometrics and Intelligent Laboratory Systems}, 6\penalty0
  (4):\penalty0 259--272, 1989.

\bibitem[Sudret(2008)]{Sudret2008}
B.~Sudret.
\newblock Global sensitivity using polynomial chaos expansion.
\newblock \emph{Reliability Engineering \& System Safety}, 2008.

\bibitem[Sudret(2007)]{Sudret2007}
Bruno Sudret.
\newblock \emph{Uncertainty propagation and sensitivity analysis in mechanical
  models. Contributions to structural reliability and stochastic spectral
  methods}.
\newblock Habilitation thesis, Blaise Pascal University, 2007.

\bibitem[Talmon et~al.(1995)Talmon, Struiksma, and van
  Mierlo]{talmon1995laboratory}
A.~M. Talmon, N.~Struiksma, and M.~C. L.~M. van Mierlo.
\newblock Laboratory measurements of the direction of sediment transport on
  transverse alluvial-bed slopes.
\newblock \emph{Journal of Hydraulic Research}, 33\penalty0 (4):\penalty0
  495--517, 1995.

\bibitem[Tassi et~al.(2023)Tassi, Benson, Delinares, Fontaine, Huybrechts, and
  Kopmann]{Gaia2023}
P.~Tassi, T.~Benson, M.~Delinares, J.~Fontaine, N.~Huybrechts, and R.~Kopmann.
\newblock Gaia - a unified framework for sediment transport and bed evolution
  in rivers, coastal seas, and transitional waters in the telemac-mascaret
  modelling system.
\newblock \emph{Environmental Modelling \& Software}, 159:\penalty0 105544,
  2023.

\bibitem[Vauchel et~al.(2022)Vauchel, Garnier, and Gomez]{Vauchel2022}
N.~Vauchel, E.~Garnier, and T.~Gomez.
\newblock Estimation of sobol indices with a multi-element polynomial chaos
  model: Application to flight dynamics.
\newblock In \emph{Proceedings of the 25ème Congrès Français de Mécanique},
  August 2022.

\bibitem[Wagner et~al.(2024)Wagner, Marelli, and Sudret]{UQdoc_21_118}
P.-R. Wagner, S.~Marelli, and B.~Sudret.
\newblock {UQLab user manual -- Stochastic Spectral Embedding}.
\newblock Technical report, Chair of Risk, Safety and Uncertainty
  Quantification, ETH Zurich, Switzerland, 2024.
\newblock Report UQLab-V2.1-118.

\bibitem[Wan and Karniadakis(2005)]{Wan2005}
Xiaoliang Wan and George~Em Karniadakis.
\newblock An adaptive multi-element generalized polynomial chaos method for
  stochastic differential equations.
\newblock \emph{Journal of Computational Physics}, 209\penalty0 (2):\penalty0
  617--642, 2005.

\bibitem[Wiener(1938)]{wiener1938homogeneous}
Norbert Wiener.
\newblock The homogeneous chaos.
\newblock \emph{American Journal of Mathematics}, 60\penalty0 (4):\penalty0
  897--936, 1938.

\bibitem[Wong and Parker(2006)]{WongParker2006}
M.~Wong and G.~Parker.
\newblock Reanalysis and correction of bed-load relation of meyer-peter and
  müller using their own database.
\newblock \emph{Journal of Hydraulic Engineering}, 132\penalty0 (11):\penalty0
  1159--1168, 2006.

\bibitem[Xiu and Karniadakis(2002)]{xiu2002wiener}
Dongbin Xiu and George~Em Karniadakis.
\newblock The {W}iener--{A}skey polynomial chaos for stochastic differential
  equations.
\newblock \emph{SIAM Journal on Scientific Computing}, 24\penalty0
  (2):\penalty0 619--644, 2002.

\bibitem[Xiu and Karniadakis(2003)]{XiKa}
Dongbin Xiu and George~Em Karniadakis.
\newblock Modeling uncertainty in flow simulations via generalized polynomial
  chaos.
\newblock \emph{Journal of Computational Physics}, 187\penalty0 (1):\penalty0
  137--167, 2003.
\newblock ISSN 0021-9991,1090-2716.
\newblock \doi{10.1016/S0021-9991(03)00092-5}.
\newblock URL \url{https://doi.org/10.1016/S0021-9991(03)00092-5}.

\bibitem[Yen and Lee(1995)]{Yen1995}
C.~L. Yen and K.~T. Lee.
\newblock Bed topography and sediment sorting in channel bend with unsteady
  flow.
\newblock \emph{Journal of Hydraulic Engineering}, 121\penalty0 (8):\penalty0
  591--599, 1995.

\bibitem[Zech and Soares-Frazão(2007)]{Zech2007}
Y.~Zech and S.~Soares-Frazão.
\newblock Dam-break flow experiments and real-case data. a database from the
  european impact research.
\newblock \emph{Journal of Hydraulic Research}, 45\penalty0 (S1):\penalty0
  5–7, 2007.

\end{thebibliography}

\end{document}